\documentclass{nic-series}

\DeclareMathAlphabet{\mathsfsl}{OT1}{cmss}{m}{sl}

\begin{document} 

%---------------------------------------------------------------------------
% Title, etc.
\title{Monte Carlo Simulation of Polymers:\\ Coarse-Grained Models}

\authortoc{Jörg Baschnagel \and Joachim Wittmer \and Hendrik Meyer}
\author{J. Baschnagel \and J. P. Wittmer \and H. Meyer}
\institute{Institut Charles Sadron,\\
           6, rue Boussingault, 67083 Strasbourg Cedex, France\\
          \email{\{baschnag, jwittmer, hmeyer\}@ics.u-strasbg.fr}
          }
\maketitle

% Avoid clipping of EPS files
\epsfclipon 
% Slanted Sans Serif fonts
\font\grg=eurm10
% Define upright \mu: This should be used for units.
\def\umu{{\hbox{\grg\char22}}}

%---------------------------------------------------------------------------
% Abstract
\begin{abstracts}
A coarse-grained simulation model eliminates microscopic degrees of freedom and represents a polymer by a simplified structure.  A priori, two classes of coarse-grained models may be distinguished:  those which are designed for a specific polymer and reflect the underlying atomistic details to some extent, and those which retain only the most basic features of a polymer chain (chain connectivity, short-range excluded-volume interactions, etc.).  In this article we mainly focus on the second class of generic polymer models, while the first class of specific coarse-grained models is only touched upon briefly.  Generic models are suited to explore general and universal properties of polymer systems, which occur particularly in the limit of long chains.  The simulation of long chains represents a challenging problem due to the large relaxation times involved.  We present some of the Monte Carlo approaches contrived to cope with this problem.  More specifically, our review contains two main parts.  One part (Sec.~\ref{sec:is}) deals with local and non-local updates of a polymer.  While local moves allow to extract information on the physical polymer dynamics from Monte Carlo simulations, the chief aim of non-local moves is to accelerate the relaxation of the polymers.  We discuss some examples for such non-local moves: the slithering-snake algorithm, the pivot algorithm, and its recently suggested variant, the double-pivot algorithm, which is particularly suited for the simulations of concentrated polymer solutions or melts.  The second part (Sec.~\ref{sec:rosenbluth}) focuses on modern Monte Carlo methods that were inspired by the Rosenbluth-Rosenbluth algorithm proposed in the 1950s to simulate self-avoiding walks.  The modern variants discussed comprise the configuration-bias Monte Carlo method, its recent extension, the recoil-growth algorithm, and the pruned-enriched Rosenbluth method, an algorithm particularly adapted to the simulation of attractively interacting polymers.  
\end{abstracts}
 
%---------------------------------------------------------------------------
% Beginning of text
%---------------------------------------------------------------------------
\section{Introduction}
\label{sec:intro}
%%%%%%%%%%%%%%%%%%%%%%%%%%%%%%%%%%%%%%%%%%%%%%%%%%%%%%%%%%%%%%%%%%%%%%%%
Polymers are macromolecules in which $N$ monomeric repeat units are connected to form long chains.\footnote{More precisely, this definition refers to ``linear homopolymers'', i.e., linear chain molecules consisting of one monomer species only. By contrast, polymer chemistry can nowadays synthesize various other topologies, such as cyclic, star- or H-polymers.  For a very commendable review on the physical chemistry of polymers see Ref.~\citen{LodgeMuthu:JPC1996}.}  Experimentally the chain length $N$ is large, typically $10^3 \lesssim N \lesssim 10^5$.  The size of a chain ($\sim\! 10^3${\AA}) thus exceeds that of a monomer ($\sim\! 1${\AA}) by several orders of magnitude.  However, contrary to granular materials,\cite{JaegerEtal:RMP1996} the chain is not so large that thermal energy\footnote{Here, we mean the thermal energy supplied at ambient temperature, i.e., $k_\text{B}T = 4.1\cdot 10^{-21}$ J for $T=300$ K.} would be unimportant.  Not at all!  Thermal energy is the important energy scale for polymers.  It provokes conformational transitions so that the polymer can assume a multitude of different configurations at ambient conditions.\footnote{Polymers are a paradigm for ``soft matter'' materials or ``complex fluids''.  Roughly speaking, ``soft matter'' consists of materials whose constituents have a mesoscopic size (microscopic scale $\sim 1${\AA} $\ll$ mesoscopic object $\sim 10^2\! - \! 10^4${\AA} $\ll$ macroscopic scale $\sim 1$mm) and for which $k_\text{B}T$ is the important energy scale (whence the softness at ambient conditions).  Examples other than polymers are colloidal suspensions, liquid crystals, or fluid membranes.\cite{Jones:Book2002}}  

Changes of the configurations occur on very different scales, ranging from the local scale of a bond to the global scale of the chain.\cite{Binder_MCMD1995,Kremer:MacroChemPhys2003}  This separation of length scales entails simplifications and difficulties.  Simplifications arise on large scales where the chain exhibits universal behavior.  That is, properties which are independent of chemical details.\cite{deGennesBook,DoiEdwards}  These properties may be studied by simplified, ``coarse-grained'' models, e.g.\ via computer simulations.  For simulations the large-scale properties, however, also give rise to a principal difficulty.  Long relaxation times are associated with large chain lengths.\cite{DoiEdwards,BinderPaul:1997,McLeish:AdvPhys2002}

The present chapter focuses on some of the Monte Carlo approaches to cope with this difficulty.  Why Monte Carlo?  Within a computational framework it appears natural to address dynamical problems via the techniques of Molecular Dynamics 
%(see Chap.~{\bf Zitat auf Mike-Allens-Einf\"uhrung-in-MD}).  
(see Ref.~\citen{allen2004}).  
A Molecular Dynamics (MD) simulation numerically integrates the equations of motion of the (polymer) system, and thereby replicates, authentically, its (classical) dynamics.  As the polymer dynamics ranges from the (fast) local motion of the monomers to (slow) large-scale rearrangements of a chain, there is a large spread in time scales.  The authenticity of MD thus carries a price:  Efficient equilibration and sampling of equilibrium properties becomes very tedious --sometimes even impossible-- for long chains.  At that point, Monte Carlo simulations may provide an alternative.  Monte Carlo moves are not bound to be local.  They can be tailored to alter large portions of a chain, thereby promising efficient equilibration.  The discussion of such moves is one of the gists of this review.      

\paragraph{Outline and Prerequisites.}  The plan of the chapter is as follows:  We begin by gathering necessary background information, both as to polymer physics (Sec.~\ref{sec:polymerphysics}) and as to the Monte Carlo method (Sec.~\ref{sec:mc}).  Then, we present the simulation models (Sec.~\ref{sec:models}), which have been used to develop and to study various Monte Carlo algorithms.  The discussion of these algorithms (Secs.~\ref{sec:is} and \ref{sec:rosenbluth}) represents the core of the chapter.  Section~\ref{sec:is} deals with local moves, allowing to study the physical polymer dynamics via Monte Carlo, and non-local moves (slithering-snake algorithm, pivot algorithm, double-pivot algorithm), aiming at speeding up the relaxation of the chains.  Section~\ref{sec:is} discusses the Rosenbluth-Rosenbluth method for simulating self-avoiding walks and some of its modern variants (pruned-enriched Rosenbluth method, configuration-bias Monte Carlo, recoil-growth algorithm).  The last section (Sec.~\ref{sec:conc}) briefly recapitulates the different methods and gives some advice when to employ which algorithm.  Finally, the appendix~\ref{app:cg} reviews a recently proposed approach to systematically derive coarse-grained models for specific polymers.  

Our presentation is based on the following prerequisites:
\begin{itemize}
\item We will restrict our attention to homopolymers, i.e., to polymers consisting of one monomer species only.  However, (some of) the algorithms discussed may also be applied e.g.\ to polymer blends or block-copolymers 
%(see Chap.~{\bf  Zitat auf Marcus-M\"ullers-Vortrag}).
(see Ref.~\citen{mueller2004}).  
\item The chains are monodisperse, i.e, $N$ is constant.     
\item We do not consider long-range (e.g., electrostatic) or specific (e.g., H-bonds) interactions between the monomers.  These interactions are treated in other chapters 
%(e.g., in Chaps.~{\bf Zitate auf Christian-Holms-Vortrag, Alan-Marks-Vortrag}).
 (e.g., see Refs.~\citen{holm2004, mark2004}).  

\item We do not treat the solvent molecules explicitly.  They are indirectly accounted for by the interactions between the monomers.  The neglect of the solvent does not affect the static properties of chains in dilute solution.  However, it does affect their physical dynamics 
%(see Chap.~{\bf Zitat auf Burkhard-D\"unwegs-Vortrag}). 
(see Ref.~\citen{duenweg2004}). 
\end{itemize}

\section{A Primer to Polymer Physics}
\label{sec:polymerphysics}
%%%%%%%%%%%%%%%%%%%%%%%%%%%%%%%%%%%%%%%%%%%%%%%%%%%%%%%%%%%%%%%%%%%%%%%%

\subsection{A Polymer in Good Solvent}  
\label{subsec:IsolatedChain}

\begin{figure}[t]
\unitlength=1mm
\begin{picture}(180,80)
\put(0,30){
\epsfysize=55mm
\epsfbox{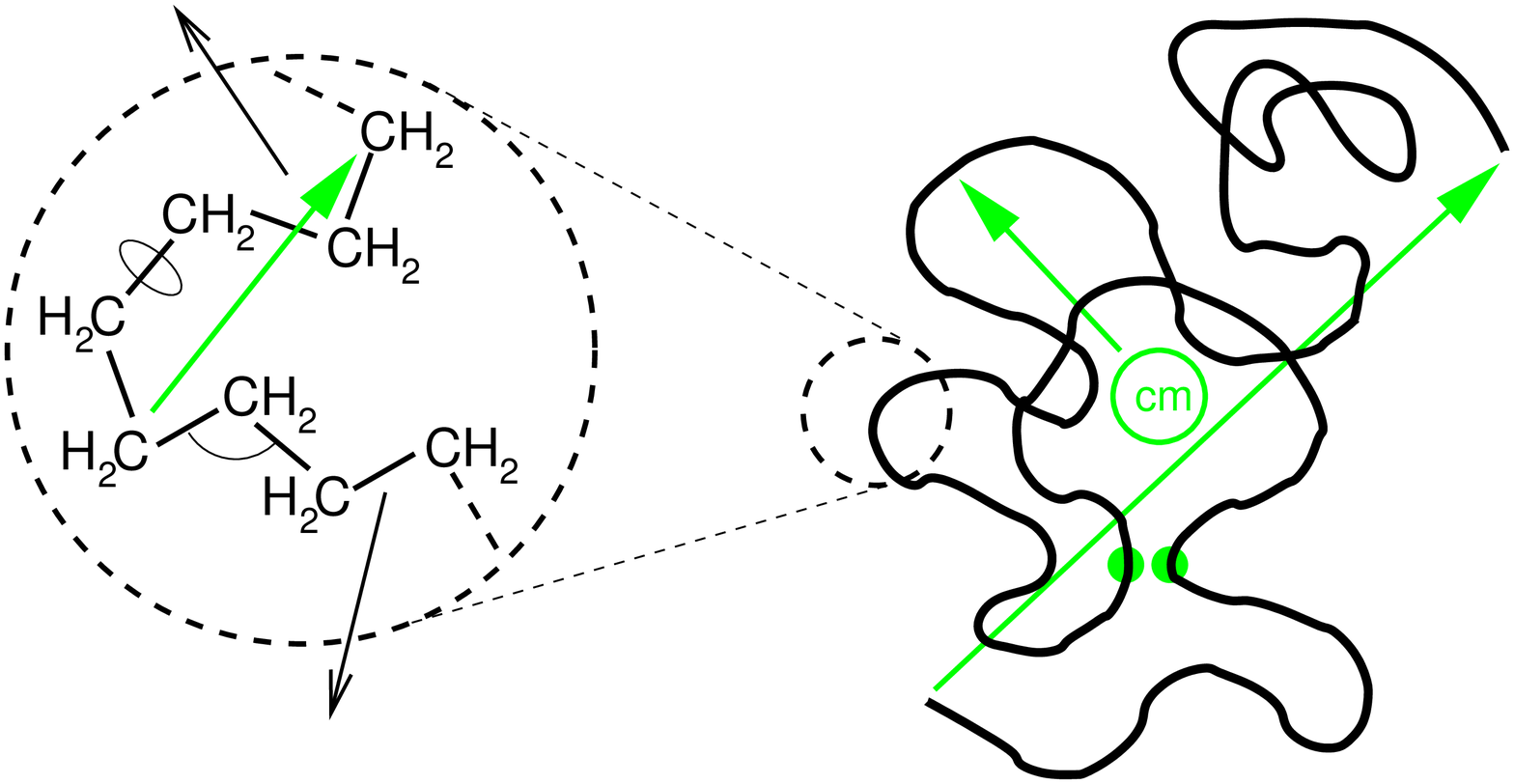}
% legend for the chemical scale
\put(-100,55){\makebox(0,0)[lb]{persistence length $\ell_\text{p}\sim 5${\AA}}}
\put(-100,-1){\makebox(0,0)[lb]{bond length $\ell_0 \sim 1${\AA}}}
\put(-92,19.5){\makebox(0,0)[lb]{$\theta$}}
\put(-100.5,38){\makebox(0,0)[lb]{$\phi$}}
\put(-125,-15){\makebox(0,0)[lb]{%
\begin{minipage}[t]{85mm}%
\begin{center}%
{\bf local properties}\\%
{\bf depend on chemistry}%
\end{center}%
\end{minipage}%
}}
% legend for the global scale
\put(-9,30){\makebox(0,0)[lb]{end-to-end distance}}
\put(-60,47){\makebox(0,0)[lb]{radius of gyration $R_\text{g}$}}
\put(-10,25){\makebox(0,0)[lb]{$N\! = \!10^4$: $R_\text{e}\!\sim \!10^3${\AA}}}
\put(-26.5,11.5){\makebox(0,0)[lb]{$v$}}
\put(-60,-22){\makebox(0,0)[lb]{%
\begin{minipage}[t]{88mm}%
\begin{center}
{\bf global properties = universal:}\\[-6mm]
\begin{align*}
\text{polymer} & \; \leftrightarrow\; \text{critical system}\\%
1/{N}    & \; \leftrightarrow\; (T-T_\text{c})/{T_\text{c}} = \tau\\%
R_\text{e} \propto R_\text{g} \sim N^\nu & \; \leftrightarrow \;\xi \sim \tau^{-\nu}
\end{align*}
\end{center}
\end{minipage}%
}}
}
\end{picture}
\vspace{-10mm}
\caption[]{Schematic illustration of polyethylene.  The local properties of the polymer depend on its microscopic degrees of freedom: the bond length $\ell$, the bond angle $\theta$, and the torsional angle $\phi$.  Because the potential of the bond length is fairly ``stiff'', $\ell$ may be kept fixed at its equilibrium value $\ell_0$ in a modeling approach.   By contrast, the potential of the torsional angle is much ``softer''.  Thus, $\phi$, which characterizes rotations about a middle C-C bond, mainly determines the local conformation of the chain.  All degrees of freedom ($\ell,\theta,\phi$) determine the intrinsic stiffness of the chain.  The stiffness reflects the persistence of orientational correlations along the backbone of the chain.  Orientational correlations decouple on the length scale of the ``persistence length'' $\ell_\text{p}$.  For typical chain lengths, $N \sim 10^4$, $\ell_\text{p}$ is much smaller than the end-to-end distance $R_\text{e}$ or the radius of gyration $R_\text{g}$.  ($R_\text{g}$ measures the average distance of a monomer from the center of mass (cm) of the chain.)  Thus, the chain appears flexible on length scales larger than $\ell_\text{p}$.  If the polymer is dissolved in a good solvent, distant monomers (filled grey circles) repel each other when they come in contact.  That is, the excluded-volume parameter $v$, measuring the effective interaction between distant monomers along the chain, is positive.  Under these conditions (i.e., linear polymer with some flexibility and repulsive monomer-monomer interactions) a correspondence between the large-scale properties of the polymer and a critical system close to its phase transition can be established:\cite{deGennesBook,Schaefer_RGBook} $1/N$ may be identified with the reduced distance, $\tau$, to the critical temperature $T_\text{c}$ of the phase transition, and $R_\text{e}$ or $R_\text{g}$ scale with $N$ as the correlation length $\xi$ of the order parameter  does with $\tau$.  $\nu$ is a universal critical exponent, often called ``Flory exponent'' in polymer science.}
\label{fig:map_pe}
\end{figure}

To substantiate the remarks of the introduction about the large-scale properties of polymers let us consider a specific example, a dilute solution of polyethylene.  Polyethylene consists of $\text{CH}_2$-monomers which are joined to form a linear polymer (Fig.~\ref{fig:map_pe}).  A configuration of the chain may be specified by the positions of the monomers\footnote{Here, we adopt a description in terms of a so-called ``united atom model''.  The united atom model represents a $\text{CH}_2$-group by a single, spherical interaction site and does not distinguish between inner ($\text{CH}_2$) and end monomers ($\text{CH}_3$).\cite{SmithEtal:ChemPhys2000}  Furthermore, we neglect the momenta of the monomers to specify the configuration, as we assume the observables and interaction potentials to depend on positions only.} $\boldsymbol{x}=(\vec{r}_1,\ldots,\vec{r}_N)$.  Thermodynamic properties are calculated by averaging an observable $A$ over all configurations
\begin{equation}
\langle A \rangle = \frac{1}{{\cal Z}} \int \text{d}\boldsymbol{x}\,
A(\boldsymbol{x}) \exp \big [ -\beta U(\boldsymbol{x}) \big] \;.
\label{eq:Adef}
\end{equation}
\hspace*{-1.75mm} Here $\beta=k_\text{B}T$, ${\cal Z}$ is the partition function and $U(\boldsymbol{x})$ the interaction potential.  We assume that $U(\boldsymbol{x})$ can be split into two parts:\footnote{Equation~\eqref{eq:Adef} does not contain the degrees of freedom of the solvent.  They are assumed to be integrated out.  Thus, $U(\boldsymbol{x})$ is an effective potential --in fact, a free energy-- depending on the properties of the solvent.}
\begin{equation}
U(\boldsymbol{x}) = \sum_{i=1}^{N-1} \underbrace{U_0
(\vec{b}_i,\ldots,\vec{b}_{j},\ldots,\vec{b}_{i+i_\text{max}})}_{\text{
``short-range'':}\ \ell,\theta,\phi,\ldots} + 
\underbrace{U_1(\boldsymbol{x},\text{solvent})}_{\text{``long-range''}} \;,
\label{eq:Udef}
\end{equation}   
where $\vec{b}_i=\vec{r}_{i+1}-\vec{r}_{i}$ denotes the bond vector from the $i$th to the $(i+1)$th monomer.  

The first term of Eq.~\eqref{eq:Udef}, $U_0$, depends on the chemical nature of the polymer.  It comprises the potentials of the bond length $\ell$, the bond angle $\theta$, the torsional angle $\phi$, etc.\ (Fig.~\ref{fig:map_pe}).\cite{Flory:StatMech}  These potentials lead to correlations between the bond vectors $\vec{b}_i$ and $\vec{b}_j$.  Typically, the correlations are of short range: they only extend up to some bond vector $b_{j=i+i_\text{max}}$ with $i_\text{max} \ll N$.  

Although distant monomers along the backbone of the chain are thus orientationally decorrelated, they can still come close in space.  The resulting interaction is long-range along the chain backbone (Fig.~\ref{fig:map_pe}).  In Eq.~\eqref{eq:Udef}, it is accounted for by the second term $U_1$.\cite{deGennesBook,DoiEdwards}  $U_1$ depends strongly on the quality of the solvent.\footnote{In Eq.~\eqref{eq:Udef} we assume that $U_0$ is independent of the solvent quality.}  In good solvents the monomers effectively repel one another (they want to be surrounded by solvent molecules), whereas they attract each other if the solvent cannot dissolve the polymer (bad solvent).    

Due to its long-range character, one intuitively expects $U_1$ to influence the large-scale behavior of the chain more strongly than $U_0$.  A possible test of this idea is to estimate how the size of a chain scales with $N$.  Common measures of the chain size are the mean-square end-to-end distance $R_\text{e}^2$ or the radius of gyration $R^2_\text{g}$ (Fig.~\ref{fig:map_pe})
\begin{equation}
R^2_\text{e} = \Big \langle (\vec{r}_N - \vec{r}_1)^2 \Big \rangle \,,
\qquad
R^2_\text{g} = \frac{1}{N} \sum_{i=1}^N \Big \langle 
\big(\vec{r}_i - \vec{R}_\text{cm}\big)^2 \Big \rangle \;,
\label{eq:def_rerg}
\end{equation}
where $\vec{R}_\text{cm}$ is the position of the chain's center of mass.  Because $R_\text{e} \propto R_\text{g}$ we focus on $R_\text{e}$ in the sequel to illustrate the role played by $U_0$ and $U_1$.  

Let $\hat{b}_i$ denote the unit vector associated with the bond $\vec{b}_i$ of fixed length $\ell_0$ (Fig.~\ref{fig:map_pe}).  Then, quite generally, we may write $R^2_\text{e}$ as
\begin{equation}
R_\text{e}^2 = \ell_0^2 \sum_{i=1}^{N-1} \sum_{j=1}^{N-1} \langle \hat{b}_i 
\cdot \hat{b}_j \rangle 
= 2 \ell_0^2 \sum_{i=1}^{N-1}\sum_{k=0}^{N-1-i} \langle \hat{b}_i \cdot 
\hat{b}_{i+k} \rangle - (N-1) \ell_0^2 \;.
\label{eq:re2}
\end{equation}
Apparently, the large-scale behavior of $R_\text{e}$ depends on the range of orientational correlations between bond vectors.  Two cases may be distinguished:\footnote{In part, the subsequent discussion closely follows that on p.~148 of Ref.~\citen{BouchaudGeorges:1990}.}
\begin{enumerate}
\item If $\langle \hat{b}_i \cdot \hat{b}_{i+k} \rangle$ is ``short-range'', i.e., if it decays more rapidly than $1/k$ for large $k$, the second term converges in the large-$N$ limit.  Then,
\begin{equation}
R_\text{e}^2 = N\ell_0^2 \bigg [ 2 \sum_{k=0}^\infty \langle \hat{b}_1 \cdot 
\hat{b}_{1+k} \rangle -1 \bigg ] 
=:  N\ell_0^2 \bigg [ 2 \frac{\ell_\text{p}}{\ell_0} -1  \bigg ]
\qquad (N \rightarrow \infty) \;,
\label{eq:re2_1case}
\end{equation}
where we introduce the persistence length $\ell_\text{p}$ in the last term. ($\ell_\text{p}$ measures the ``persistence'' of orientational correlations along the backbone and thus the intrinsic stiffness of the chain; see Fig.~\ref{fig:map_pe}).  Equation~\eqref{eq:re2_1case} shows that short-range orientational correlations only affect the prefactor --they renormalize the bond length to $b=\ell_0 [2(\ell_\text{p}/\ell_0) - 1]^{1/2}$ ($b$ is called ``effective bond length'' \cite{DoiEdwards})-- but they do not  change the scaling of $R_\text{e}$ with $N$.  The scaling is always ``random-walk-like'':\footnote{By the term ``random-walk-like'' we mean the diffusional motion of a Brownian particle.  This motion can be thought of as resulting from the addition of many small displacements in random directions so that the overall mean-square displacement of the particle in time $t$, $R^2(t)=\langle [\vec{r}(t) - \vec{r}(0)]^2\rangle$, scales as $R^2 \sim t$.  This allows for the following identifications in regard to polymer physics: $R \leftrightarrow R_\text{e}$ and $t \leftrightarrow N$.}  $R_\text{e} \sim N^{1/2}$.\cite{DoiEdwards} In polymer science, a chain exhibiting this random-walk-like behavior is commonly referred to as an ``ideal chain''.  

Of course, the finite-range correlations, assumed for $U_0$ in Eq.~\eqref{eq:Udef}, are also of short range.  Thus, provided $U_1=0$, the end-to-end distance of a (long) chain is given by $R_\text{e}=bN^{1/2}$, irrespective of the precise form of $U_0$.  The microscopic degrees of freedom, $\ell,\theta,\phi$, determine the prefactor, the effective bond length $b$, but not the scaling with $N$.  Therefore, if we are interested in studying large-scale properties, we can replace a chemically realistic model for polyethylene by a much simpler ``coarse-grained model'', which is microscopically unrealistic, but correctly reproduces the large-$N$ behavior.  An example for such a coarse-grained model is a ``bead-spring model'', where $N$ effective monomers (``beads'') are connected by harmonic springs of average length $b$ (Fig.~\ref{fig:BeadSpring}).
\begin{figure}[t]
\unitlength=1mm
\begin{picture}(180,50)
\put(10,20){
\epsfysize=30mm
\epsfbox{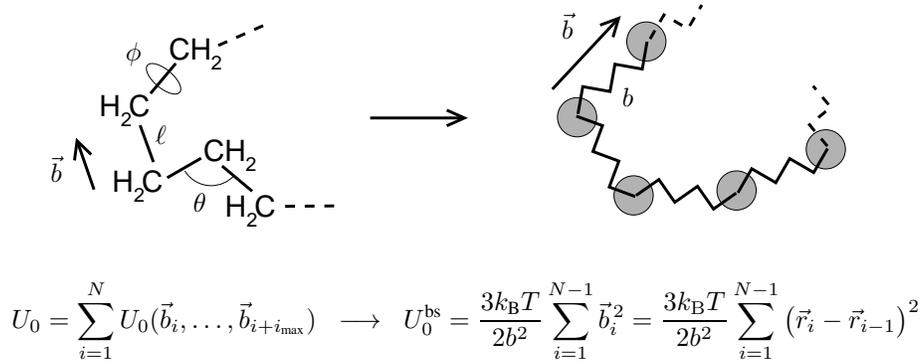}
% legend for the chemical scale and the bead-spring model
\put(-92,11){\makebox(0,0)[lb]{$\ell$}}
\put(-106,6){\makebox(0,0)[lb]{$\vec{b}$}}
\put(-87,2){\makebox(0,0)[lb]{$\theta$}}
\put(-96,21.5){\makebox(0,0)[lb]{$\phi$}}
\put(-30,16){\makebox(0,0)[lb]{$b$}}
\put(-38,25){\makebox(0,0)[lb]{$\vec{b}$}}
% place the equations
\put(-114,-18){\makebox(0,0)[lb]{%
\begin{minipage}[t]{\textwidth}%
\begin{center}
\begin{equation*}
U_0=\sum_{i=1}^N U_0(\vec{b}_i,\ldots,\vec{b}_{i+i_\text{max}}) 
\; \; \longrightarrow \; \; 
U_0^\text{bs} = \frac{3 k_\text{B}T}{2b^2} \sum_{i=1}^{N-1} \vec{b}_i^{\,2}
= \frac{3k_\mathrm{B}T}{2b^2} \sum_{i=1}^{N-1} \big
(\vec{r}_i - \vec{r}_{i-1} \big)^2
\end{equation*}
\end{center}
\end{minipage}%
}}
}
\end{picture}
\vspace{-5mm}
\caption[]{From a chemically realistic model to a coarse-grained bead-spring model.  Local properties of the realistic model are determined by its microscopic degrees of freedom: $\ell$, $\theta$, and $\phi$.  On the global level of the chain, however, the influence of the microscopic degrees of freedom can be lumped into one parameter, the effective bond length $b$. The microscopic degrees of freedom do not determine the scaling of the end-to-end distance, which, under the sole effect of $U_0$, is given by $R_\text{e}=bN^{1/2}$ (``ideal chain'').  This behavior may be recovered from Eq.~\eqref{eq:Adef} when calculating $R_\text{e}$ with the potential $U_0^\text{bs}$ of a coarse-grained bead-spring model.  This model identifies the monomers with spherical ``beads'' which are bound to one another by harmonic springs with force constant $3k_\text{B}T/b^2$. (This bead-spring model is often called ``Gaussian chain'' model in the polymer literature.\cite{DoiEdwards})}
\label{fig:BeadSpring}
\end{figure}
\item However, if $\langle \hat{b}_i \cdot \hat{b}_{i+k} \rangle$ decays as $1/k$ or more slowly (as $1/k^y$ with $y<1$) due to long-range correlations, the scaling behavior of $R_\text{e}^2$ is changed.  Instead of $R^2_\text{e} \sim N$ we find
\begin{equation}
R^2_\text{e} \sim N\ell_0^2 \int^N\!\! \text{d}k \, 
\langle \hat{b}(k) \cdot \hat{b}(0) \rangle
\sim \left\{
\begin{array}{ll}
N^{2-y} & \; (y<1) \;,\\
N\ln N  & \; (y=1) \;.
\end{array}
\right.
\label{eq:re2_2case}
\end{equation}
Thus, long-range correlations lead to a ``swelling'' of the chain size with respect to a pure random walk. 

Such long-range correlations are embodied in the potential $U_1$ in Eq.~\eqref{eq:Udef}.  For a polymer in a good solvent a swelling of the chain dimension is physically reasonable.  As soon as two (distant) monomers come close in space, they repel each other.  On the level of the coarse-grained bead-spring model we can incorporate this repulsive interaction by writing $U_1$ as (see e.g.\ Ref.~\citen{DoiEdwards} or the lucid discussion on pp.~16--20 of Ref.~\citen{Schaefer_RGBook})
\begin{equation}
U_1(\vec{r}^{\,N}) = \int \text{d}^3 \vec{r} \ \frac{1}{2} k_\text{B} T v 
\rho(\vec{r})^2 
\quad \mbox{with} \quad 
\rho(\vec{r}) = \sum_{i=1}^N \delta \big(\vec{r} - \vec{r}_i\big) \;.
\label{eq:U1_gs}
\end{equation}
Here, $\rho(\vec{r})$ is the monomer density at point $\vec{r}$ and $v$ ($>\!\!0$) is the excluded-volume parameter.  $v$ measures the strength of the repulsion of a binary contact between two beads.  Because a binary contact occurs with probability $\rho(\vec{r})^2$, Eq.~\eqref{eq:U1_gs} expresses the total energy penalty resulting from the repulsive contacts of all beads in the chain. 
\end{enumerate}
From the previous discussion of $U_0$ and $U_1$ the following conclusion may be drawn:  When focusing on the large-scale properties of linear polymers with some flexibility and predominantly repulsive interactions we may forego a microscopic description in favor of a coarse-grained model.  An example is the bead-spring model introduced above (Fig.~\ref{fig:BeadSpring}), which is characterized by two parameters, $b$ and $v$.  Another possibility is a self-avoiding walk (SAW) on a (hyper-cubic) lattice.  That is, a random walk which is not allowed to visit an already occupied lattice site again (see Sec.~\ref{subsec:lattice}).  The replacement ``microscopic model $\rightarrow$ SAW'' is permissible because a linear polymer in good solvent can be shown to correspond to a critical system which undergoes a phase transition for $N\rightarrow \infty$ (Fig.~\ref{fig:map_pe}).  It belongs to the universality class of the $n$-vector model in the limit $n\rightarrow 0$.\cite{deGennesBook,Schaefer_RGBook}  This implies that the large-$N$ behavior is determined by critical exponents.  For instance,
\begin{equation}
R_\text{g} \propto R_\text{e} = b N^\nu \quad \mbox{or} \quad
{\cal Z} \sim \mu^N N^{\gamma-1} \qquad (N\rightarrow \infty) \;,
\label{eq:rerg_Z}
\end{equation}
where the partition function ${\cal Z}$ counts the number of $N$-step SAW's starting at the origin and ending anywhere.  The connectivity constant $\mu$ and the bond length $b$ are non-universal.  They depend on the polymer and the external conditions (temperature, solvent, etc.).  By contrast, the critical exponents $\nu$ and $\gamma$ are universal.  They only depend on the dimension of space.\footnote{In the course of the research on critical phenomena it has become clear that all systems with short-range, isotropic interactions, the same dimension of space $d$, and the same dimensionality $n$ of the order parameter ($n=1$: scalar, $n\geq 2$: $n$-dimensional vector) have critical exponents which depend only on ($d,n$) and take the same values as those of the $n$-vector model.\cite{Fisher:1974,DombBook}}  Thus, they can be determined for all polymers by studying this (simple) model.  In fact, the currently most precise values of $\nu$ and $\gamma$ (see footnote on page~\pageref{pg:gamma}) have been obtained from high-precision Monte Carlo simulations of SAW's.\cite{CaraccioloEtal:PRE1998,PelissettoVicari:PhysRep2002}

\subsection{Phase Diagram of a Polymer Solution}  
\label{subsec:solutions_and_melts}

\begin{figure}[t]
\begin{center}
\epsfysize=85mm
\epsfbox{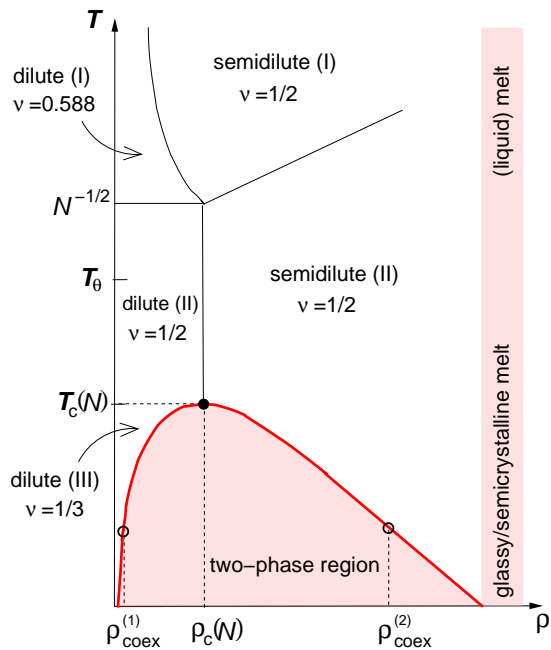}
\end{center}
\caption[]{Schematic phase diagram of flexible polymers (see Chap.~9 of Ref.~\citen{Schaefer_RGBook} or Chap.~4 of Ref.~\citen{GrosbergKhokhlovBook}).  For small monomer density $\rho$ the solution is dilute.  Three different regimes may be distinguished according to the temperature $T$: swollen chains [Eq.~\eqref{eq:rerg_Z}, $T > T_\Theta$: dilute (I)], nearly ideal chains [Eq.~\eqref{eq:nu_theta_solvent}, $T \approx T_\Theta$: dilute (II)], and collapsed chains [Eq.~\eqref{eq:nu_bad_solvent}, $T < T_\Theta$: dilute (III)].  There is an interval $\Delta T$ around the $\Theta$-point of order $\Delta T/T_\Theta \sim 1/\sqrt{N}$, where the chains are nearly ideal.  Whereas the chains may be considered as being isolated in dilute solution, they strongly overlap in the semidilute regimes.  For $T \leq T_\text{c}(N)$ phase separation in a dilute phase of collapsed chains and a semidilute solution of nearly ideal chains occurs.  If the monomer density approaches 1, we obtain a polymer melt.  At high $T$ the melt is a (viscous) liquid, whereas at low $T$ it may become a glassy\cite{BinderEtal:ProgPolPhys2003} or a semicrystalline\cite{CrystBook:ReiterSommer} solid, depending on the ability of the polymer to form ordered structures or not.}
\label{fig:phaseDiagram}
\end{figure}

The utility of coarse-grained models to investigate the statistical physics of polymer systems is not limited to the previous example.  A dilute solution in a good solvent is just one region in the phase diagram.  The phase diagram of flexible polymers is schematically shown in Fig.~\ref{fig:phaseDiagram}.
Out of the various regimes we choose to discuss two cases in more detail, a chain in another than good solvent and (high-temperature) polymer melts.  In the following sections we concentrate on those cases because novel Monte Carlo approaches have been applied to them. 

\paragraph{A Chain in a $\Theta$-Solvent or a Bad Solvent.}  To extend the discussion of the good solvent to other solvents let us reconsider Eq.~\eqref{eq:U1_gs}.  This equation corresponds to the first term of a virial expansion in the monomer density $\rho(\vec{r})$.  That is,\cite{DoiEdwards,Schaefer_RGBook}
\begin{equation}
U_1 = \int \text{d}^3 \vec{r} \, \bigg [\frac{1}{2}k_\text{B}T v 
\rho(\vec{r})^2 + \frac{1}{6}k_\text{B}T w \rho(\vec{r})^3 + \ldots \bigg ]\;.
\label{eq:U1_virial}
\end{equation}
This identifies the excluded-volume parameter $v$ with the second virial coefficient.  In general, the virial coefficients depend on temperature $T$.  The second virial coefficient vanishes at some temperature, called ``$\Theta$-temperature $T_\Theta$'' in polymer science, and behaves as $v = v_0(1 - {T_\Theta}/{T})$ close to the $\Theta$-point ($v_0=\text{const.}>0$).  This implies that we can tune the solvent quality by temperature.  In addition to the case of a good solvent ($T > T_\Theta$) two further cases may be distinguished:
\begin{enumerate}
\item {\em $\Theta$-solvent ($T=T_\Theta$)}: Since binary interactions are absent [but ternary interactions are present: $w>0$ in Eq.~\eqref{eq:U1_virial}], the polymer behaves nearly as an ideal chain:\cite{Schaefer_RGBook}
\begin{equation}
R_\text{e} \propto R_\text{g} \sim \sqrt{N} \quad (+ \ln N \;\mbox{corrections}) \;.
\label{eq:nu_theta_solvent}
\end{equation}
\item {\em Bad solvent ($T<T_\Theta$)}: Since the binary interactions are attractive, the polymer is collapsed to a dense sphere of monomers, implying that the average monomer density ${\rho}$ inside the sphere is of order 1.  Thus,
\begin{equation}
{\rho} \approx \frac{N}{R^3_\text{g}} \sim 1 \quad \Rightarrow \quad
R_\text{e} \propto R_\text{g}
\sim N^{\nu} \quad \mbox{with} \quad \nu=\frac{1}{3} \;.
\label{eq:nu_bad_solvent}
\end{equation}
The simulation of this situation is complicated because the time to equilibrate the chain and to sample equilibrium properties from many independent configurations becomes exceedingly long.  Two factors are responsible for that.  On the one hand, the local dynamics is sluggish (maybe even glass-like) due to the dense packing of monomers that strongly attract each other.  On the other hand, the polymer is entangled with itself.  Bonds cannot pass through each other.  These topological constraints may also lead to slow dynamics for long chains.  
\end{enumerate}

\begin{figure}
\begin{center}
\epsfysize=70mm
\epsfbox{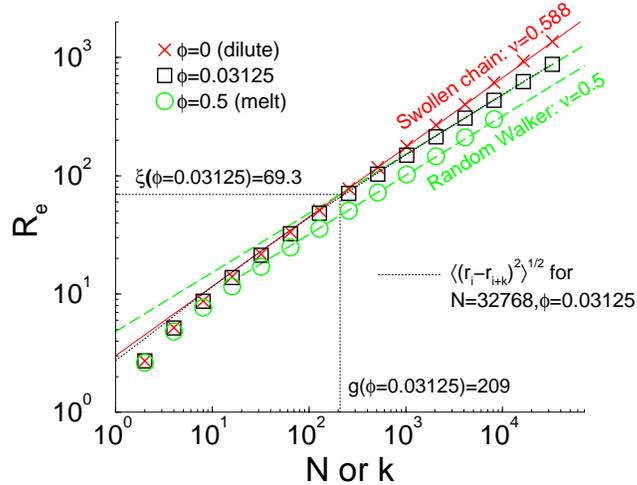}
\end{center}
\vspace{-5mm}
\caption[]{End-to-end distance $R_\text{e}$ versus chain length $N$ for the (athermal) bond-fluctuation model which will be discussed in more detail in Secs.~\ref{subsec:lattice} and Sec.~\ref{sec:is}.  Results for three volume fractions (of occupied lattice sites) are given, illustrating the dilute ($\phi=0$), the semidilute ($\phi=0.03125$) and the melt ($\phi=0.5$) limits of the schematic phase diagram (Fig.~\ref{fig:phaseDiagram}).  Using the slithering-snake algorithm (Sec.~\ref{subsubsec:snake}) it is possible to simulate chains containing up to $N=32768$ monomers for $\phi < 0.5$.  Since the slithering-snake algorithm becomes less efficient at high densities (Sec.~\ref{subsubsec:snake}), the recently proposed double-pivot algorithm, described in Sec.~\ref{subsubsec:pivot}, was harnessed to probe systems of higher densities ($\phi=0.5$).  Periodic boxes of linear size $L=512$ and containing up to $2^{22}$ monomers are required to eliminate finite-size effects.  Such periodic boundary conditions are not needed for single chains.  Here, an infinite box was used ($L(\phi=0)=\infty$).  As only excluded-volume interactions are taken into account, good solvent statistics applies in dilute solution.  The chains are swollen, as indicated by the exponent $\nu=0.588$ (solid line), which fits the data over three orders of magnitude.   In the opposite (so-called) melt limit long-range correlations appear to be screened down to small chain lengths of about $N\approx 10$ (grey dashed line).\cite{wp_entangle}  Both chain statistics are visible for the intermediate density ($\phi=0.03125$):  Small chains ($N \ll g$, $R_\text{e} \ll \xi$) are swollen (solid line) and long chains are Gaussian (dashed line).  The intercept of both lines defines the size $\xi$ of the ``excluded volume blob''\cite{deGennesBook,DoiEdwards} and the number of monomers $g$ that the blob contains.  The indicated numbers are specific to the volume fraction (and persistence length) given, but are independent of chain length.  For a given density $\xi$ corresponds to the chain size where the coils start to overlap.  Also presented in the figure is the spatial distance $\langle (\vec{r}_i-\vec{r}_{i+k})^2 \rangle^{1/2}$ along the longest chain for $\phi=0.03125$ (dotted line).  With the exception of small $N$ or $k$ (i.e., $N,k \ll 10$) this distance is, within the numerical accuracy of the data, identical to $R_\text{e}(N)$ with $N=k$.  This agreement also demonstrates that the difference between a segment of a long chain and a chain having the same length as the segment becomes irrelevant for distances larger than $\xi$. In precisely this sense the (long-range) excluded volume interactions are screened in semidilute solutions and in melts.  Mean-field descriptions become appropriate on the level of coarse-grained (Gaussian) chains of blobs.\cite{deGennesBook,DoiEdwards}}
\label{fig:rexi}
\end{figure}

\paragraph{The Size of a Chain in a Polymer Melt.}  In a good solvent a chain expands with respect to the ideal state, owing to long-range monomer-monomer repulsions.  This is peculiar to dilute solutions.  In a dense liquid of chains, a ``polymer melt'', the situation is quite different.  One can show\cite{deGennesBook,DoiEdwards,Edwards:JPA1975} that the intra-chain excluded-volume interactions are screened by the presence of the surrounding polymers.  Thus, a chain in a melt behaves on large scales as an ideal chain, implying $R_\text{e}\propto R_\text{g} \sim N^{1/2}$ (see Fig.~\ref{fig:rexi}).  

This ideality, first proposed by Flory,\cite{Flory:StatMech} appears fairly unexpected.  Some feeling why this should be so may be obtained from the following argument:  Inside the volume of a chain ($\sim\! R_\text{g}^3$) the monomer density resulting from the $N$ monomers of the chain is very small.  For ideal chains it is of order $N/R_\text{g}^3 \sim N^{-1/2}$, whereas it scales as $\sim\! N^{-0.764}$ under good solvent conditions (using Eq.~\eqref{eq:rerg_Z} and $\nu=0.588$).  We see that in dilute solution, swelling reduces the monomer density inside the chain and thus the total interaction energy [see Eq.~\eqref{eq:U1_gs}].  However, no energetic advantage may be gained in a melt because the overall monomer density is $\rho \sim 1$.  Swelling would reduce the number of intra-chain contacts, but this reduction must be compensated by inter-chain contacts to keep $\rho$ constant. Thus, a chain has to have $N^{1/2}$ contacts with other chains, which is huge in the large-$N$ limit.  This strong interpenetration of the chains suppresses the expansion of an individual chain.

\subsection{Dynamics of Polymer Melts:  Rouse and Reptation Models}
\label{subsec:PolymerDynamics}

\begin{figure}[t]
\unitlength=1mm
\begin{picture}(90,55)
\put(3,0){
\epsfysize=55mm
\epsfbox{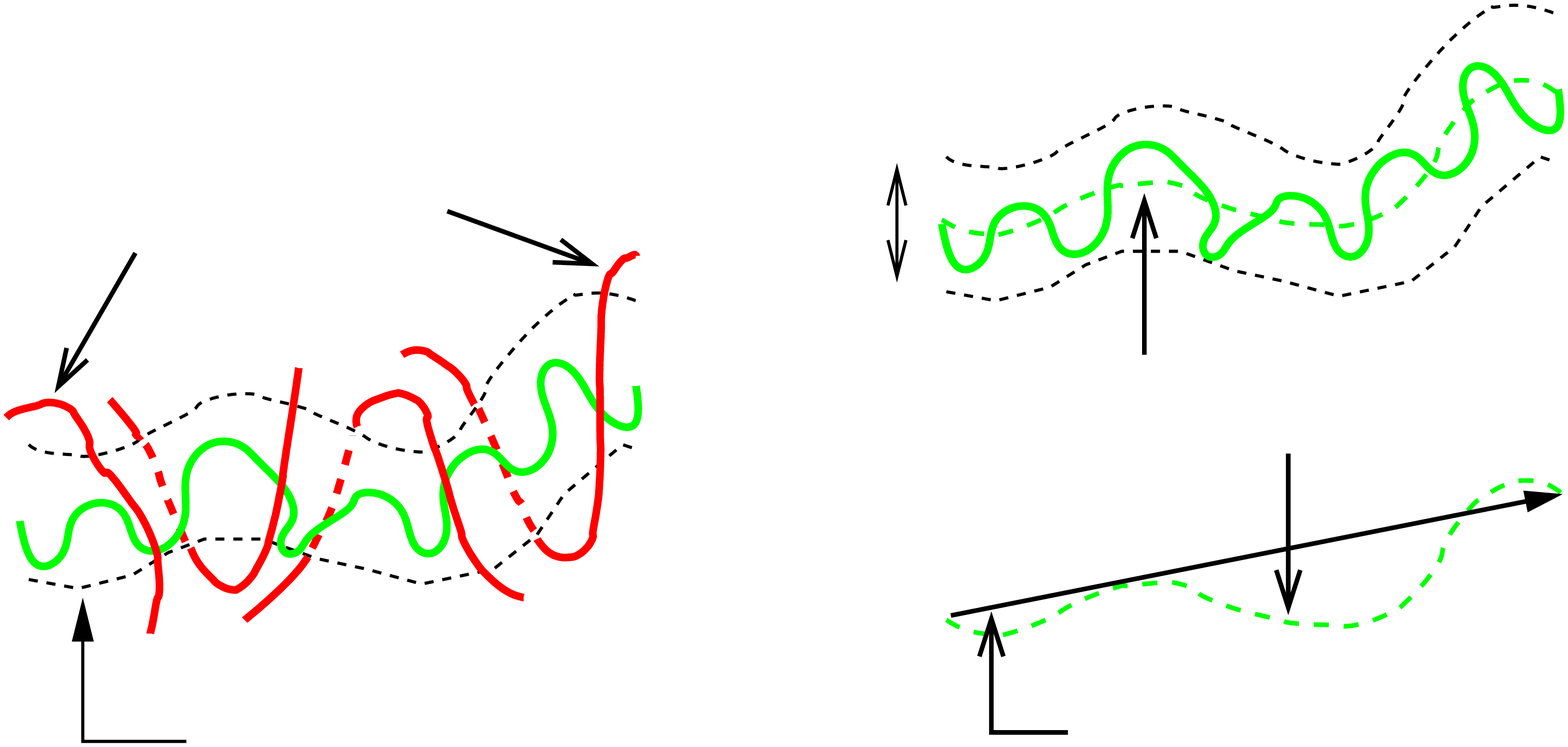}
\put(-100,-3.5){\makebox(0,0)[lb]{\shortstack{tube $\hat{=}$\\[1mm]chain topology}}}
\put(-107,38){\makebox(0,0)[lb]{neighbor chains}}
\put(-55,37){\makebox(0,0)[lb]{$d_\text{T}$}}
\put(-35,0.5){\makebox(0,0)[lb]{end-to-end distance}}
\put(-35,24){\makebox(0,0)[lb]{primitive path}}
}
\end{picture}
\vspace*{3mm}
\caption[]{Sketch of the reptation concept for the dynamics of long-chain polymer melts.\cite{DoiEdwards,McLeish:AdvPhys2002}  The chain is supposed to be enclosed in a ``tube'' formed by its neighbors.  The tube may be characterized by an axis, the primitive path.  The tube confines the motion of the enclosed chain:  It predominantly moves along the primitive path.  Perpendicular excursions are suppressed beyond the tube diameter $d_\text{T}$.  The tube diameter is larger than the effective bond length $b$: $d_\text{T}=b\sqrt{N_\text{e}}$, where the ``entanglement length'' $N_\text{e} \gg 1$.  The primitive path represents the shortest connection between the chain ends, which respects the topology imposed on the enclosed chain by the entanglements with its neighbors.  The length $L$ of the primitive path is thus larger than $R_\text{e}$, which is the shortest connection between the chain ends in space.  $L$ varies linearly with $N$: $Ld_\text{T}=R_\text{e}^2$ so that $L=d_\text{T} (N/N_\text{e})$.}
\label{fig:reptation}
\end{figure}

\paragraph{The Rouse Model.}  As a monomer in a dilute solution moves, it creates a vortex (``wave'') in the solvent.  The solvent transports the ``wave'' which is transported to other monomers of the chains so that a coupling between the motion of (distant) monomers arises 
%(see Chap.~{\bf Zitat auf Burkhard-D\"unwegs-Vortrag}).  
(see Ref.~\citen{duenweg2004}).
This long-range hydrodynamic interaction becomes screened by other chains when the concentration of the solution increases.\cite{DoiEdwards}  In a dense melt, hydrodynamic interactions are completely suppressed.  Thus, it is generally believed that the Rouse theory\cite{deGennesBook,DoiEdwards} provides a viable description for the long-time behavior of polymer dynamics in a melt, provided entanglements with other chains, giving rise to reptation dynamics,\cite{DoiEdwards,McLeish:AdvPhys2002} are not important (see Fig.~\ref{fig:reptation} and also below).\footnote{To our knowledge, there is no established derivation of the Rouse model from a microscopic theory.  For a recent attempt see Ref.~\citen{ChongFuchs:PRL2002}.}

The Rouse theory assumes the chains to be ideal and models them as a sequence of Brownian beads, connected by harmonic springs and subjected to a local random force and a local friction.\cite{deGennesBook,DoiEdwards}  This bead-spring model is characterized by two parameters:  the effective bond length $b$ and the monomer mobility $m$.  The mobility, or more precisly $1/m$, measures the time it takes a bead to diffuse over the distance $b$.  Thus, the diffusion coefficient of a monomer is proportional to $mb^2$.  As the center of mass (CM) of a chain does not experience any external force other than the antagonistic friction and random forces, the theory predicts that the CM diffuses freely at all times
\begin{equation}
g_3(t)= \Big \langle \big [\vec{R}_\text{cm}(t)-
\vec{R}_\text{cm}(0)\big]^2 \Big \rangle = 6D_Nt \;,
\label{eq:g3rouse}
\end{equation}
where $\vec{R}_\text{cm}(t)$ denotes the position of the CM at time $t$.  The diffusion coefficient of a chain is by a factor of $N$ slower than that of a monomer, i.e., 
\begin{equation}
D_N \sim \frac{mb^2}{N} \;.
\label{eq:DRouse}
\end{equation}
From Eqs.~(\ref{eq:g3rouse},\ref{eq:DRouse}) the longest relaxation time $\tau_N$ can be obtained.  Arguing that a chain is relaxed when its CM has diffused over a distance of the order of its own size, we find
\begin{equation}
g_3(\tau_N) \sim D_N \tau_N \stackrel{!}{\sim} 
R_\text{g}^2 \sim b^2 N 
\quad \Rightarrow \quad \tau_N \sim \frac{N^2}{m} \;,
\label{eq:TauRouse}
\end{equation}
where the ideality of the chain was exploited.

\paragraph{Strongly Entangled Polymers and Reptation Model.}  The single-chain picture proposed by the Rouse theory is supposed to be valid as long as entanglements with other chains do not dominate the polymer dynamics.  This is believed to be the case for short chains, for which $N$ is smaller than the entanglement length $N_\text{e}$ (Fig.~\ref{fig:reptation}).  For $N \gg N_\text{e}$ the prevailing picture is that a chain is enclosed in a temporary ``tube'' formed by its neighbors.  Entanglements force the enclosed chain to diffuse along the contour of the tube having a length of $L \sim N$ (``reptation''; see Fig.~\ref{fig:reptation}).\cite{deGennesBook,DoiEdwards}  Because the curvilinear diffusion through the tube is presumed to be Rouse-like, reptation theory predicts the relaxation time of the chain to scale with $N$ as
\begin{equation}
\tau_N \sim \frac{L^2}{(mb^2/N)} \sim N^3 
\label{eq:TauRep}
\end{equation}
so that the diffusion coefficient of the CM in space is given by [Eq.~\eqref{eq:TauRouse}]
\begin{equation}
g_3(\tau_N) \sim D_N \tau_N \stackrel{!}{\sim} 
R_\text{g}^2 \sim N \quad \Rightarrow \quad D_N \sim \frac{1}{N^2} \;.
\label{eq:DNRep}
\end{equation}
Experimentally, one finds a still stronger dependence: $\tau_N \sim N^{\approx 3.4}$ and $D_N \sim N^{\approx -2.4}$.\footnote{This exponent varies very little --if at all-- with the chemical properties of the (linear) polymer.\cite{LodgeEtal:AdvChemPhys1990,TaoEtal:Macro2000}}  

Clearly, simulation methods which attempt to model the true physical dynamics, such as Molecular Dynamics or Monte Carlo algorithms employing local random moves, must suffer from these long relaxation times.  Various alternative Monte Carlo methods have been proposed to efficiently equilibrate dense polymer melts.  We will present some of these approaches (Sec.~\ref{sec:is} and Sec.~\ref{sec:rosenbluth}).

\section{Monte Carlo Methods: A Brief Overview}
\label{sec:mc}
%%%%%%%%%%%%%%%%%%%%%%%%%%%%%%%%%%%%%%%%%%%%%%%%%%%%%%%%%%%%%%%%%%%%%%%%
In equilibrium statistical mechanics thermodynamic properties are calculated as ensemble averages over all points $\boldsymbol{x}$ in a high-dimensional configuration space $\Gamma$.\footnote{We assume that the momenta can be integrated out, since the observables only depend on the positions of the particles.}   In the canonical ensemble the average of an observable $A(\boldsymbol{x})$ is given by
\begin{equation}
\langle A \rangle 
= \int \text{d}\boldsymbol{x}\, A(\boldsymbol{x}) P_\text{eq}(\boldsymbol{x})
= \frac{1}{{\cal Z}} \int \text{d}\boldsymbol{x}\, A(\boldsymbol{x})
\exp \big [ -\beta U(\boldsymbol{x}) \big] \;.
\label{eq:Acanonic}
\end{equation}
In general, the integral cannot be solved analytically.  Monte Carlo (MC) simulations provide a numerical approach to this problem by generating a random sample of configuration-space points $\boldsymbol{x}_1,\ldots,\boldsymbol{x}_m,\ldots,\boldsymbol{x}_M$ according to some distribution $P_\text{s}(\boldsymbol{x})$.  $\langle A \rangle$ is then estimated by\cite{FrenkelSmitBook,BinderHeermann:2002,LandauBinder:2000}
\begin{equation}
\overline{A} = \frac{\sum\limits_{m=1}^M A(\boldsymbol{x}_m) 
\text{e}^{-\beta U(\boldsymbol{x}_m)}/P_\text{s}(\boldsymbol{x}_m)}
{\sum\limits_{m=1}^M \text{e}^{-\beta U(\boldsymbol{x}_m)}/
P_\text{s}(\boldsymbol{x}_m)}
=\frac{\sum\limits_{m=1}^M A(\boldsymbol{x}_m) W(\boldsymbol{x}_m)}
{\sum\limits_{m=1}^M W(\boldsymbol{x}_m)} \; ,
\label{eq:Aoverline}
\end{equation}
where we introduced the ``weight'' $W(\boldsymbol{x})=P_\text{eq}(\boldsymbol{x})/P_\text{s}(\boldsymbol{x})$.  Note that, while $\langle A \rangle$ is a number, $\overline{A}$ is still a random variable.  Whether $\overline{A}$ represents a good estimate for $\langle A \rangle$ depends on on the total number $M$ of configurations used and, for a given $M$, on the choice of $P_\text{s}(\boldsymbol{x})$.  

To see this in more detail\footnote{In part, our discussion closely follows Sec.~2.3 of Ref.~\citen{Sokal_MCMD1995}.} let us define the mean value with respect to $P_\text{s}$ by
\begin{equation}
\big \langle (\cdot) \big \rangle_\text{s} = \int \text{d}\boldsymbol{x}\, 
(\cdot) P_\text{s}(\boldsymbol{x}) \;.
\label{eq:def_ave_Ps}
\end{equation}
For large $M$, the average of $\overline{A}$ and its variance $\text{var}_\text{s}(\overline{A})$ may be estimated from the small-fluctuation approximations\cite{Sokal_MCMD1995}
\begin{align}
\bigg \langle \frac{Y}{Z} \bigg \rangle_\text{s} & \approx 
\frac{\langle Y \rangle_\text{s}}{\langle Z \rangle_\text{s}} \bigg [
1 - \frac{\langle Y Z\rangle_\text{s}}{\langle Y \rangle_\text{s}\langle Z
\rangle_\text{s}} + \frac{\langle Z^2 \rangle_\text{s}}{\langle Z
\rangle_\text{s}^2} \bigg ] \;, \label{eq:sfa_ave} \\
\text{var}_\text{s}\bigg(\frac{Y}{Z} \bigg ) & \approx
\frac{1}{\langle Z \rangle_\text{s}^2} \bigg [\langle Y^2 \rangle_\text{s} 
- 2\,\frac{\langle Y \rangle_\text{s}\langle Y Z\rangle_\text{s}}
{\langle Z \rangle_\text{s}} + \frac{\langle Y\rangle_\text{s}^2
\langle Z^2 \rangle_\text{s}}{\langle Z
\rangle_\text{s}^2} \bigg ] \;.\label{eq:sfa_var}
\end{align}
This gives\footnote{Note that $\langle W \rangle_\text{s}=1$, $\langle W A \rangle_\text{s}= \langle A \rangle$, etc.  In Eq.~\eqref{eq:sfa_A_and_varA} the $\approx$-sign means that there are corrections of $O(1/M^2)$ which we have neglected.}
\begin{equation}
\big \langle \overline{A}\big \rangle_\text{s} \approx  
\langle A \rangle - \frac 1M \Big [ \langle WA \rangle - \langle W \rangle 
\langle A \rangle \Big ] \,,
\quad
\text{var}_\text{s}\big (\overline{A} \big ) \approx \frac 1M
\Big \langle W \big (A - \langle A \rangle \big )^2 \Big \rangle \;.
\label{eq:sfa_A_and_varA}
\end{equation}
Equation~\eqref{eq:sfa_A_and_varA} shows that $\overline{A}$ provides an unbiased estimate of $\langle A \rangle$ in the limit $M \gg 1$ unless $\langle W \rangle \gg 1$, i.e., unless $P_\text{s}(\boldsymbol{x})$ is very different from $P_\text{eq}(\boldsymbol{x})$.  When $P_\text{s}(\boldsymbol{x})$ deviates considerably from $P_\text{eq}(\boldsymbol{x})$, it predominantly samples configuration-space points, which are not representative of the thermal equilibrium.  One could try to compensate this inefficient sampling by making $M$ larger and larger.  However, on the one hand this renders the simulation very time-consuming.  On the other hand, there is no guarantee that the maximum $M$ one is willing (or able) to simulate suffices to outweigh the error incurred by the inadequate choice of $P_\text{s}(\boldsymbol{x})$.

Thus, $P_\text{s}(\boldsymbol{x})$ should approximate $P_\text{eq}(\boldsymbol{x})$ as closely as possible to obtain meaningful results from MC simulations.  To this end, two approaches may be distinguished:\cite{Sokal_MCMD1995,Sokal:QFT1992}
\begin{enumerate}
\item {\em Static MC methods}:  Static methods generate a sequence of {\em statistically independent} configuration-space points from the distribution $P_\text{s}(\boldsymbol{x})$.  In this case one has to tune the algorithm cleverly so that the weights $W(\boldsymbol{x})$ do not get out of hand.  Examples how to achieve this will be discussed in Sec.~\ref{sec:rosenbluth}.  
\item {\em Dynamic MC methods}: Dynamic methods generate a sequence of {\em correlated } con\-fi\-gu\-ra\-tion-space points via some stochastic process which has $P_\text{eq}(\boldsymbol{x})$ as its unique equilibrium distribution.  In practice, this process is always taken to be a Markov process.\cite{BinderHeermann:2002,LandauBinder:2000}  The defining property of a Markov process is that it has no ``memory''.  That is, the probability for the occurrence of the future configuration $\boldsymbol{x}$ depends only on the present configuration $\boldsymbol{x'}$ and not on the other configurations that the process visited in the past.
\end{enumerate}
Dynamic MC methods have become a widely used simulation technique, to which we will also heavily refer in the following sections.  So, we provide a brief introduction here 
%(many more details may be found in Chap.~{\bf Zitat auf das Kapitel von Daan Frenkel}).  
(many more details may be found in Ref.~\citen{frenkel2004}).

Let us assume that the configuration space is discrete and that the Markov process evolves in this space in discrete time steps $\Delta t$ ($=1$).  The time evolution of this Markov chain may be characterized by the ``master equation'' for the probability $P(\boldsymbol{x},t)$ to find the system in the state $\boldsymbol{x}$ at time $t$
\begin{equation}
P(\boldsymbol{x},t+1)-P(\boldsymbol{x},t) = \sum_{\boldsymbol{x} \neq 
\boldsymbol{x'}} \Big [ w(\boldsymbol{x}|\boldsymbol{x'}) P(\boldsymbol{x'},t)
- w(\boldsymbol{x'}|\boldsymbol{x}) P(\boldsymbol{x},t) \Big ] \;.
\label{eq:master}  
\end{equation}
Here,  $w(\boldsymbol{x}|\boldsymbol{x'})$ denotes the transition probability from $\boldsymbol{x'}$ to $\boldsymbol{x}$ which is independent of time.  (In the continuous time limit ($\Delta t \rightarrow 0$) it becomes a ``transition rate'', i.e., a transition probability per unit time.)  Equation~\eqref{eq:master} expresses the balance between the flux of all other states $\boldsymbol{x'}$ towards $\boldsymbol{x}$ (first term of the rhs), leading to an increase of $P(\boldsymbol{x})$, and the flux away from $\boldsymbol{x}$ (second term of the rhs) which diminishes $P(\boldsymbol{x})$.  Note that only terms with $\boldsymbol{x} \neq \boldsymbol{x'}$ contribute.  We can rewrite Eq.~\eqref{eq:master} by including the missing term for $\boldsymbol{x} = \boldsymbol{x'}$ if we take into account that $w(\boldsymbol{x}|\boldsymbol{x'})$ is normalized.  Since a transition from $\boldsymbol{x'}$ to some state $\boldsymbol{x}$, including $\boldsymbol{x'}$, will occur with certainty, $w(\boldsymbol{x}|\boldsymbol{x'})$ satisfies   
\begin{equation}
\sum_{\boldsymbol{x'}} w(\boldsymbol{x}|\boldsymbol{x'}) = 1 \;.
\label{eq:norm_w}
\end{equation}
Inserting Eq.~\eqref{eq:norm_w} into Eq.~\eqref{eq:master} the master equation takes the following form
\begin{equation}
P(\boldsymbol{x},t+1)=\sum_{\boldsymbol{x'}} w(\boldsymbol{x}|\boldsymbol{x'}) 
P(\boldsymbol{x'},t) \;.
\label{eq:master2}
\end{equation}
For the application of these results to statistical physics it is necessary that $P(\boldsymbol{x},t)$ converges to a unique stationary distribution, irrespective of the initial configuration of the system, in the long-time limit and that this distribution is the (canonical) equilibrium distribution $P_\text{eq}(\boldsymbol{x})$.  Thus, the right-hand side of Eq.~\eqref{eq:master} must vanish for $P(\boldsymbol{x'},t)=P_\text{eq}(\boldsymbol{x'})$.  Certainly, this is the case if each term of the sum vanishes separately.  This leads to the condition of ``detailed balance'' 
%(see Chap.~{\bf Zitat auf das Kapitel von Daan Frenkel} 
%or Refs.~\citen{FrenkelSmitBook,BinderHeermann:2002,LandauBinder:2000})
(see Refs.~\citen{FrenkelSmitBook,BinderHeermann:2002,LandauBinder:2000,frenkel2004})
\begin{equation}
w(\boldsymbol{x}|\boldsymbol{x'}) P_\text{eq}(\boldsymbol{x'}) = 
w(\boldsymbol{x'}|\boldsymbol{x}) P_\text{eq}(\boldsymbol{x}) \;.
\label{eq:detbal}
\end{equation}
To exploit this condition in MC algorithms the transition probability may be split into two independent parts:  First, we propose a transition from $\boldsymbol{x'}$ to $\boldsymbol{x}$ according to some probability $P_\text{pro}(\boldsymbol{x'}\rightarrow \boldsymbol{x})$.  Then, this move will be accepted or rejected with probabilities $\text{acc}(\boldsymbol{x'}\rightarrow \boldsymbol{x})$ and $1-\text{acc}(\boldsymbol{x'}\rightarrow \boldsymbol{x})$, respectively.  So, we have
\begin{equation}
\frac{w(\boldsymbol{x}|\boldsymbol{x'})}{w(\boldsymbol{x'}|\boldsymbol{x})}
=\frac{P_\text{pro}(\boldsymbol{x'}\rightarrow \boldsymbol{x}) \,
\text{acc}(\boldsymbol{x'}\rightarrow \boldsymbol{x})}
{P_\text{pro}(\boldsymbol{x}\rightarrow \boldsymbol{x'}) \,
\text{acc}(\boldsymbol{x}\rightarrow \boldsymbol{x'})}
=\text{e}^{-\beta\left[U(\boldsymbol{x}) - U(\boldsymbol{x'})\right]} \;.
\label{eq:detbal2}
\end{equation}
To solve this equation for $\text{acc}(\boldsymbol{x'}\rightarrow \boldsymbol{x})$ we set
\begin{equation}
\text{acc}(\boldsymbol{x'}\rightarrow \boldsymbol{x}) = 
F \bigg(\frac{P_\text{pro}(\boldsymbol{x}\rightarrow \boldsymbol{x'})
\,\text{e}^{-\beta U(\boldsymbol{x})}}{P_\text{pro}(\boldsymbol{x'}\rightarrow 
\boldsymbol{x})\,\text{e}^{-\beta U(\boldsymbol{x'})}}\bigg) \;.
\label{eq:defF}
\end{equation}
From Eq.~\eqref{eq:detbal2} we see that the function $F(x)$ satisfies $F(x)/F(1/x) = x$.  One solution to this equation was proposed by Metropolis {\em et al.}:\cite{Metropolis:JCP1953} $F(x)=\min(1,x)$.  This leads to the ``Metropolis criterion'' for the acceptance probability 
\begin{equation}
\text{acc}(\boldsymbol{x'}\rightarrow \boldsymbol{x}) 
= \min \bigg (1,\frac{P_\text{pro}(\boldsymbol{x}\rightarrow \boldsymbol{x'})}
{P_\text{pro}(\boldsymbol{x'}\rightarrow \boldsymbol{x})}\,
\text{e}^{-\beta\left[U(\boldsymbol{x}) - U(\boldsymbol{x'})\right]} \bigg)\;.
\label{eq:metropolis}
\end{equation} 
The Metropolis criterion is the core of essentially all dynamic MC algorithms.  It embodies detailed balance which guarantees that the simulation, irrespective of the initial configuration, converges to the canonical equilibrium distribution, provided the set of chosen Monte Carlo moves leads to ergodic sampling.\footnote{By ``ergodic'' sampling we mean that the probability of finding the system in configuration $\boldsymbol{x}$, starting from some state $\boldsymbol{x'}$ (including $\boldsymbol{x}$), is non-zero for all $\boldsymbol{x}$ after a sufficiently long time.\cite{NarayanYoung:PRE2001}  This definition is a bit dangerous because it conflicts with others in the literature.  For instance, in mathematical texts on Markov chains (= discrete-time Markov processes with a discrete configuration space) our definition rather corresponds to an ``irreducible and aperiodic'' chain (there, ``ergodic'' is a synonym for ``irreducible'').\cite{Sokal_MCMD1995,Sokal:QFT1992,FellerBook}  In Ref.~\citen{ManoDeem:JCP1999} our definition would be termed ``regular sampling''.}    

\paragraph{Detailed Balance versus Stationarity.}  Detailed balance is an important, but very strict criterion.  Less stringent is the condition of stationarity [Eq.~\eqref{eq:master2}] 
\begin{equation}
P_\text{eq}(\boldsymbol{x})=\sum_{\boldsymbol{x'}} 
w(\boldsymbol{x}|\boldsymbol{x'}) P_\text{eq}(\boldsymbol{x'}) \;,
\label{eq:master2_eq}
\end{equation}
implying that $P_\text{eq}(\boldsymbol{x})$ remains invariant under the Markov dynamics.  Stationarity in conjunction with the ergodicity of chosen set of MC moves ensures a valid simulation.\cite{NarayanYoung:PRE2001,ManoDeem:JCP1999}  

In practice, this milder condition may be important.  Imagine that we want to update a polymer chain consisting of $N$ monomers and that each monomer can be displaced in $N_\text{dis}$ directions.  One possibility is to select a monomer and a direction randomly.  Thus, $P_\text{pro}(\boldsymbol{x'}\rightarrow \boldsymbol{x})=1/(NN_\text{dis})=P_\text{pro}(\boldsymbol{x}\rightarrow \boldsymbol{x'})$.  This procedure obeys detailed balance:  In the next move the same monomer and the reverse displacement may be chosen with the same a priori probability.  On the other hand, one could also attempt to move one monomer after the other, proceeding regularly from monomer 1 to monomer $N$.  This sequential updating scheme violates detailed balance:  The next step never selects again the monomer whose displacement has just been attempted.  So, the probability for the reverse move is zero.  

However, sequential updating is a valid scheme if the individual steps obey detailed balance\cite{ManoDeem:JCP1999} or at least stationarity.  To see that we can write the transition probability from $\boldsymbol{x'}$ to $\boldsymbol{x}$ for sequential updating as
\begin{equation}
w(\boldsymbol{x}|\boldsymbol{x'}) = 
\sum_{\boldsymbol{z}_N} \cdots \sum_{\boldsymbol{z}_2} 
\sum_{\boldsymbol{z}_1} w^{(N)}(\boldsymbol{x}|\boldsymbol{z}_N) \cdots
w^{(2)}(\boldsymbol{z}_2|\boldsymbol{z}_1) w^{(1)}(\boldsymbol{z}_1|
\boldsymbol{x'}) \;.
\label{eq:wsequential}
\end{equation}   
This means that the process passes sequentially first with probability $w^{(1)}(\boldsymbol{z}_1|\boldsymbol{x'})$ from $\boldsymbol{x'}$ to $\boldsymbol{z}_1$ by attempting to move the first monomer, then from $\boldsymbol{z}_1$ to $\boldsymbol{z}_2$ by attempting to move the second monomer, and so on until configuration $\boldsymbol{x}$ is reached.  Multiplying Eq.~\eqref{eq:wsequential} by $P_\text{eq}(\boldsymbol{x'})$ and summing over all $\boldsymbol{x'}$ we find
\begin{eqnarray}
\lefteqn{\sum_{\boldsymbol{x'}} w(\boldsymbol{x}|\boldsymbol{x'}) 
P_\text{eq}(\boldsymbol{x'})} \nonumber \\
&=& \sum_{\boldsymbol{z}_N} \cdots \sum_{\boldsymbol{z}_2} 
\sum_{\boldsymbol{z}_1} w^{(N)}(\boldsymbol{x}|\boldsymbol{z}_N) \cdots
w^{(2)}(\boldsymbol{z}_2|\boldsymbol{z}_1) \underbrace{\sum\nolimits_{\boldsymbol{x'}}
w^{(1)}(\boldsymbol{z}_1| \boldsymbol{x'})P_\text{eq}(\boldsymbol{x'})}_
{=\,P_\text{eq}(\boldsymbol{z}_1)} \nonumber \\
&=& \ldots = P_\text{eq}(\boldsymbol{x}) \;,
\label{eq:wseq_stat}
\end{eqnarray}
i.e., sequential sampling preserves the stationarity of the equilibrium distribution.  Thus, it is a correct simulation procedure.  This conclusion is important for a variety of MC methods which perform different trial moves in a fixed order.

\section{Some Coarse-Grained Simulation Models}
\label{sec:models}
%%%%%%%%%%%%%%%%%%%%%%%%%%%%%%%%%%%%%%%%%%%%%%%%%%%%%%%%%%%%%%%%%%%%%%%%
In Sec.~\ref{sec:intro} we introduced the term ``coarse-grained model''.  This was defined as a model which associates a group of chemical monomers with a ``bead'' (effective monomer) in order to eliminate microscopic degrees of freedom (bond length vibrations, etc.).  Here, we refine our definition and distinguish between two types of coarse-grained models:
\begin{enumerate}
\item  {\em The coarse-grained model is derived from a specific polymer.}  In practice, this usually implies that the properties of the model (potential parameters, density, etc.) have to be adjusted to results from atomistic simulations of the polymer under consideration (see Appendix~\ref{app:cg} for an example).  The incentive to devise such models rests upon the fact that they may be simulated much more efficiently than their atomistic counterpart.  Thus, it is tempting to split the simulation into two levels:  First, one uses the coarse-grained model for equilibration and for the determination of large-scale properties.  Then, atomistic details may be reinserted to allow for a thorough comparison with experiments.  Recent attempts to perform such multi-scale approaches are described in Refs.~\citen{MullerPlathe:2002,JBETal:AdvPolySci2000} (see also Appendix~\ref{app:cg}).
\item {\em The coarse-grained model has no direct connection to any specific polymer.}  It is a generic model retaining only features common to all polymers of the same chain topology.  For (uncharged) linear polymers these features are chain connectivity, excluded-volume interactions, and, additionally, monomer-monomer attractions if one wants to simulate $\Theta$- or bad-solvent conditions (see Fig.~\ref{fig:phaseDiagram}). Many of these generic models, be it lattice or continuum models, have been introduced in the literature (see Refs.~\citen{KremerBinder1988,Binder:Review1992} for comprehensive overviews).  In the following we present those models in more detail, which will be discussed in Secs.~\ref{sec:is},\ref{sec:rosenbluth}.  
\end{enumerate}

\subsection{Lattice Models}
\label{subsec:lattice}

\begin{figure}[t]
\unitlength=1mm
\begin{picture}(126,40)
\put(0,12){
\epsfysize=25mm
\epsfbox{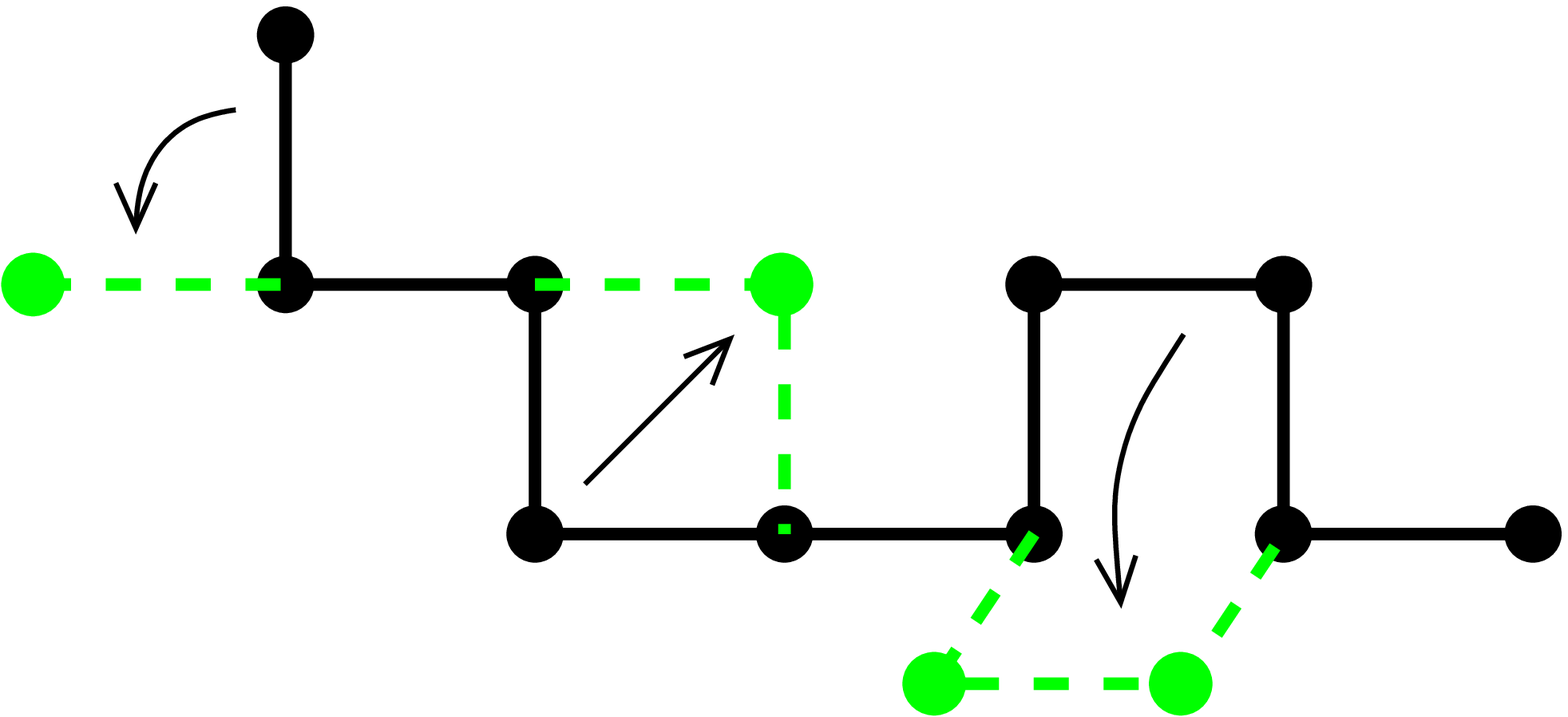}
\put(-55,10){\makebox(0,0)[lb]{{\small end rotation}}}
\put(-32,18){\makebox(0,0)[lb]{{\small kink jump}}}
\put(-24,-4){\makebox(0,0)[lb]{{\small crankshaft}}}
}
\put(67,0){
\epsfysize=45mm
\epsfbox{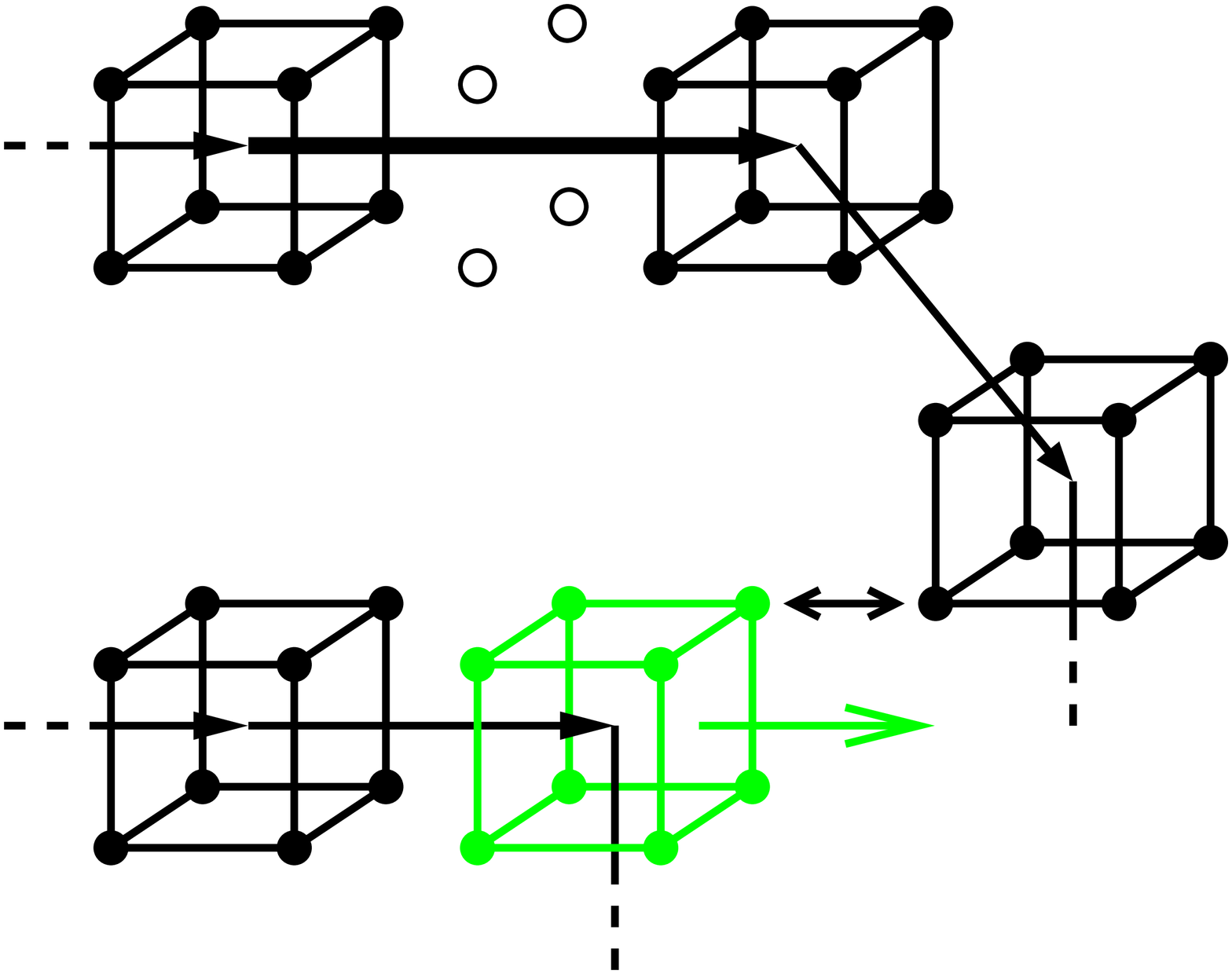}
\put(-18,6){\makebox(0,0)[lb]{{\small jump forbidden}}}
}
\end{picture}
\caption[]{{\em Left figure}:  Single-site self-avoiding walk (SAW) of chain length $N=10$ on a simple cubic lattice (solid lines and black dots).  The grey dots and the grey dashed lines indicate the moves discussed in the text: end-bond rotation, kink jumps and $90^\circ$ crankshaft motion.  {\em Right figure}:  Sketch of a possible configuration of monomers in the 3D bond-fluctuation model (BFM). (A vectorized version of the BFM algorithm can be found in Ref.~\citen{HPW:BFM}.) The bond vector $(3,0,0)$ (thick black arrow) blocks four lattice sites (marked by $\bigcirc$) that are no longer available to other monomers due to the excluded volume interaction.  This interaction also prevents the jump of the grey monomer in the direction of the large arrow ($\longrightarrow$), since the corners of the monomers, indicated by $\leftrightarrow$, would then occupy the same lattice site.}
\label{fig:latticeModels}
\end{figure}

\paragraph{The Self-Avoiding Walk.}  About 50 years ago Orr and Montroll\cite{OrrMontroll} proposed the self-avoiding walk (SAW) as a model for a linear polymer in a good solvent.   The SAW is defined on a discrete lattice, often on a square or simple cubic lattice (Fig.~\ref{fig:latticeModels}).  Each monomer occupies one lattice site, the bond length equals the lattice constant, and the bond angles are restricted by the lattice geometry and by the repulsive hard-core monomer-monomer interaction (e.g.\ $90^\circ$ and $180^\circ$ for the cubic lattice, as immediate backfolding is forbidden).  This model can be complemented by attractive interactions if, for instance, an energy gain $-\epsilon$ is associated with every occupied nearest neighbor pair.\cite{Grassberger_PERM:PRE1997}  In addition to excluded volume interactions the simulation then also has to take account of the Boltzmann factor $\exp(n_\text{nn} \epsilon/k_\text{B}T)$, where $n_\text{nn}$ is the number of nearest neighbors.

To simulate the SAW by dynamic Monte Carlo one must first decide about the elementary moves that propose a new SAW configuration $\boldsymbol{x}$ from an old one $\boldsymbol{x'}$.  The earliest suggestion\cite{VerdierStockmayer:JCP1962} comprised one-bead excitations\cite{Sokal_MCMD1995,KremerBinder1988,Binder:Review1992} (Fig.~\ref{fig:latticeModels}).  In these algorithms, one chooses a monomer at random.  If the monomer is at the chain end, the bond to its neighbor is turned to a randomly selected lattice direction.  Due to the fixed bond length an inner monomer is only mobile if its bond angle is $90^\circ$ on the square or simple cubic lattice.  In this case, one attempts a ``kink-jump'' motion, i.e., a one-bead flip to the opposite lattice site.  End-bond rotation and kink jumps are accepted according to the Metropolis criterion if the target sites are empty. 

These moves are special examples of the class of ``local $N$-conserving moves''.\cite{Sokal_MCMD1995}  Quite generally, a ``local move'' alters the configuration of a small piece of the original SAW while leaving the remaining monomers unchanged.  This definition opens the possibility to invent moves comprising more than one bead, such as two-bead or three-bead excitations.  Figure~\ref{fig:latticeModels} shows a common example, the $90^\circ$ crankshaft motion (only possible in 3D).  The crankshaft motion removes an important drawback of kink jumps.  It introduces new bond vectors, whereas a kink jump does not.  Therefore, if only end-bond rotations and kink jumps are allowed, new bond orientations have to diffuse from the ends toward the interior part of the chains.  This algorithm is not very efficient in reshuffling the bond vectors and so in preparing independent configurations.  The inclusion of crankshaft motions remedies this problem.    

However, even then a disturbing feature remains.  It has been proved that all local $N$-conserving algorithms for two- and three-dimensional SAW's are not ergodic for large $N$.\cite{MadrasSokal:JStatPhys1987}  There are dense configurations (``double cul-de-sac'' in 2D, ``knots'' in 3D; see Ref.~\citen{Sokal_MCMD1995}) which are completely frozen:  They can neither be transformed into nor reached from other configurations.  Whether this problem is serious in practice is a question that, to our knowledge, is not fully settled (see e.g.\ Ref.~\citen{KremerBinder1988} or footnote~9 of Ref.~\citen{CaraccioloEtal:JStatPhys2000}).  One can argue that, if one starts from an extended configuration --for instance, from a straight rod-- and if one is interested in high-$T$ properties only, non-ergodicity effects due to compact structures should be small.  This argument may be true for short chains,\footnote{Here, it is not clear what ``short'' really means.  For $N \lesssim 10^2$ the error incurred by using local $N$-conserving algorithms seems to be small (see Ref.~\citen{KremerBinder1988} and the footnote~9 of Ref.~\citen{CaraccioloEtal:JStatPhys2000}).} but should fail for long ones, since it has been proved that the fraction of SAW's belonging to the ergodicity class\footnote{By ``ergodicity class of a straight rod'' we mean all mutually accessible configurations, one of which is the rod.} of the straight rod is exponentially small in the large-$N$ limit.  Of course, if one is interested in low-$T$ properties, problems with non-ergodicity might be sizable, even for small chain length.\cite{CaraccioloEtal:PRE2002}

\paragraph{The Bond-Fluctuation Model.}  The bond-fluctuation model (BFM) was proposed\cite{CarmesinKremer:1988,CarmesinKremer:1990} as an alternative to a (single-site) SAW model, which retains the computational efficiency of the lattice without being plagued by severe ergodicity problems.  The key idea is to increase the size of a monomer which now occupies, instead of a single site, a whole unit cell of the lattice (e.g.\ a square for the 2D- or a cube for the 3D hyper-cubic lattice; see Fig.~\ref{fig:latticeModels}).  This enlarged monomer size has two important consequences:
\begin{enumerate}
\item A priori, many different bond vectors can occur.  This multitude is restricted by two conditions.  First, adjacent monomers may not overlap.  This limits the bond length to $\ell \geq \ell_\text{min}=2$ (in units of the lattice constant).  Second, the hard-core monomer-monomer interaction should suffice to prevent two bonds from intersecting each other in the course of the simulation.  In 2D this only imposes an upper bound on the bond length, $\ell \leq \ell_\text{max}=\sqrt{13}$,\cite{CarmesinKremer:1988,CarmesinKremer:1990} whereas in 3D, in addition to $\ell \leq \ell_\text{max}=\sqrt{10}$, some smaller bond vectors also have to be excluded.\cite{DeutschBinder:JCP1991}  The resulting sets of allowed bond vectors are:
\begin{equation}
\begin{gathered}
\{\vec{b}\}= [2,0], [2,1], [2,2], [3,0], [3,1], [3,2] \quad (\text{2D}) \;,\\
\{\vec{b}\}= [2,0,0], [2,1,0], [2,1,1], [2,2,1], [3,0,0], [3,1,0] 
\quad (\text{3D}) \;,
\label{eq:BFMbonds}
\end{gathered}
\end{equation}      
where $[\,\cdot\,]$ denotes a class of bond vectors sharing the same length, but differing in direction.  For instance, the class $[2,0]$ ($[2,0,0]$) comprises all vectors with a length of 2 and direction along the lattice axis (4 directions in 2D, 6 in 3D).  Equation~\eqref{eq:BFMbonds} gives rise to 41 bond angles in 2D \cite{LobeBaschnagel:JCP1994} and to 87 bond angles in 3D.\cite{DeutschBinder:JCP1991}  This has to be compared to 3 (2D) or 5 (3D) bond angles for the SAW model on the hypercubic lattice where a monomer is associated with a lattice site.  Due to the multitude of different bond lengths and bond angles the BFM is much closer to continuous-space behavior than the single-site lattice model\footnote{The main advantage of lattice models is their computational efficiency.  Longer length and times scales may be probed.  However later on, the results of the simulation shall be compared to theories or experiences, which ``live'' in continuous space.  So, the important question arises of how well the lattice algorithm approximates continuum properties.  A general, intuitive answer is:  The finer the lattice, i.e, the more sites are occupied by one particle, the closer the continuum limit should be realized.  Recently, this statement was made more precise by the example of monatomic fluids interacting via a Lennard-Jones or a Buckingham potential.\cite{Pana:JCP2000}}.\cite{DeutschDickman:JCP1990} 
\item Ergodicity problems are much less severe than for the single-site SAW.  For the BFM a local $N$-conserving move consists of selecting a monomer at random and of attempting a displacement by one lattice constant in a randomly chosen lattice direction.  As these local jumps\footnote{Larger jumps distances were also tested (in 3D), but found less efficient in concentrated solutions.\cite{DeutschBinder:JCP1991}} permit transitions between different vectors, the algorithm can escape from configurations where a single-site model would be frozen in.\cite{CarmesinKremer:1988}  If the attempted displacement satisfies both the bond vectors constraints of Eq.~\eqref{eq:BFMbonds} and the excluded volume interaction, the move is accepted.  Of course, it is also possible to include a finite interaction energy.  Then, the move is accepted according to the Metropolis criterion.  A possible choice\footnote{Another choice uses a discretization of the Lennard-Jones potential.\cite{Wittkop:PRE1996}} is to work with an energy $-\epsilon$ between pairs of monomers with distance $2 \leq r \leq \sqrt{6}$.  This interval comprises all neighbors which contribute to the first peak of the pair-distribution function\cite{HansenMcDonaldBook} in a dense polymer system.\cite{DeutschBinder:JCP1991}  This choice was made in studies of the $\Theta$-point\cite{WildingEtal:JCP1996} and of the phase transition in binary polymer blends 
%(see Chap.~{\bf Zitat auf Marcus-M\"ullers-Vortrag}).    
(see Ref.~\citen{mueller2004}).    

\end{enumerate}

\subsection{Continuum Models}
\label{subsec:continuum}

\begin{figure}[t]
\epsfysize=60mm
\epsfbox{LJ+FENE_pots.eps}
\begin{picture}(50,50)
\unitlength=1mm
\put(5,20){
\epsfysize=32mm
\epsffile{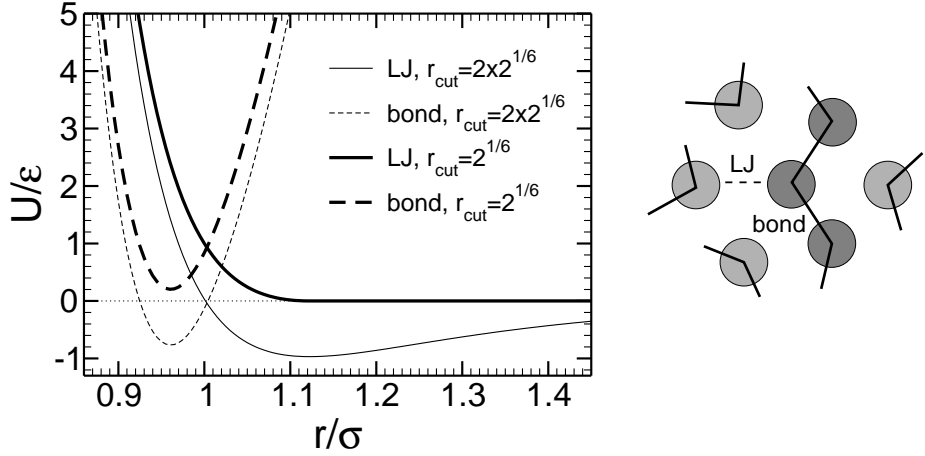}
\put(-26,17){\makebox(0,0)[lb]{{\small \sf LJ}}}
\put(-23,9){\makebox(0,0)[lb]{{\small \sf bond}}}
}
\end{picture}
\caption[]{Bond and Lennard-Jones potentials versus the distance $r$ between two monomers of the bead-spring model (for the bond potential $\vec{r}=\vec{b}$).  The bond potential results from the superposition of Eqs.~(\ref{eq:FENE},\ref{eq:LJ12-6TS}).  For both cut-off parameters $r_\text{cut}$ the bond potential was shifted by $-20$ to show it on the same scale as the LJ-potentials.  The LJ-potential with $r_\text{cut}=2^{1/6}$ is purely repulsive, whereas the potential with $r_\text{cut}=2 \times 2^{1/6}$ has an attractive minimum at $r_\text{min}=2^{1/6}$.}  
\label{fig:continuumModels}
\end{figure}

\paragraph{Two Bead-Spring Models.}  A widely used continuum model is the bead-spring model introduced by Grest and Kremer.\cite{KremerGrest:Beadspring}  In this model nearest-neighbor monomers along the backbone of the chain are bonded to each other by a FENE (finitely extendible non-linear elastic) potential
\begin{equation}
U_\text{F}(\ell) =  \left \{
\begin{array}{ll}
- \frac{1}{2}k \ell^2_\text{max} \ln \left [1 - (\ell/\ell_\text{max})^2 
\right] & \quad \ell \leq \ell_\text{max} \;, \\
\infty  & \quad \mbox{else} \;,
\end{array}
\right .
\label{eq:FENE}
\end{equation}
whereas all monomers, bonded and non-bonded ones, interact via a truncated and shifted Lennard-Jones (LJ) potential
\begin{equation}
U_\text{LJ}^\text{ts}(r) = \left \{
\begin{array}{ll}
4 \epsilon \left [(\sigma/r)^{12} - (\sigma/r)^{6} \right] + 
C (r_\text{cut}) & \quad \mbox{for} \; r \leq r_\text{cut}\;, \\
0 & \quad \mbox{else} \;,
\end{array}
\right .
\label{eq:LJ12-6TS}
\end{equation}
where $C(r_\text{cut})$ ensures that the potential vanishes at the cut-off parameter $r_\text{cut}$.  Such a cut-off is commonly employed to render the interaction short-ranged (Fig.~\ref{fig:continuumModels}).\footnote{From a computational point of view short-range interactions are convenient because the simulation can be speeded up by neighbor lists.\cite{FrenkelSmitBook,AllenTildesleyBook}  However, as the truncation ignores the contribution of the tail of the potential, the error incurred must be corrected before comparing with results for the full potential.  For instance, the truncation shifts the location of the critical point of the liquid-gas transition in a LJ-liquid (see Ref.~\citen{Smit:JCP1992} or Sec.~3.2.2 of Ref.~\citen{FrenkelSmitBook} for details).  To avoid these truncation effects some authors prefer to work with the full LJ-potential.\cite{YongEtal:JCP1996}}  The parameter $\epsilon$ defines the energy scale and $\sigma$ the length scale of the system.  That is, we set $\epsilon=\sigma=1$ (LJ units) in the following.  

For small values of the bond length the FENE potential is harmonic (``elastic behavior''), i.e., $U_\text{F}(\ell) = k\ell^2/2$ for $0 \leq \ell \ll \ell_\text{max}$, whereas the logarithmic divergence imposes $\ell < \ell_\text{max}$ (``finite extensibility'').  The parameters $\ell_\text{max}$ and $k$ have to be chosen such that the possibility of bond crossing becomes so unlikely that it never occurs.  Reference~\citen{KremerGrest:Beadspring} suggests $k=30$ and $\ell_\text{max}=1.5$ (in LJ units).  This has become a standard choice.   

The FENE potential alone does not prevent monomers from overlapping.  To realize excluded volume the LJ-interaction has to be taken into account also between bonded monomers.  The superposition of the FENE- and the LJ-potentials yields a steep effective bond potential with a minimum at $\ell_0 \simeq 0.96$ (Fig.~\ref{fig:continuumModels}).  The shape of the bond potential depends on the cut-off parameter of the LJ-interaction:
\begin{itemize}
\item If one takes $r_\text{cut} = r_\text{min}= 2^{1/6}$ ($C(r_\text{cut})=1$), i.e., as the minimum of the LJ-potential, the monomer-monomer interaction becomes purely repulsive.  This model is commonly called ``Kremer-Grest model''.\cite{KremerGrest:Beadspring}  For isolated chains it realizes good solvent conditions.  
\item The simulation of $\Theta$- or bad solvents requires to incorporate part of the LJ-attraction by increasing $r_\text{cut}$.  Obviously, there is freedom where to cut off the attractive part.  One possibility is $r_\text{cut} =  2 \times 2^{1/6}$ $(C(r_\text{cut})=127/4096)$.\cite{BennemannEtal:PRE1998}  This choice is a compromise between the wish to include the major part of the attractive interaction and the need to keep the potential short-ranged.  The resulting phase diagram was studied in Ref.~\citen{MullerGonzalez:Macro2000}.
\end{itemize}

\paragraph{(Yet) Another Bead-Spring Model.}  If we recall the idea of the coarse-graining --a coarse-grained monomer stands for a group of chemical monomers-- it appears plausible that coarse-grained monomers are softer than their chemical counterparts.  Thus, an exponent smaller than 12 in Eq.~\eqref{eq:LJ12-6TS} may be better suited to represent their repulsion.  In fact, such an observation was made in a recent effort to develop a coarse-grained model for poly(vinyl alcohol) (see Ref.~\citen{MeyerEtal:JCP2000} and Appendix~\ref{app:cg}).  This study also suggests the following generic model which may be considered as a variant of the Kremer-Grest model.

In this (``Kremer-Grest-like'') model non-bonded monomers interact via
a purely repulsive 9-6 LJ-potential,
\begin{equation}
U_\text{9-6}^\text{rep}(r) = \left \{
\begin{array}{ll}
\epsilon_0 \left [(\sigma_0/r)^{9} - (\sigma_0/r)^{6} \right] +
C (r_\text{min}) & \quad \mbox{for} \; r \leq r_\text{min}=(3/2)^{1/3}\,
\sigma_0 \;, \\
0 & \quad \mbox{else} \;,
\label{eq:cgpvanum}
\end{array}
\right .
\end{equation}
where $\epsilon_0=1.511$ and $C (r_\text{min})=4\epsilon_0/27$.  These non-bonded interactions are excluded between nearest neighbors in the chain, which are connected to each other by a harmonic potential  
\begin{equation}
U_\text{bond}(\ell) = \frac{1}{2}k(\ell-\ell_0)^2 \quad
(k=2141.84\,\sigma_0^{-2},\;\; \ell_0 = 0.97 \, \sigma_0) \;.
\label{eq:cgpvabond}
\end{equation} 
The equilibrium bond length $\ell_0$ agrees with that of the Kremer-Grest model.  The spring constant $k$ has to be chosen so large to inhibit bond crossings (see Ref.~\citen{GrestMurat:MCMD1995} for further discussion).  A similar bond potential, in conjunction with Eq.~\eqref{eq:LJ12-6TS} and $r_\text{cut} = 2^{1/6}$, has recently been used to study the effect of the bond length on the structure and dynamics of polymer melts.\cite{AbramsKremer:JCP} 
 
\paragraph{Local Moves for Continuum Models.} The continuum models are constructed for use in Molecular Dynamics simulations. However, simulation within Monte Carlo schemes is also possible. Similarly to the lattice models a local updating scheme can be realized by selecting a monomer and a direction at random and by attempting a displacement in the chosen direction.  This proposition is again accepted according to the Metropolis criterion.  

\label{pg:Delta}The size $\Delta$ of the displacement is a tunable parameter.  It should neither be too small nor too large.  If $\Delta$ is too small, many moves may be accepted, but the system advances only slowly in configuration space.  Many displacements are thus needed to obtain well decorrelated configurations.  On the other hand, if $\Delta$ is too large, many moves will be rejected and the decorrelation is also slow.  A scheme how to optimize the choice for $\Delta$ is explained in Sec.~3.3 of Ref.~\citen{FrenkelSmitBook}.

%%%%%%%%%%%%%%%%%%%%%%%%%%%%%%%%%%%%%%%%%%%%%%%%%%%%%%%%%%%%%%%%%%%%%%%%
\section{Monte Carlo Methods for Polymers: From Local to Non-Local Moves}
\label{sec:is}

The method of importance sampling is based on a Markov process in configuration space.  A priori, this stochastic dynamics is merely a numerical algorithm, aiming at an efficient sampling according to $P_\text{eq}(\boldsymbol{x})$.  It need not correspond to the physical dynamics of the (polymer) system under consideration.  An appealing consequence of this feature is the freedom to invent clever MC moves which decorrelate the configurations in the smallest (CPU) time possible.  These non-physical moves serve to rapidly equilibrate the system and to produce statistically independent equilibrium configurations for the study of structural and thermodynamic properties.  We will pursue this idea in Secs.~\ref{subsubsec:snake},\ref{subsubsec:pivot}.  In the following section we rather want to concentrate on local moves and the ensuing dynamic interpretation of the MC method.

\subsection{Local Moves: Studying Dynamic Properties with Monte Carlo}
\label{subsubsec:localmoves}

By employing non-local moves we can explore the statics of the system, but information about its dynamic properties is lost.  Of course, the equilibrated configurations could be used in a Molecular Dynamics (MD) simulation to analyze the dynamic properties.  However, if one is not willing to do that, the question arises of under which conditions the MC dynamics can be realistic.   The answer to this comprises two parts: 
\begin{enumerate}
\item Certainly, one can only expect the MC dynamics to become reliable on length and time scales where the deterministic motion of the monomers has been damped by the interaction with the surrounding (other monomers and/or solvent).  For instance, in a (classical) MD simulation the monomers move ballistically at early times, i.e., their displacement is proportional to $t$.  This is a consequence of the underlying Newtonian dynamics in the limit of vanishing force.  At short times the monomers behave as if they did not ``feel'' the bonding to their neighbors and the presence of other particles,  that is, as if they were free particles.  As time increases, the interaction with the surrounding becomes important.  The motion of the monomers is then a result of a multitude of individual collisions.  This ``averaging'' over fast degrees of freedom gradually lends a stochastic character to the dynamics which ultimately becomes diffusive in the long-time limit.       
\item The moves should be ``physical''.  Usually, this implies that they are local.\footnote{Examples for local moves of lattice models are given in Fig.~\ref{fig:latticeModels}.  See the very end of Sec.~\ref{subsec:continuum} for a brief discussion of local moves in continuum models.} Furthermore, the dynamics should not be dominated by the momenta which are absent in Monte Carlo.  The latter condition is satisfied in dense melts, but not in dilute solution.  In dilute solution the motion of distant monomers along the chain backbone are coupled via hydrodynamic interactions 
%(see Chap.~{\bf Zitat auf Burkhard-D\"unwegs-Vortrag} 
(see Ref.~\citen{duenweg2004}).    
and Sec.~\ref{subsec:PolymerDynamics}).  Thus, we might expect that a local Monte Carlo algorithm reproduces Rouse dynamics where these long-range interactions are neglected.     
\end{enumerate}
This expectation can be verified by estimating the scaling of local MC algorithms with $N$.  To this end, let us assume that the center of mass (CM) of an isolated chain, be it on a lattice or in the continuum, may be considered as a free Brownian particle.   This is reasonable, since the CM does not experience any external force other than the random force of the heat bath (resulting from the compound effect of the random monomer hops and the acceptance criterion).  So, it should diffuse freely [Eq.~\eqref{eq:g3rouse}].  The corresponding diffusion constant $D_N$ depends on chain length.  To estimate this dependence we can argue that the center of mass is displaced by $\sim b/N$, if one monomer moves over a distance of order $b$ while the other monomers remain fixed.  This elementary motion takes on average the time $1/m$ with $m$ denoting again the (temperature, density, etc.\ dependent) mobility of the monomer.  For the CM to diffuse over the distance $b$, $N$ such random motions are needed.  This take the time $m \times N$, which we use as our time unit here.\footnote{\label{defMCS}This statement introduces the time unit $\tau_\text{mcs}$ of a ``Monte Carlo step (MCS)''.  A MCS is defined as the time it takes to give each of the $N$ monomers the possibility to move once.\cite{BinderHeermann:2002,LandauBinder:2000}  Thus, we measure time in units of attempted elementary moves per monomer.}  Utilizing Eq.~\eqref{eq:g3rouse} we then find $g_3(t=1) \sim (mN)\times (b/N)^2 \sim D_N$.  So, 
\begin{equation}
D_N \sim \frac{m b^2}{N^{\approx 1}} \; .
\label{drouse}
\end{equation}
Inserting this result in Eq.~\eqref{eq:TauRouse} we obtain the relaxation time of a chain
\begin{equation}
\tau_{N} \sim \frac{N^{\approx 1+2\nu}}{m} 
\sim \left \{
\begin{array}{ll}
N^{\approx 2} & \quad \mbox{(ideal chain: $\nu=0.5$)}\;, \\
N^{\approx 2.176} & \quad \mbox{(3D excluded-volume chain: $\nu=0.588$)}\;.
\end{array}
\right .
\label{taurouse}
\end{equation}
Equations~(\ref{drouse},\ref{taurouse}) agree with the predictions of the Rouse theory [Eqs.~(\ref{eq:DRouse},\ref{eq:TauRouse})].

\paragraph{Monte Carlo Dynamics versus Molecular Dynamics: An Example.}  The previous arguments suggest that the MC dynamics, based on local moves, becomes realistic for time and length scales outside the microscopic regime (of a bond).  We want to support this assertion by a comparison between MC and MD simulations.  

\begin{figure}
\begin{center}
\epsfysize=65mm
\epsfbox{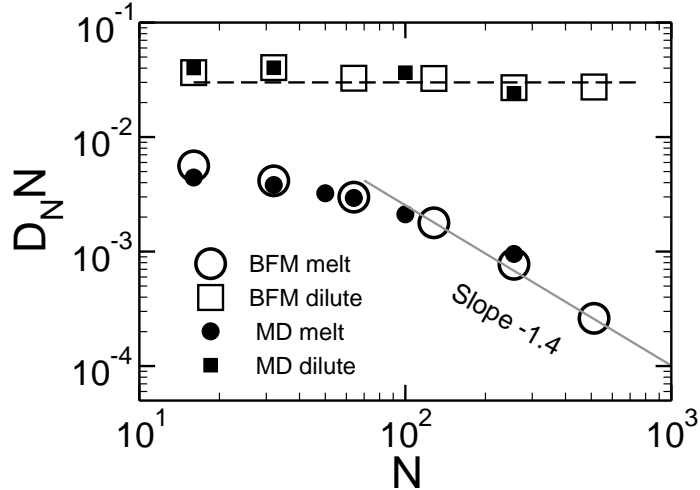}
\end{center}
\caption[]{Diffusion coefficient $D_N$ versus $N$.  Two simulation methods are compared:  The open symbols represent MC data of the (lattice) BFM, the filled symbols were obtained from MD simulations of a (continuum) Kremer-Grest-like model [Eqs.~(\ref{eq:cgpvanum},\ref{eq:cgpvabond})].  For both models results obtained in 3D for a dilute solution and for a melt are shown.  For the BFM this corresponds to the following volume fraction $\phi$: $\phi=0.0078$ (dilute), $\phi=0.5$ (melt).  For the Kremer-Grest-like model this corresponds to the following monomer densities $\rho$: $\rho=0.0835$ (dilute), $\rho=0.835$ (melt).  Qualitatively, the MD simulations yield the same dependence of $D_N$ on $N$.  To illustrate this agreement the MD data were vertically shifted by an amount that optimizes the agreement with the MC results.  (The shift factors are different for the dilute solution and the melt.)  In dilute solution, we find Rouse behavior [Eq.~\eqref{eq:DRouse}] for both methods.  In the melt, the chains diffuse more slowly.  The dependence of $D_N$ on $N$ is qualitatively compatible with the Rouse-to-reptation crossover when $N$ passes the threshold $N_\text{e}$ (Sec.~\ref{subsec:PolymerDynamics}).  Quantitatively however, there are deviations.  Particularly for large $N$, the decrease of $D_N$ appears to be stronger than predicted by reptation theory [Eq.~\eqref{eq:DNRep}].  Roughly, we find $D_N \sim N^{\approx -2.4}$.  Adapted from Ref.~\citen{MattioniEtal:Preprint}.}
\label{fig:DN_BFM_MD}
\end{figure}

Figure~\ref{fig:DN_BFM_MD} shows the diffusion coefficient $D_N$ of a chain versus chain length.  $D_N$ was derived from the long-time limit of Eq.~\eqref{eq:g3rouse} for both the BFM, simulated via MC, and the Kremer-Grest-like continuum model of Eqs.~(\ref{eq:cgpvanum},\ref{eq:cgpvabond}), simulated via MD.  The figure displays the results of the simulations for a dilute solution and a dense melt.\footnote{\label{pg:BFM}In the BFM, density is commonly specified in terms of the volume fraction $\phi$ of lattice sites occupied by monomers.  As a monomer comprises all sites of a unit cube, the monomer density $\rho$ is smaller than $\phi$ by a factor 8, $\rho=\phi/8$.  Although the value $\phi=0.5$ appears small, the work by Paul {\em et al.}\cite{wp_entangle} established that the chains have melt-like properties at this density (see also Ref.~\citen{Binder_MCMD1995}).  Since then, $\phi=0.5$ has become a standard choice (in 3D).  For the Kremer-Grest model, the work of Ref.~\citen{KremerGrest:Beadspring} showed that a monomer density of $\rho=0.85$, or a value close to this, is a good choice to realize melt-like behavior.  We adopted this choice in our MD study.  The MD simulations were done at constant volume and constant temperature (Langevin thermostat\cite{AllenTildesleyBook}).}   Clearly, there is a high degree of accord between the results, illustrating that the BFM with local moves reproduces the realistic dynamics of the MD simulations.  Thus, MC simulations can be more than just a versatile tool to calculate high-dimensional integrals.  They may provide information on the dynamics of a system.\footnote{There is further ample evidence for the correctness of this statement from other studies.  For simple liquids of LJ-particles see e.g.\ Ref.~\citen{HuitemaEerden:JCP1999}.  For polymers see the review in Ref.~\citen{BinderPaul:1997} or the comparison of MD simulations for polybutadiene and polyethylene with MC simulations of the BFM.\cite{Paul:ChemPhys2001}  Furthermore, Monte Carlo methods have been applied to simulate dynamic processes in such diverse fields as relaxation phenomena in spin and structural glasses, spinodal decomposition of mixtures, nucleation processes, diffusion-limited aggregation, etc. (see e.g.\ the textbooks of Binder and Heermann\cite{BinderHeermann:2002} or Landau and Binder\cite{LandauBinder:2000}).}

\paragraph{Relaxation Time and Computational Complexity.}\label{pg:compcomp}  An important issue in any algorithm is its ``computational complexity''.  Quite generally, the computational complexity may be defined as the time required to solve a computational problem.\cite{HostingDictionnary}  Here, the computational problem is to decorrelate chain configurations.  According to Eq.~\eqref{taurouse} this takes a relaxation time $\tau_N \sim N^{1+2\nu}$ in units of the Monte Carlo step (MCS; see footnote on page~\pageref{defMCS}).  As a MCS comprises $N$ attempted moves of a monomer, the computational complexity $\tau_\text{cc}$ scales with $N$ as $\tau_\text{cc} = N \tau_N \sim N^{2+2\nu}$.  

This rapid increase of $\tau_\text{cc}$ with chain length --called ``critical slowing-down''\cite{LandauBinder:2000}-- makes it difficult in practice to efficiently decorrelate configurations of long chains by local moves.  In order to be able to simulate large chains with sufficient statistics, moves have to be implemented, which reduces ($\tau_\text{cc} \sim N^\alpha$ with $\alpha < 2 + 2 \nu$) or even eliminate ($\tau_\text{cc} \sim N^0$) the critical slowing-down.  These moves cannot be local, they have to act, in some way, on all monomers of the chain.  In the following we want to discuss two examples of such global updates: bilocal moves and the pivot algorithm.

\subsection{Bilocal Moves: The Slithering-Snake and the Extended Reptation Algorithms}
\label{subsubsec:snake}

\begin{figure}
\begin{center}
\epsfysize=65mm
\epsfbox{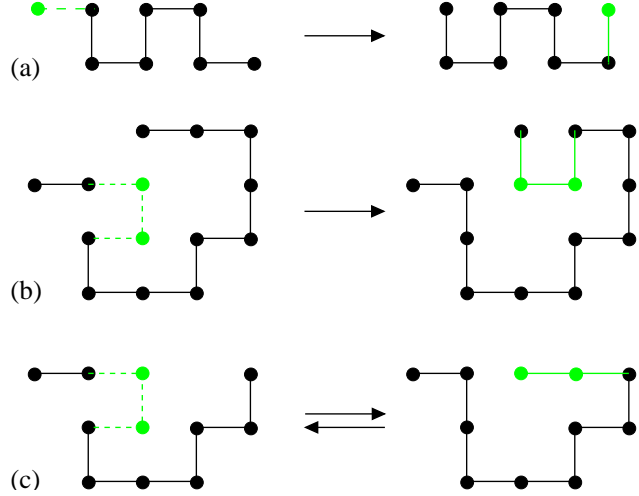}
\end{center}
\caption[]{Slithering-snake move (a) and general reptation moves (b,c).  Both moves are illustrated by the shrinkage-growth implementation.  For the slithering-snake algorithm, a randomly chosen end bond (dashed line) is removed and then a new bond vector (also randomly chosen) is attached to the other chain end.  For the general reptation algorithm, three moves are shown:  Kink-kink transport (b) and kind-end/end-kink reptation (c).  Kink-kink transport implies that a randomly chosen kink is shrunk to a bond and a new kink is inserted somewhere else along the chain.  Kink-end reptation ($\rightarrow$) amounts to replacing a randomly chosen kink by a bond and to appending two new bond vectors (also randomly chosen) to the other chain end.  End-kink reptation ($\leftarrow$) corresponds to the reverse ``reaction''.}
\label{fig:bilocal}
\end{figure}

A bilocal $N$-conserving move consists in altering the configuration of two small groups of consecutive monomers.  The groups are usually far from one another along the backbone of the chain.  Typical examples are the slithering-snake and the extended reptation algorithm:
\begin{itemize}
\item The {\em slithering-snake} (or {\em reptation}) algorithm removes a bond from one chain end, adds a new one to the other end and shifts the inner monomers one bond up along the chain in direction of the new bond (Fig.~\ref{fig:bilocal}a).  As the positions of the inner monomers remain unchanged, the chains  ``slithers'' along its contour during the MC move (whence the name of the algorithm).\footnote{The slithering-snake algorithm was invented by Kron in the 1960's and later independently by Wall and Mandel.\cite{WallMandel:Snake}  For an overview of applications to SAW's see e.g.\ Ref.~\citen{KremerBinder1988} and to off-lattice models see e.g.\ Ref.~\citen{LeontidisEtal:1995}.}
%
% Reinitialize footnote counter due to overflow
% ---------------------------------------------
\setcounter{footnote}{0}
%----------------------------------------------
%
\item The {\em extended reptation} algorithm transports a kink or an end group via a slithering motion along the chain.\footnote{This generalization of the slithering-snake algorithm was first discussed in detail by Reiter.\cite{Reiter:Macro1990}  More recently, algorithmic and statistical properties of extended reptation moves were analyzed and their implementation was discussed in Refs.~\citen{CaraccioloEtal:JStatPhys2000,CaraccioloEtal:PRE2002}.}  Commonly utilized moves are: (1) ``Kink-kink reptation'', which deletes a kink at some position along the chain and inserts a new one at another position (Fig.~\ref{fig:bilocal}b). (2) ``Kink-end reptation'', which removes a kink somewhere along the chain and adds two new bonds at one of the chain ends (Fig.~\ref{fig:bilocal}c$\rightarrow$).  (3) ``End-kink reptation'', the inverse of ``kink-end reptation'' (Fig.~\ref{fig:bilocal}c$\leftarrow$).
\end{itemize}
In the remainder of this section we will concentrate on the slithering-snake algorithm.  Extended reptation is only discussed in comparison to the slithering-snake algorithm. 

\paragraph{Implementation and Ergodicity.}  The slithering-snake and the extended reptation algorithms can be implemented in two ways: in a shrinkage-growth or a growth-shrinkage fashion.  As growth-shrinkage is just the inverse of shrinkage-growth, we illustrate the procedure for the latter via the example of an isolated chain.\footnote{For the multi-chain system the only difference to the isolated chain is that additionally one chain out of the $n$ chains in the systems has to be chosen at random.}  For the slithering-snake algorithm one chain end is selected at random, the bond to its neighbor is cleaved, and a randomly chosen new bond vector is attached at the other end.  If this move respects the excluded-volume condition in the athermal case and additionally passes the Metropolis test in the thermal case, it is accepted.  Otherwise the old configuration is recounted.  For the extended reptation algorithm the procedure is more complicated.  Details may be found for the SAW on a hypercubic lattice in Refs.~\citen{CaraccioloEtal:PRE2002,Reiter:Macro1990} and for a continuum bond-fluctuation model in Ref.~\citen{ReiterDuering:MacTheoSim1995}.

\begin{figure}
\begin{center}
\epsfysize=65mm
\epsfbox{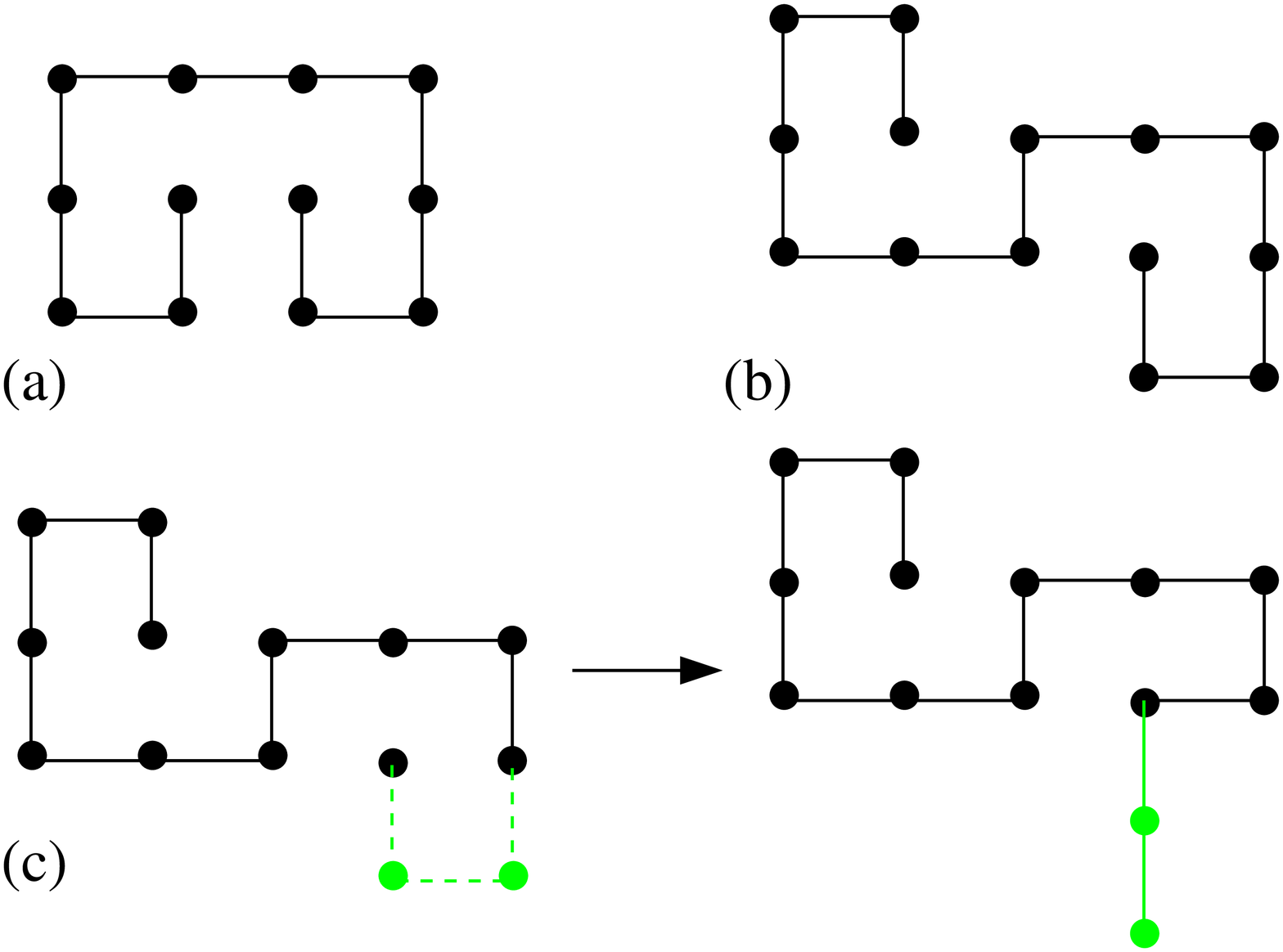}
\end{center}
\caption[]{Configurations of 2D SAW's to illustrate the ergodicity problem of the slithering-snake algorithm (a,b) and its solution via extended reptation moves (c).  Panel (a) shows a configuration that cannot be moved by slithering-snake moves, if chain growth is attempted first.  However, it is not blocked in the shrinkage-growth scheme.  By contrast, the configuration of panel (b) is frozen for both growth-shrinkage and shrinkage-growth moves.  Panel (c) shows that this configuration may be dissolved by extended reptation moves, e.g.\ by kink-end reptation if the chain end, where the kink is, happens to be selected for the attachment of the two bonds.}
\label{fig:snake}
\end{figure}

Usually, shrinkage-growth is preferred to the growth-shrinkage procedure because it is computationally more efficient.  The reason for this is illustrated in Fig.~\ref{fig:snake}.  The nested configuration of Fig.~\ref{fig:snake}a would be frozen, if a new bond had to be appended before an end bond may be removed.  However, it can be unraveled when shrinkage is attempted first.  Thus, the shrinkage-growth algorithm is less plagued by --though not exempt of-- non-ergodicity effects.  An example is provided by the double cul-de-sac configuration of Fig.~\ref{fig:snake}b.\footnote{Non-ergodicity effects are less severe for the slithering-snake algorithm than for the $N$-conserving local moves discussed before.  For the slithering-snake algorithm the ergodicity class of a straight rod contains at least a fraction of $N^{-(\gamma-1)/2}$ of all SAW configurations, whereas this fraction is exponentially small for the local algorithms.\cite{Sokal_MCMD1995,CaraccioloEtal:PRE2002}}  It is frozen in the shrinkage-growth procedure, but not for the kink-end reptation move shown in Fig.~\ref{fig:snake}c.  In fact, kink-end/end-kink moves are known to be ergodic\cite{Reiter:Macro1990} (as well as other bilocal algorithms; see Ref.~\citen{CaraccioloEtal:JStatPhys2000} for a thorough discussion).

Should one thus abandon the slithering-snake algorithm in favor of extended reptation?  Usually, the answer is ``No''.  For the SAW on the hypercubic lattice problems with ergodicity arise to the constraints imposed by the small coordination number of the lattice.  If many more bond vectors are a priori possible, as for the bond-fluctuation model or for (typical) continuum models, non-ergodicity should not represent a problem.\footnote{See pp.~283/284 of Ref.~\citen{KremerBinder1988} for further discussion of that point.  Contrary to SAW's, the equilibrium configurations of collapsed chains are typically (very) dense.  Quite generally with increasing density, the slithering-snake or the extended reptation algorithm become less efficient, as the ``free volume'' to add new bonds decreases (see e.g.\ Ref.~\citen{ReiterEtal:JCP1990} for a comparison of various algorithms to simulate high-density polymer systems and the subsequent discussion).  However, the recent study of Ref.~\citen{CaraccioloEtal:PRE2002} for 2D SAW's with $N \leq 3200$ at the $\Theta$-point suggests that extended reptation is almost as efficient as for pure SAW's with no attractive interactions.}  

\paragraph{Relaxation Time and Computational Complexity: Isolated Chains.}  One expects that the slithering-snake algorithm is able to decorrelate configurations more efficiently than a local updating scheme, the speed-up factor being roughly of order $N$.  This expectation results from the following heuristic argument:  The elementary move of the algorithm may be interpreted as a shift of all monomers along the contour of the chain.  For the CM this curvilinear motion has two consequences:  (1) The curvilinear diffusion coefficient $D_\text{c}$ should not depend on $N$, since all monomers are always shifted at once, irrespective of chain length.  Thereby, the slithering-snake algorithm gains a factor of $N$ in regard to the physical reptation dynamics, in which the curvilinear displacement is Rouse-like (Sec.~\ref{subsec:PolymerDynamics}). (2) An elementary move displaces the CM by $\sim\!b$ along the chain backbone.  After $N$ such moves, the CM has diffused curvilinearly a distance of the order of the contour length $L \propto Nb$.  Thus, the relaxation time $\tau_N$ should be given by 
\begin{equation}
\tau_N = \frac{L^2}{D_\text{c}} \quad \Rightarrow \quad 
\tau_N \sim N^{\approx 2} \;.
\label{eq:tauN_snake}
\end{equation}
With respect to the computational complexity (page~\pageref{pg:compcomp}) one expects $\tau_\text{cc} \propto \tau_N$.  There is no extra factor of $N$, as in the case of local moves, for the slithering-snake algorithm.  The algorithm is bilocal.  It takes a time of order 1 to check and update the chain ends.\footnote{Here, we assume that the time to shift the monomer index along the chain is implemented in a way that it also takes a time of order 1 only.}
 
Note that Eq.~\eqref{eq:tauN_snake} is independent of the conformational properties of the chain, contrary to Eq.~\eqref{taurouse} (which depends on $\nu$).  Thus, it should be valid for both 2D and 3D dilute polymer solutions as well as for dense melts.  While for the slithering-snake algorithm\cite{KremerBinder1988} and for some extended reptation algorithms\cite{CaraccioloEtal:PRE2002,Reiter:Macro1990} the scaling found for $\tau_N$ is very close to Eq.~\eqref{eq:tauN_snake}, the behavior in dense systems is quite different.\cite{Deutsch:ARH,MattioniEtal:EPJE2003}  The influence of density on the slithering-snake dynamics\footnote{For local moves the set of allowed bond vectors automatically prevents bonds from crossing each other in the course of the simulation (see Sec.~\ref{subsec:lattice}).  If slithering-snake moves are considered, the uncrossability of the bonds has to be checked explicitly to avoid configurations which cannot be attained or unraveled by local updates.  Bond crossing can occur if vectors from the classes $[2,2,1]$ or $[3,1,0]$ are selected.\cite{HWeber:Diss}} has recently been studied by the bond-fluctuation model.\cite{MattioniEtal:EPJE2003}  The following paragraphs briefly summarize some results of this work.

\begin{figure}
\begin{center}
\epsfysize=60mm
\epsfbox{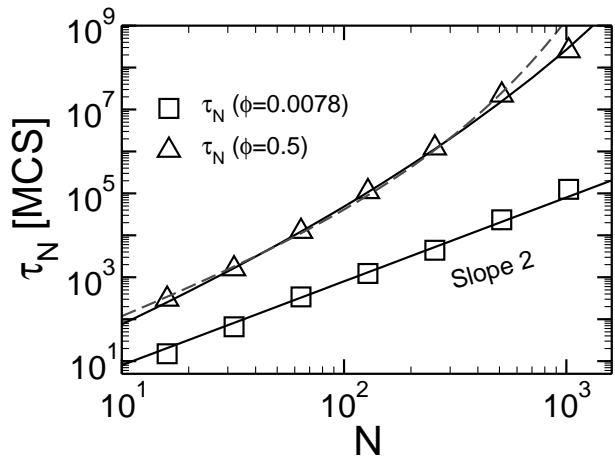}
\end{center}
\caption[]{Relaxation time $\tau_N$ versus $N$. $\tau_N$ is defined by $g_3(\tau_N)=R^2_\text{g}$.  In the dilute limit ($\phi=0.0078$) $\tau_N \sim N^{\approx 2}$, as expected from Eq.~\eqref{eq:tauN_snake}.  In the melt ($\phi=0.5$), the increase of $\tau_N$ with $N$ is stronger.  The stretched exponentials are motivated by the activated reptation hypothesis:\cite{Deutsch:ARH,Semenov:ARH_review,SemenovRubinstein:EPJB1998}  $\tau_N \approx N^2 \exp(0.8N^{1/3})$ (bold line) provides a better description than $\tau_N \approx N^2 \exp(0.074N^{2/3})$ (dashed grey line).  Adapted from Ref.~\citen{MattioniEtal:EPJE2003}.}
\label{fig:Tau_snake}
\end{figure}

\paragraph{From Dilute Solutions to Dense Melts: A Case Study by the BFM.}  Reference~\citen{MattioniEtal:EPJE2003} describes simulations for athermal systems containing chains of length $16 \leq N \leq 1024$ at different volume fractions $\phi$.  $\phi$ ranges from dilute solutions to dense melts ($\phi \approx 0.5$; see footnote on page~\pageref{pg:BFM}).  Figure~\ref{fig:Tau_snake} compares the relaxation time $\tau_N$ in dilute solution with that in the melt.  In dilute solution, the simulation results agree with the prediction of Eq.~\eqref{eq:tauN_snake}, $\tau_N \sim N^{\approx 2}$.  This implies that the assumption of independent, free diffusive motion, which underlies Eq.~\eqref{eq:tauN_snake}, is well borne out.  If this assumption was also true in the melt, the sole effect of density would be to slow down the monomer mobility $m$.  However, the dependence of $\tau_N$ on $N$ should not change.  Figure~\ref{fig:Tau_snake} shows that this is not true.  At $\phi=0.5$, $\tau_N$ increases exponentially with $N$.  This strong slowing-down of the dynamics reflects correlations between the motion of the chains.  

The importance of such intermolecular interactions for the polymer dynamics was first discussed by Deutsch.\cite{Deutsch:ARH}  However, Deutsch goes beyond a mere interpretation of the dynamic properties of the slithering-snake algorithm.  He identifies the slithering dynamics with the physical dynamics along the primitive path in the reptation model (see Fig.~\ref{fig:reptation}).  This suggests an attractive application:  The slithering-snake algorithm mimics the back and forth reptation motion of real chains without modeling the (time consuming) local monomer fluctuations around the primitive path.  It focuses on the long time behavior of very large chains, where all of these local motions have already relaxed.  This suggests that the slithering-snake dynamics may be interpreted in terms of theories proposed for the dynamics of strongly entangled polymer melts, such as the one of Deutsch.\cite{Deutsch:ARH}

The main results of this theory may be summarized as follows:  A chain can reptate through the network of its neighbors only as long as the end monomer does not enter a dense region which prohibits any further forward move.  The only way out of the trap is to partially retract and to explore the environment for new pathways.  These intermolecular interactions create a free energy barrier which temporarily localizes the chain in the region it initially occupied, and protracts the relaxation.  Further relaxation in a dense region could only occur if the chain end encounters another end which moves out of its way.  This implies that the portion of the chain, which altered its initial configuration while exploring the environment, should span the typical distance between chain ends $d_\text{end}$.  Let there be $g$ monomers in this portion.  Then, by exploiting the ideality of the chains in the melt, we have $g = (d_\text{end}/b)^2 \sim N^{2/3}$ because the density of chain ends scales as $\rho/N \propto  d_\text{end}^{-3}$.  Thus, $g$ is large for long chains.  If we now assume that the monomers have to overcome the free energy barrier $g\Delta \mu$, where $\Delta \mu$ is the difference in the monomer chemical potential between the newly explored environment and the region of the initial chain configuration, the barrier is large and the relaxation dynamics should be activated.  Thus, $\tau_N \propto N^2 \exp [ \text{const} \, N^{2/3}]$.  This is the main prediction by Deutsch.  The assumption of a finite $\Delta \mu$ was challenged by Semenov\cite{Semenov:ARH_review} who suggested that the barrier is due to fluctuations of the molecular field rather than to a permanent chemical potential difference (see also Ref.~\citen{SemenovRubinstein:EPJB1998}).  This picture implies that the barrier should be proportional to $\sqrt{g}$ so that $\tau_N \propto N^2 \exp [ \text{const}\,  N^{1/3}]$.  

The simulation data of Fig.~\ref{fig:Tau_snake} appear to agree with the latter prediction better than with the original one of Ref.~\citen{Deutsch:ARH} (at least for the chain lengths simulated up to now).  Certainly, more work is needed to test these predictions.  

\begin{figure}
\begin{center}
\epsfysize=65mm
\epsfbox{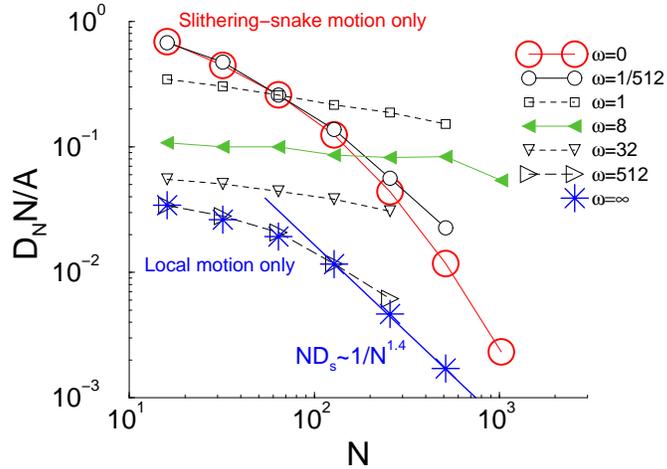}
\end{center}
\caption[]{Spatial diffusion coefficient $D_N$ versus $N$ for the BFM at $\phi=0.5$.  Different ratios $\omega$ of local to slithering-snake moves are compared:  $\omega=0$ corresponds to pure slithering-snake dynamics, $\omega=\infty$ to the pure local dynamics.  The diffusion coefficient is scaled by $N/A$, where $A$ is the acceptance rate.  For both local and slithering-snake dynamics the acceptance rate is roughly $A \approx 0.1$ for all $N$ at $\phi=0.5$.  The data for $\omega \ll 1$ and for $\omega \gg 1$ are very similar to the pure slithering-snake ($\omega=0$) and the pure local limit ($\omega=\infty$), respectively.  For $\omega \approx 8$, $ND_N/A$ is approximately independent of $N$.  This may define a reasonable choice of $\omega$ for efficient equilibration of longer chains by local and slithering-snake moves.  Adapted from Ref.~\citen{MattioniEtal:EPJE2003}.}
\label{figDomegaN}
\end{figure}

\paragraph{Slithering-Snake versus Local Moves.}  From a merely computational point of view Fig.~\ref{fig:Tau_snake} appears to indicate that the slithering-snake dynamics is not very efficient in equilibrating dense melts.  Its relaxation time increases with $N$ more strongly than a power law which is typically found for local updating schemes.\cite{BinderPaul:1997}  Nevertheless, simulations of the BFM for short chains ($N=10,20$) suggest that the slithering-snake algorithm decorrelates melt configurations ($\phi \approx 0.5$) very efficiently.\cite{WolfgardtEtal:JPhys1995,TriesEtal:JCP1997}  

This point certainly needs more studies.  Work in this direction was done in Ref.~\citen{MattioniEtal:EPJE2003}.  Figure~\ref{figDomegaN} shows a preliminary result for the diffusion coefficient $D_N$ as a function of chain length.  $D_N$ was obtained from simulations employing a mixture of local and slithering-snake moves.  This introduces, as an additional parameter, the ratio $\omega$ of local to slithering-snake moves.  The figure indicates that pure slithering-snake dynamics ($\omega=0$) equilibrates short chains more efficiently than pure local dynamics ($\omega=\infty$), in accord with the observations made in Refs.~\citen{WolfgardtEtal:JPhys1995,TriesEtal:JCP1997}.  By contrast, with increasing chain length $D_N$ slows down exponentially for the slithering-snake algorithm, as expected due to $D_N \sim R^2_\text{g}/\tau_N$, whereas the local dynamics exhibits a crossover from Rouse-like, $D_N \sim 1/N$, to reptation-like behavior, $D_N \sim 1/N^{\approx 2.4}$ (see Sec.~\ref{subsec:PolymerDynamics}).  If this trend persists, the pure slithering-snake algorithm will become inefficient to equilibrate long chains.  However, one can speculate that the addition of local moves weakens the confinement imposed by neighboring chains on the slithering-snake dynamics in the large-$N$ limit.  Indeed, this seems to be borne out by the data.  For short chains ($N < 64$) $D_N$ decreases monotonously with increasing $\omega$, since local moves are less efficient in exploring the configuration space and the confinement is negligible.  As $N$ increases, one finds a non-monotonous behavior.  The dynamics first becomes more rapid, as local moves are added.  This effect appears to saturate at $\omega \approx 10$.  Larger values of $\omega$ causes the diffusion coefficient to decrease again strongly (at fixed $N$).  This implies that a judicious (model-dependent) choice of $\omega$ is crucial if one wants to equilibrate a melt of long chains efficiently by mixing local and slithering-snake moves.

\paragraph{Remark.}  The efficiency of the slithering-snake algorithm (with or without local moves) or of the extended reptation algorithm deteriorates considerably as $\phi$ approaches 1, since there is not sufficient space for the growth step.  If one is interested in these high densities,\footnote{Note that $\phi \approx 1$ means that the polymer melt has zero compressibility $\kappa_T$.  By contrast, real polymer melts are compressible, with $k_\text{B}T\rho \kappa_T$ being of the order $10^{-1}$ for temperatures above the glass transition or crystallization temperatures.\cite{CurroEtal}  For the BFM\cite{BinderEtal:ProgPolPhys2003} at $\phi\approx 0.5$, $k_\text{B}T\rho \kappa_T \approx 0.2$ and for the bead-spring models of Sec.~\ref{subsec:continuum} at $\rho\approx 0.9$,  $10^{-2} < k_\text{B}T\rho \kappa_T < 10^{-1}$ (see Refs.~\citen{BinderEtal:ProgPolPhys2003,CurroEtal:JCP1999}).  Thus, it appears that the limit $\phi \rightarrow 1$ is not needed to model dense polymer melts.}  an alternative simulation method may be provided by reptation moves including a ``walker''.\cite{ReiterDuering:MacTheoSim1995,ReiterEtal:JCP1990}  A ``walker'' is defined as an isolated monomer (or as a small group of monomers).  In the MC move, the ``walker'' attaches to a chain in its neighborhood, which then releases a monomer somewhere along its backbone, yielding a new ``walker'' at a different position than the original one.  Since the walker can be created by cleaving a monomer from a chain, the algorithm works even at $\phi=1$.  In order to preserve monodispersity the update of the configuration is finished if the ``walker'' attaches again to the chain it was originally cleaved from.

\subsection{Non-Local Moves: The Pivot Algorithm}
\label{subsubsec:pivot}

A non-local $N$-conserving move attempts to update a chain portion of order $N$.  If successful, it drastically modifies the chain dimension.  Global properties, such as the end-to-end distance, should therefore relax within a few ($N$-independent) steps so that the critical slowing-down is largely moderated --if not removed.\footnote{This idea is related to that of cluster algorithms\cite{Sokal:QFT1992,Janke:spinreview1996} employed in MC simulations of spin systems near criticality.  Close to the phase transition, the spins are strongly correlated.  They form cluster of size $\xi$ (= correlation length, corresponding to $R_\text{e}$ in the polymer problem, see Sec.~\ref{sec:intro}).  A cluster algorithm finds one cluster (Wolff algorithm) or all of them (Swendsen-Wang algorithm) and updates all spins of the cluster(s) at once.  This strongly reduces or even eliminates, in favorable cases, the critical slowing-down.\cite{Sokal:QFT1992,Janke:spinreview1996}}  

This appealing feature makes the search for appropriate non-local moves very attractive.  However, not every conceivable move turns out to be efficient.  This is mainly due to two reasons:
\begin{enumerate}
\item A drastic modification of the chain configuration is much more likely to violate the excluded-volume constraint than local or bilocal moves do.  So, we expect the acceptance rate of non-local updates to decrease with $N$.  The challenge consists in inventing moves whose acceptance rate does not rapidly vanish as $N \rightarrow \infty$.
\item A non-local move typically requires a CPU time of order $N$ (checking self-avoidance, updating the configuration if accepted), in contrast to order 1 for a local or a bilocal move.  The extra factor $N$ must be compensated by a very efficient decorrelation of the configurations to justify the use of non-local moves.
\end{enumerate} 
A paradigm for a non-local algorithm, which satisfies these criteria, is the ``pivot algorithm''.\footnote{The pivot algorithm was invented by Lal in 1969. A comprehensive discussion of the algorithm may be found in Refs.~\citen{Sokal_MCMD1995,MadrasSokal:JStatPhys1988}.}

\begin{figure}
\begin{center}
\epsfysize=45mm
\epsfbox{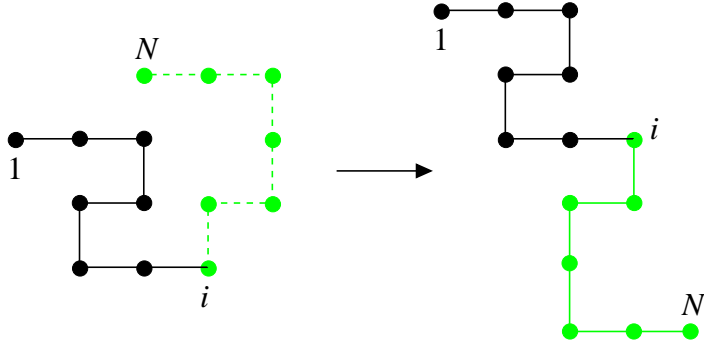}
\end{center}
\caption[]{Illustration of the pivot algorithm.  A monomer $i$ (= pivot point) is chosen at random.  It divides the chain into two pieces:  The monomers $1,\ldots,i$ remain fixed, while the monomers $i+1,\ldots,N$ are translated to a new position via a randomly chosen symmetry operation (rotation, reflection, etc.).  In the example of the figure, a $180^\circ$ rotation around the monomer $i$ is shown.}
\label{fig:pivot}
\end{figure}

\paragraph{Pivot Algorithm for the SAW.}  The elementary move of the pivot algorithm works as follows (Fig.~\ref{fig:pivot}):  First, one randomly chooses a monomer $i$ and a symmetry operation (e.g.\ a rotation, a reflection, etc.).  The monomer serves as a ``pivot point'' for the symmetry operation which turns the chain portion comprising the monomers $i+1,\ldots, N$ to a new position, while the other piece of the chain (monomers $1,\ldots,i$) remains unchanged.  The proposed move is accepted if the resulting configuration is self-avoiding.  Otherwise, it is rejected and the old configuration is recounted.    

Qualitatively, the pivot move resembles an attempt to construct a SAW of $\sim\!\! N$ mo\-no\-mers by joining two SAW's of $\sim\!\! N/2$ monomers at the pivot point.  The probability for the result to be self-avoiding should scale as ${\cal Z}(N)/{\cal Z}^2(N/2) \sim N^{-(\gamma-1)}$ [see Eq.~\eqref{eq:rerg_Z}].  This heuristic argument suggests that the acceptance rate vanishes as $N^{\approx -0.344}$ in 2D ($\gamma=43/32$)\cite{CaraccioloEtal:PRE1998} and as $N^{\approx -0.158}$ in 3D ($\gamma\simeq 1.158$)\cite{CaraccioloEtal:PRE1998}.  Although these estimates are quantitatively not very accurate, they correctly predict the qualitative trends:  The acceptance rate decreases with increasing chain length as a power law $N^{-\alpha}$ and the exponent $\alpha$ is larger in 2D than in 3D (2D: $\alpha \approx 0.19$, 3D: $\alpha \approx 0.11$).\cite{MadrasSokal:JStatPhys1988}

\paragraph{Relaxation Time and Computational Complexity.}  Fortunately, the numerical value of $\alpha$ is small, implying that even for long chains, e.g.\ $N = 10^5$, every $N^{0.11} \approx 3.5$th move is accepted.  Since a successful move implies a huge modification of the conformation, one can expect global properties to relax after a few steps.  So, the relaxation time $\tau_N$ scales as $\tau_N \sim N^\alpha$.  This increase is distinctly slower than that of all algorithms discussed so far.  Due to this property and due to the fact that the pivot algorithm is known to be ergodic\cite{MadrasSokal:JStatPhys1988} it has become very popular (see e.g.\ the compilation of references in Ref.~\citen{Kennedy:JStatPhys2002}).  Currently, the pivot algorithm is considered to be the most efficient algorithm for studying configurational properties\footnote{By configurational properties we mean quantities characterizing the chain dimension, such as $R_\text{e}$, $R_\text{g}$, etc.  A very accurate estimate of the critical exponent $\nu$ in 3D ($\nu=0.5877\pm 0.0006$) was obtained by the pivot algorithm.\cite{LiEtal:JStatPhys1995}  By contrast, it appears difficult to measure precisely the partition function ${\cal Z}(N)$, and so the exponent $\gamma$ with the pivot algorithm.  For this purpose, other algorithms, such as the ``join-and-cut'' algorithm\cite{CaraccioloEtal:PRE1998} or chain-growth algorithms\cite{GrassbergerEtal:JPMG1997}, are better suited.  The current best estimate for $\gamma$ in 3D is $\gamma=0.1575\pm 0.0006$.\cite{CaraccioloEtal:PRE1998}\label{pg:gamma}} of isolated SAW's.\cite{CaraccioloEtal:PRE1998,MadrasSokal:JStatPhys1988,GrassbergerEtal:JPMG1997} 

The pivot algorithm quickly decorrelates global quantities, such as the end-to-end distance.  However, it is not as efficient for local properties:  The conformation of a specific monomer is only altered if this monomer is selected as a pivot point and if the move is successful.  A successful moves takes a time of order $N^\alpha$ and, as the chain consists of $N$ monomers, the decorrelation time of a local observable should scale as $\tau_\text{loc} \sim N^{1+\alpha}$.  This extra factor $N$ is felt if one starts from an arbitrary initial configuration.  Full equilibration on all length scales is required before large-scale equilibrium properties may be sampled.  The equilibration time must be longer than the longest relaxation time in the system, i.e, than $\tau_\text{loc}$.  

For a non-local algorithm the computational complexity is a particularly important quantity because inefficient implementations may ruin the advantage gained by fast decorrelation.  The most naive check for self-avoidance would take a time of order $N^2$ so that $\tau_\text{cc}=N^2 \tau_N \sim N^{2+\alpha}$, comparable to the slithering-snake algorithm [Eq.~\eqref{eq:tauN_snake}].  Obviously, a faster check is called for. In Ref.~\citen{MadrasSokal:JStatPhys1988} it was argued that, by starting at the pivot point and working outwards, self-intersections may be detected in a time of order $N^{1-\alpha}$.  This procedure must be repeated $\sim N^\alpha$ times to obtain one accepted pivot.  So, the time required per accepted pivot scales as $\sim N$.\footnote{The pivot algorithm may be implemented so that the time required to obtain an accepted move is of order $N^q$ with $q<1$.\cite{Kennedy:JStatPhys2002}}  Once the pivot is accepted, we still have to update the monomer positions which also takes a time of order $N$.  So, in total we find $\tau_\text{cc} \sim N\tau_N \sim N^{1+\alpha}$ for global properties and $\tau_\text{cc}^\text{loc} \sim N\tau_\text{loc} \sim N^{2+\alpha}$ for local properties.  The estimate for $\tau_\text{cc}^\text{loc}$ is again comparable to the slithering-snake algorithm [Eq.~\eqref{eq:tauN_snake}]\label{pg:snake_for_pivot}. Therefore, one could also use slithering-snake moves to engender an initial, equilibrated configuration for the pivot algorithm.

\subsection{Non-Local Moves in the Melt: The Double-Pivot Algorithm}
\label{subsubsec:doublepivot}

Due to its efficiency in decorrelating SAW configurations it is tempting to apply the pivot algorithm also to other situations, such as the collapse transition, SAW's in confined geometry, or dense polymer melts.  However, in these cases, the algorithm becomes inefficient.  The non-local moves either lead to large energy differences (collapse transition) or violate the excluded-volume condition (SAW's in confined geometry, dense melts) so that they are rejected.

Should one therefore give up the idea of using pivot-like moves, say, for dense melts?  Recent work suggests that this conclusion might be wrong.  Instead of pivoting a piece of one chain to a new position the MC move can involve two chains.  Such a move was termed ``double-pivot (DP)'' algorithm.\footnote{The double-pivot algorithm has been proposed recently in Ref.~\citen{AuhlEtal}.  In part, this work was motivated by a novel chain-bridging algorithm which was successfully employed in atomistic simulations of long polyethylene chains (see Ref.~\citen{KarayiannisEtal:double_bridging}).
% and Chap.~{\bf Zitat auf Doros-Theodorous-Vortrag}).  
Our presentation of the DP algorithm is inspired by the discussion of Ref.~\citen{KarayiannisEtal:double_bridging}.  Due to the newness of the algorithm the implementation that we propose might turn out not to be the most efficient one.}  The basic idea of the algorithm is to cleave simultaneously a bond in a chain and in one of its neighbor chains, and to reconnect the monomers such that the chains remain monodisperse.\footnote{In general, connectivity-altering moves between arbitrary monomer pairs of neighboring chains lead to a distribution of chain lengths, i.e., to polydispersity.  The width of the distribution can be controlled by introducing a chemical potential.  See Sec.~3.3 of Ref.~\citen{KremerBinder1988} for further discussion in the context of lattice models and e.g.\ Ref.~\citen{KarayiannisEtal:double_bridging} for an application to atomistic MC simulations of polyethylene.}  The algorithm works as follows (Fig.~\ref{fig:doublepivot}):
\begin{figure}
\begin{center}
\epsfysize=45mm
\epsfbox{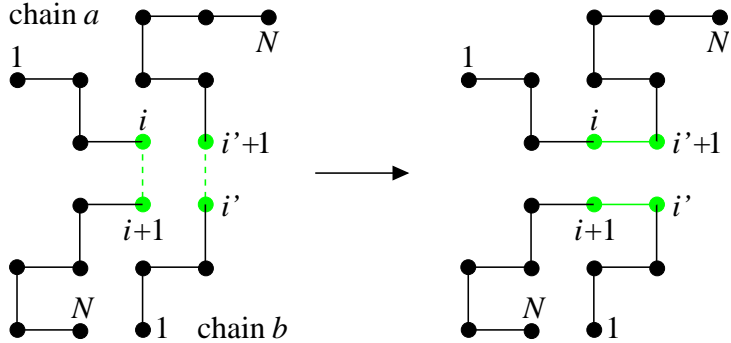}
\end{center}
\caption[]{Illustration of a double-pivot (DP) move on a square lattice.  A DP move flips bonds between two adjacent chains $a$ and $b$.  To this end, the neighborhood of monomer $i$ in chain $a$ is inspected to find potential bridging sites on chain $b$.  To preserve monodispersity a potential bridging site has to satisfy the following conditions: (a) The $(i+1)$th monomer of chain $b$, denoted by $i'+1$ in the figure, must be separated by a distance of the bond length from monomer $i$.  (b) The same condition must also hold for the distance of the $i$th monomer of chain $b$, denoted by $i'$, from monomer $i+1$ of chain $a$.  Between these four monomers a connectivity-altering move is attempted.  The bonds from $i$ to $i+1$ and from $i'$ to $i'+1$ are broken, and new bonds, $i$ with $i'+1$ and $i'$ with $i+1$, are created.  The proposed move is accepted or rejected according to Eq.~\eqref{eq:metropolis_DP}.  From the figure it is clear that the DP algorithm can only be carried out if there are matching monomers on a neighbor chain within the distance of a bond.  This is the more likely, the higher the concentration of the solution.  Therefore, the algorithm (presumably) works best in concentrated solutions or melts.  If successful, the move entails a drastic change of the chain configuration.}
\label{fig:doublepivot}
\end{figure}
\begin{enumerate}
\item A monomer, say monomer $i$ in chain $a$, is chosen at random.  Around this monomer the neighborhood is inspected to find bridgeable neighbors on other chains $b$.  A bridge\-able neighbor is defined by the following requirements:
\begin{enumerate}
\item It must be possible to connect the neighbor to $i$ by a bond vector.  This imposes a restriction on the intermolecular distance between $i$ and its neighbor.  In the example of Fig.~\ref{fig:doublepivot} it must coincide with the lattice constant.  For the BFM it must be among the set of allowed bond vectors [Eq.~\eqref{eq:BFMbonds}], whereas, for a continuum model, the bond energy resulting from taking the intermolecular distance as a bond vector should not be so large that the proposed bond would never occur in equilibrium.  In the latter case, it might be necessary to reduce the strength of the bond potential, e.g., the force constant of Eq.~\eqref{eq:FENE}.\cite{AuhlEtal}
\item To maintain monodispersity the neighbor must be either the $(i-1)$th or the $(i+1)$th monomer of the chain $b$.  We distinguish the monomers of chain $b$ by a ``prime'', e.g.\ $i'$, from those of chain $a$.
\item If it is monomer $i'\pm 1$, monomer $i'$ must be separated from monomer $i\pm 1$ of chain $a$ by a distance which satisfies condition (a).
\end{enumerate}
Using these three criteria we determine the total number of bridgeable neighbors of monomer $i$, $N_\text{DP}(i,\boldsymbol{x'})$.  If $N_\text{DP}(i,\boldsymbol{x}')=0$ for all $i$, the configuration $\boldsymbol{x'}$ must be updated by local (or bilocal) moves to bring the monomers in more favorable positions for bridging.\cite{AuhlEtal,KarayiannisEtal:double_bridging}  
\item A double-pivot move is initiated by randomly selecting one of the bridgeable neighbors, say $i'+1$ in chain $b$, from $N_\text{DP}(i,\boldsymbol{x'})$. Then, the bonds between $i$ and $i+1$ and between $i'$ and $i'+1$ are cleaved, and one attempts to create new bonds between $i$ and $i'+1$ and between $i'$ and $i+1$.  This move just switches the connectivity while preserving the chain length, and is proposed with probability 
\begin{equation}
P_\text{pro}(\boldsymbol{x'} \rightarrow \boldsymbol{x})
= \frac{1}{N_\text{DP}(i,\boldsymbol{x'})} \;.
\label{eq:propose_DP}
\end{equation}
\item To satisfy detailed balance we have to determine $P_\text{pro}(\boldsymbol{x} \rightarrow \boldsymbol{x'})$ of the reverse move.  As the forward move only alters the connectivity between the chains, but does not displace the monomers, the number of bridgeable neighbors of a specific monomer remains unchanged.  That is, $N_\text{DP}(i,\boldsymbol{x})=N_\text{DP}(i,\boldsymbol{x'})$.  To reverse the forward move, we have to select monomer $i'+1$ on chain $b$ and its bridgeable neighbor $i$ on chain $a$.  This occurs with probability
\begin{equation}
P_\text{pro}(\boldsymbol{x} \rightarrow \boldsymbol{x'})
= \frac{1}{N_\text{DP}(i'+1,\boldsymbol{x'})}
\label{eq:propose_DP_back}
\end{equation}
so that the Metropolis criterion reads
\begin{equation}
\text{acc}(\boldsymbol{x'}\rightarrow \boldsymbol{x}) 
= \min \bigg (1,\frac{N_\text{DP}(i,\boldsymbol{x'})}
{N_\text{DP}(i'+1,\boldsymbol{x'})}\;
\text{e}^{-\beta\left[U(\boldsymbol{x}) - U(\boldsymbol{x'})\right]} \bigg)\;.
\label{eq:metropolis_DP}
\end{equation}
The difference $U(\boldsymbol{x}) - U(\boldsymbol{x'})$ is the local change in energy due to the switching of the bonds between chains $a$ and $b$.
\end{enumerate}
The steps 1.--3.\ may be repeated several times.  However, the number of iterations should not be too large.  Otherwise it is likely that an accepted move annihilates one of its predecessors by performing the transition between two chains in the reverse direction.  To avoid this inefficiency it is important to mix up the local configuration of the system.  This may be achieved by e.g.\ local MC moves or by combining\cite{AuhlEtal} the DP algorithm with MD simulations.

%%%%%%%%%%%%%%%%%%%%%%%%%%%%%%%%%%%%%%%%%%%%%%%%%%%%%%%%%%%%%%%%%%%%%%%%
\section{Monte Carlo Methods for Polymers: Rosenbluth Sampling and Its Modern Variants}
\label{sec:rosenbluth}

\begin{figure}
\begin{center}
\epsfysize=35mm
\epsfbox{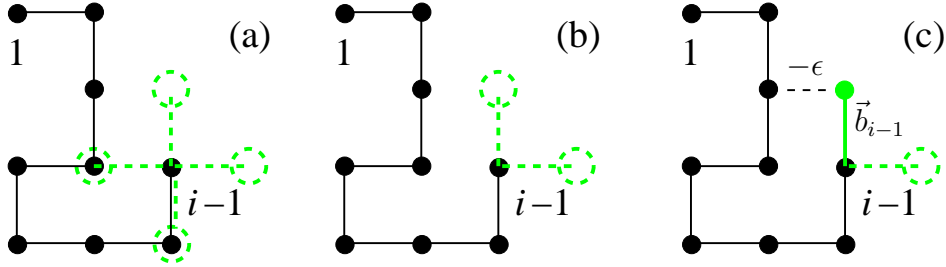}
\begin{picture}(10,10)
\unitlength=1mm
\put(52,21){\makebox(0,0)[lb]{{\large $\vec{b}_{i-1}$}}}
\put(43,29){\makebox(0,0)[lb]{{\large $-\epsilon$}}}
\end{picture}
\end{center}
\caption[]{Illustration of the simple-sampling (a) and the Rosenbluth-Rosenbluth (RR) methods [(b) and (c)] on a square lattice.  The coordination number of the lattice $z$ ($=4$) defines the number of possible bond vectors $\{\vec{b}\}$.  In simple sampling, all of these vectors have the same a priori probability.  Thus, the bond vector $\vec{b}_{i-1}$ ($=\vec{r}_i-\vec{r}_{i-1}$) from monomer $i\!-\!1$ to the new monomer $i$ can also point in the direction of already occupied lattice sites.  This leads to the attrition problem.  The RR method strongly reduces the attrition by taking $\vec{b}_{i-1}$ only from the open directions $k_i$ ($=2$ in the example of the figure).  The new bond vector can be chosen either uniformly from the $k_i$ possibilities, as indicated by the dashed lines and dashed circles in panel (b), or according to the local Boltzmann factor [Eq.~\eqref{eq:PsmethB}], if e.g.\ an attractive monomer-monomer interaction is present [filled grey circle in panel (c)].}
\label{fig:SSRR}
\end{figure}

The first MC method to simulate a SAW was ``simple sampling''.\cite{KremerBinder1988}  This static method (Sec.~\ref{sec:mc}) works as follows:  
\begin{enumerate}
\item Place the first monomer at the origin, randomly choose a bond vector, and append it to the monomer.  
\item Choose the next bond vector, again randomly, connect it to the second monomer, and check the self-avoidance (Fig.~\ref{fig:SSRR}).
\item If the chain is self-avoiding, the random growth process may be continued.  If not, the self-avoiding piece of the SAW, obtained up to this point, must be discarded, and we have to start from scratch at the first step again.
\item The steps 1.\ to 3.\ are repeated until a SAW of the desired length $N$ is obtained.  Then, data analysis may be done.
\item Repeat steps 1.\ to 4.\ to gather sufficient statistics.
\end{enumerate}
Apparently, completed SAW's are independent of one another and occur with the same probability $P_\text{s}(\boldsymbol{x})=P_\text{saw}(N)$.  This is the main advantage of a static MC method.  The main disadvantage of simple sampling is that $P_\text{saw}(N)$ becomes exponentially small for large $N$.  To see that, let us calculate $P_\text{saw}(N)$ for a SAW on a hypercubic lattice.  The hypercubic lattice has the coordination number $z=2d$.  Thus, the number of random walks (RW's) starting at the origin and having $N-1$ steps, is ${\cal Z}_{\mbox{\tiny RW}}=z^{N-1}$.  ${\cal Z}_{\mbox{\tiny RW}}$ is the partition function of the RW.  Out of these $z^{N-1}$ random-walk configurations simple sampling selects those which are self-avoiding.  As there are ${\cal Z}$ [$\sim \mu^N N^{\gamma-1}$, Eq.~\eqref{eq:rerg_Z}] such configurations, $P_\text{saw}(N)$ is given by
\begin{equation}
P_\text{saw}(N) \sim \bigg(\frac{\mu}{z}\bigg)^N N^{\gamma -1} 
= N^{\gamma -1}\, \text{e}^{-\lambda N} \sim \, \text{e}^{-\lambda N} 
\quad (N \; \mbox{large}) \; ,
\label{eq:attrition}
\end{equation}
where $\lambda=\ln(z/\mu)$ is called ``attrition constant''.  The attrition constant is $\lambda > 0$ (in 2D and 3D),\footnote{For a compilation of attrition constants see Refs.~\citen{Sokal_MCMD1995,KremerBinder1988}.  If $d\rightarrow \infty$, $\lambda$ goes to zero as $\lambda\rightarrow 1/2d$ for the hypercubic lattice.} reflecting that the monomer partition function, i.e., the number of ways to place a monomer on the lattice, of a RW ($=z$) is, due to the neglect of self-avoidance, larger than that of a SAW ($=\mu$).

Equation~\eqref{eq:attrition} illustrates that simple sampling is not an efficient simulation method for SAW's.  On average, $\text{e}^{\lambda N}$ random walks have to be constructed to obtain one SAW (``attrition problem'').  As $\lambda=0.416$ and 0.248 for the 2D and 3D hypercubic lattices, this number of constructions is prohibitively large already for $N \gtrsim 50$.  Even if one modifies the construction method by avoiding immediate backfolds, thereby replacing $z$ by $z-1$ (``Non-Reversal Random Walk (NRRW)''), the exponential attrition remains.  Generation of SAW's with $N > 10^2$ is still unfeasible.  Therefore, all alternative simulation techniques have to alleviate this attrition problem.

\subsection{Inversely Restricted Sampling: Rosenbluth-Rosenbluth Method}
\label{subsub:rosenbluth}

The attrition problem arises because simple sampling chooses blindly from the nearest-neighbor sites to place a new monomer.  A more clever algorithm could scan the local environment around the last monomer and exclude those trial directions which lead to self-intersections.  The position of the next monomer $i$ can then be chosen with equal probability from the remaining $k_i$ open directions (Fig.~\ref{fig:SSRR}).\footnote{Here, we assume $0< k_i \leq z-1$.  If all neighbor sites are blocked ($k_i=0$), the configuration obtained until then is trapped.  It must be discarded, and the construction resumes from the beginning.  The first monomer can be placed anywhere on the lattice.  If the construction always starts, say, from the origin, we set $k_1=1$ in Eq.~\eqref{eq:probRR}.}  This method, known as ``inversely restricted sampling'' or ``Rosenbluth--Rosenbluth (RR) algorithm'',\footnote{See Ref.~\citen{BatoulisKremer:JPhysA1988} for a detailed statistical analysis of the RR algorithm\cite{RR:JCP1955} in the context of simulating SAW's.  Chapter~11 of Ref.~\citen{FrenkelSmitBook} gives a discussion in the larger context of free energy calculations with applications to discrete and continuous chain models.} strongly reduces the attrition\footnote{In the RR algorithm, attrition occurs because the growing chain may be trapped, i.e., $k_i=0$ for some $i$.  In practice, this does not appear to be a serious problem in 3D for high-coordination lattices.\cite{BatoulisKremer:JPhysA1988,Guttmann:Macro1986}  It was suggested that effects of trapping should become visible only for $N \gtrsim 10^4$.\cite{Guttmann:Macro1986}  This implies that there is still exponential attrition, as $N\rightarrow \infty$, but at a much lower rate than in Eq.~\eqref{eq:attrition}.\label{pg:trapping}} at the expense of introducing a bias.  A SAW is not generated with uniform probability, but with probability
\begin{equation}
P_\text{s}(\boldsymbol{x}) = \prod_{i=1}^{N} \frac{1}{k_i} \quad (k_1=1) \;.
\label{eq:probRR}
\end{equation}
Equation~\eqref{eq:probRR} shows that configurations with small $k_i$'s have a higher probability of occurring.  This bias toward dense configurations in the production of a SAW must be corrected in its analysis by the weight $W(\boldsymbol{x}) \propto 1/P_\text{s}(\boldsymbol{x})$ when calculating observables [see Eq.~\eqref{eq:Aoverline}].  

The RR algorithm is a static MC method.  As such, it has the advantage that successively generated SAW's are independent of each other.  All problems of decorrelating configurations, discussed in Sec.~\ref{sec:is}, are absent by construction.  On the other hand, Eq.~\eqref{eq:probRR} also points to the major difficulty of the method.  The RR method favors dense configurations which are not representative of long SAW's.  Thus, $P_\text{s}(\boldsymbol{x})$ differs from $P_\text{eq}(\boldsymbol{x})$.  As $N$ increases, the difference becomes more pronounced.  To compensate the discrepancy between the two probabilities the distribution of weights must become broad:  Dense configurations have small weights and open chains have in general larger weights.  A detailed analysis of this problem was undertaken by Batoulis and Kremer.\cite{BatoulisKremer:JPhysA1988}  They showed that the distribution of weights, obtained from $M$ repetitions of the RR method ($\{W(\boldsymbol{x}_m)$, $m=1,\ldots,M$), is dominated by few configurations having the largest weights.  The most relevant SAW configurations have, however, smaller weights.  To sample this portion of the weight distribution sufficiently, $M$ has to become very large [see the discussion of Eq.~\eqref{eq:sfa_A_and_varA}].  This problem makes the RR algorithm not suitable for the simulation of long SAW's.\footnote{In the simulations on the (high-coordination) FCC lattice of Ref.~\citen{BatoulisKremer:JPhysA1988} the systematic error due to the weights rendered a precise determination of $R_\text{g}$ impossible for $N \gtrsim 200$.}

However, the RR algorithm should be well suited if the bias introduced by the sampling engenders configurations which are close to the physical ones.  That is, if the equilibrium configurations are less swollen than those of a SAW.  As this is the case close to $\Theta$-point in 3D (see Sec.~\ref{sec:intro}), one might expect the RR algorithm to be more efficient for $T \approx T_\Theta$.  In fact, this expectation is nicely borne out.  The biased sampling of the RR method produces an effective attraction between the monomers which closely resembles that of $\Theta$-chains.  An important consequence of this special property is that the weights are nearly compensated by the Boltzmann factor.\cite{Guttmann:Macro1986}  Thus, the RR method was employed to study properties near the $\Theta$-point (see e.g.\ Refs.~10,11 of Ref.~\citen{Guttmann:Macro1986}), and it also represents the core of a modern algorithm, the ``Pruned-Enriched Rosenbluth Method (PERM)\cite{Grassberger_PERM:PRE1997}''.

\subsection{Pruned-Enriched Rosenbluth Method (PERM)}
\label{subsub:PERM}

The simulation of a polymer chain close to the $\Theta$-point requires the introduction of an attraction between the monomers to compensate their mutual repulsion.  Typically, these thermal interactions are modeled by a short range inter-monomer potential.\footnote{For the SAW on the hypercubic lattice the attraction is usually implemented between non-bonded nearest neighbors.\cite{Grassberger_PERM:PRE1997}  In a simulation of the $\Theta$-transition with the BFM, a square-well potential of range $\sqrt{6}$ (in units of the lattice constant) was used.\cite{WildingEtal:JCP1996}  This choice ensures that the first peak of the pair-distribution function is encompassed by the range of the potential.  A more complicated choice was made in Ref.~\citen{WittkopEtal:JCP1996}.  In the continuum, a Gaussian chain model with a non-truncated LJ-interaction was extensively studied (see Refs.~\citen{YongEtal:JCP1996,RubioEtal:JCP1999} and references therein).}  As a chain is grown according to the Rosenbluth scheme, the presence of the potential implies that the internal energy of a chain changes:
\begin{equation}
u_i(\vec{b}_{i-1}) = U(\vec{r}_1,\ldots,\vec{r}_{i-1} + 
\vec{b}_{i-1}) - U(\vec{r}_1,\ldots,\vec{r}_{i-1}) \quad
[u_1(\vec{b}_{0}) := U(\vec{r}_1)] \;.
\label{eq:UbondRR}
\end{equation}
Here, $U(\vec{r}_1,\ldots,\vec{r}_{i-1})$ is the potential energy of chain having $(i\!-\!1)$ monomers and $\vec{b}_{i-1}$ is the bond vector from the $(i\!-\!1)$th to the $i$th monomer.  A priori, there are different ways to incorporate $u_i(\vec{b}_{i-1})$ in the construction.  Two possible choices are the following:\cite{Grassberger_PERM:PRE1997}
\label{pg:PERM}
\begin{enumerate}
\item The first method is the classical Rosenbluth scheme.  The position of the $i$th monomer is chosen from the free neighbors with uniform probability [see Eq.~\eqref{eq:probRR}].  Thus, the weight of the new chain configuration $\boldsymbol{x}$ ($=\vec{r}_1,\ldots,\vec{r}_N$) is given by~\footnote{Note that, contrary to the definition of $W(\boldsymbol{x})$ in Sec.~\ref{sec:mc}, Eq.~\eqref{eq:WmethA} does not include the factor of the (unknown) partition function.}  
\begin{equation}
W(\boldsymbol{x}) = \frac{\text{e}^{-\beta U(\boldsymbol{x})}}
{P_\text{s}(\boldsymbol{x})} = \prod_{i=1}^N \bigg [k_i 
\text{e}^{-\beta u_i(\vec{b}_{i-1})} \bigg ] \;.
\label{eq:WmethA}
\end{equation}
\item An alternative consists in including the Boltzmann factor in the probability $P_{\text{s},i}$ for placing the $i$th monomer.  Let $\{\vec{b}\}$ denote the ensemble of possible bond vectors.  For the hypercubic lattice, $\{\vec{b}\}$ coincides with the number of lattice directions $z$, for the BFM it is given by Eq.~\eqref{eq:BFMbonds}.  Then, we may write for $P_{\text{s},i}$ (Fig.~\ref{fig:SSRR})
\begin{equation}
P_{\text{s},i} = \frac{\text{e}^{-\beta u_i(\vec{b}_{i-1})}}
{\sum_{\{\vec{b}\}} \text{e}^{-\beta u_i(\vec{b})}} = 
\frac{\text{e}^{-\beta u_i(\vec{b}_{i-1})}}{\mathsfsl{w}_i}
\; \stackrel{\text{SAW}}{\longrightarrow} \;
\frac{1}{k_i} \;,
\label{eq:PsmethB}
\end{equation}
where the normalization  $\mathsfsl{w}_i$ reduces to $k_i$ if only excluded-volume interactions are taken into account (SAW limit).  This implies that the weight $W(\boldsymbol{x})$ is given by  
\begin{equation}
W(\boldsymbol{x}) = \frac{\text{e}^{-\beta U(\boldsymbol{x})}}
{P_\text{s}(\boldsymbol{x})} = \prod_{i=1}^N \bigg 
[\frac{\text{e}^{-\beta u_i(\vec{b}_{i-1})}}{P_{\text{s},i}} \bigg ] = 
\prod_{i=1}^N \mathsfsl{w}_i \;.
\label{eq:WmethB} 
\end{equation}
\end{enumerate}
Both methods have been used to simulate $\Theta$-polymers via the Rosenbluth algorithm.\cite{Grassberger_PERM:PRE1997}  

Despite the fact that, precisely at $T_\Theta$, the RR configurations more or less coincide with the equilibrium configurations, the accuracy of the method deteriorates for $N > N_\text{c}$.\footnote{The critical chain length $N_\text{c}$ depends on the model.  For a simple cubic lattice it is $N_\text{c} \approx 10^3$.\cite{Grassberger_PERM:PRE1997}}  Again, the reason is that $P_\text{s}(\boldsymbol{x})$ does not perfectly agree with $P_\text{eq}(\boldsymbol{x})$.  As a result, the weight distribution becomes so broad that chains with the biggest weights dominate the sample (see Fig.~2 of Ref.~\citen{Grassberger_PERM:PRE1997}).  Due to Eq.~\eqref{eq:Aoverline} this leads to a large variance of the computed observables.  

To improve the accuracy one has to reduce the variance.  Grassberger proposed a clever way to achieve this.\cite{Grassberger_PERM:PRE1997}  Assume that we have constructed a chain up to monomer $i$ ($1\leq i \leq N$) via the RR method.  This chain has the weight $W_i$ which we want to prevent from fluctuating too much.  That is, if $W_i$ exceeds a lower or an upper bound, we interfere in the following way:
\begin{enumerate}
\item If $W_i < W_i^-$, we ``prune'' the sample:  If $W_i$ decreases below the threshold $W_i^-$, a random number $0 \leq \zeta \leq 1$ is uniformly drawn.  If $\zeta < 1/2$, the chain is removed.  Otherwise, it is kept, its weight is doubled ($W_i \rightarrow 2 W_i$), and the step-by-step growth continues.  
\item If $W_i > W_i^+$, we ``enrich\footnote{Enrichment is a classical technique for simulating SAW's.  Briefly, it works as follows:  If a chain survives the $s$ step, $c$ copies of its configuration are made, which serve as independent starting points for further growth.  The method may be implemented in a ``breadth-first'' or a ``depth-first'' fashion.  The former implies that all copies are first grown to size $2s$ before the entire sample is copied again.  By contrast, the latter method tries to complete the construction of one copy up to chain length $N$ before passing to next one.  The pros and cons of the two implementations are discussed in the context of the PERM in Ref.~\citen{Grassberger_PERM:PRE1997}.  More details about enrichment may be found in Refs.~\citen{Sokal_MCMD1995,KremerBinder1988}.}'' the sample:  If $W_i$ exceeds the upper bound $W_i^+$, $c$ copies, typically 2,\cite{Grassberger:CPC2002} of the configuration are made, each of which is given the new weight $W_i \rightarrow W_i/c$.  These copies are then grown independently of each other.
\end{enumerate}
This control of the weight distribution within the RR algorithm was termed ``Pruned-En\-riched Ro\-sen\-bluth Method (PERM)''.\cite{Grassberger_PERM:PRE1997}  

Of course, the question arises of how to choose the bounds $W_i^\pm$.  Here, it is important to note that neither the pruning nor the enrichment step introduces any bias.  In the calculation of the sums in Eq.~\eqref{eq:Aoverline} the increase of $W_i$ by pruning is compensated by the probability $1/2$ with which the configuration is retained, and the decrease of $W_i$ in the enrichment is compensated by the number of copies $c$.  Thus, we are free to choose the bounds $W_i^\pm$.  Bad choices can ``only'' render the method inefficient, but not incorrect.  In order to determine optimum values for  $W_i^\pm$ the following procedure was proposed (for temperatures that are not too low):\cite{Grassberger_PERM:PRE1997,Grassberger:CPC2002}  
\begin{itemize}
\item First, one chooses $W_i^-=0$ and $W_i^+$ very large.  That is, one performs a simulation via the original RR method.  This simulation yields the weights $W_i$ for $i=1,\ldots, N$.  First estimates for the bounds $W_i^\pm$ are then determined by $W_i^-=C^-W_i$ and $W_i^+=C^+W_i$ with $C^+/C^- \approx$ O(1)-O(10).
\item These estimates are refined ``on the fly''.  Imagine that we have obtained $M_i$ configurations of chain length $i$ from the simulation.  Then, we first calculate the partition function by [$\boldsymbol{x}=(\vec{r}_1, \ldots, \vec{r}_i)$, see Eq.~\eqref{eq:Aoverline}]
\begin{equation}
{\cal Z}_i = \int \text{d}\boldsymbol{x}\,\text{e}^{-\beta U_i(\boldsymbol{x})}
\approx \frac{1}{M_i} \sum_{m=1}^{M_i} 
\frac{\text{e}^{-\beta U_i(\boldsymbol{x}_m)}}{P_\text{s}(\boldsymbol{x}_m)}
=\frac{1}{M_i} \sum_{m=1}^{M_i} W_i(\boldsymbol{x}_m) \;,
\label{eq:ZPERM}
\end{equation}
and from that, we determine the new bounds by $W_i^\pm = C^\pm {\cal Z}_i$.
\end{itemize}   

\paragraph{Applications of the PERM.}  The PERM was invented to simulate the transition from an excluded-volume to a collapsed chain at the $\Theta$-temperature $T_\Theta$.  Theoretically, the transition is usually identified with a tricritical point in the limit $N\rightarrow \infty$.\cite{Schaefer_RGBook}  A tricritical point exhibits mean-field behavior in 3D.  Thus, one expects that $\nu=1/2$ and $\gamma=1$ [Eq.~\eqref{eq:rerg_Z}] at $T_\Theta$ for $N\rightarrow \infty$.  This asymptotic large-$N$ behavior is supplemented by (universal) corrections of order $1/\ln N$ for finite chain length (see e.g.\ Chap.~21 of Ref.~\citen{Schaefer_RGBook} for a good discussion).  A significant test of the theory therefore requires to study very long chains.

In Ref.~\citen{Grassberger_PERM:PRE1997} such a test was attempted.  Grassberger performed a comparative study of various models employed in the literature: a SAW on a simple cubic lattice with attractive nearest-neighbor interactions, the BFM with two versions for the attractive monomer-monomer interactions,\cite{WildingEtal:JCP1996,WittkopEtal:JCP1996} and a LJ bead-spring model\cite{YongEtal:JCP1996}.  Using the PERM these models could be simulated with high precision and, partly, with much longer chain lengths than studied before.  These simulations yielded refined estimates of $T_\Theta$ for the various models, confirmed the mean-field-like asymptotic character of the $\Theta$-point, but also showed that the leading-order logarithmic corrections cannot explain the finite-$N$ behavior found, even for $N=10^4$.\footnote{These findings elicited further theoretical\cite{HagerSchaefer:PRE1999} and numerical work.\cite{RubioEtal:JCP1999,HagerSchaefer:PRE1999}  On the theoretical side, subleading corrections of order $\ln(\ln N)/(\ln N)^2$ were calculated and found to be as large as the leading $1/\ln N$ term.  On the numerical side, MC simulations\cite{HagerSchaefer:PRE1999} for $N \leq 10^4$ of a NRRW, including weakly attractive two-body, but repulsive three-body interactions, were performed.  This model shows logarithmic corrections which are much weaker than those found in Ref.~\citen{Grassberger_PERM:PRE1997} and are roughly compatible with the theoretical predictions.  However, Ref.~\citen{HagerSchaefer:PRE1999} stresses a problem in the analysis of the $\Theta$-point.  To estimate $T_\Theta$, an infinite-$N$ property, precisely from the simulations one has to rely on the theoretical predictions for the finite-$N$ corrections to extrapolate to $N\rightarrow \infty$.  To this end, the simulated chains must be long enough for these corrections to apply.  This regime appears to be very hard to attain, even for $N \sim 10^6$.}  

On the technical side, it was found that the selection of the position of the next monomer $i$ with uniform probability from the $k_i$ open directions is only the best choice for the SAW on the simple cubic lattice (first method of page~\pageref{pg:PERM}).  As alluded to at the end of Sec.~\ref{subsub:rosenbluth}, this is due to a near cancellation of the Rosenbluth weight and the Boltzmann factor:  Many nearest-neigbor contacts lead to a low Rosenbluth weight, but to a large Boltzmann factor, and vice versa.  For more long-range or more complicated interactions the degree of cancellation need not necessarily be the same.  In fact, for the BFM it was found in Ref.~\citen{Grassberger_PERM:PRE1997} that a selection of the next monomer position according to the Boltzmann factor [Eq.~\eqref{eq:PsmethB}] is more efficient.  Similar approaches were also used to study e.g.\ simple models of proteins.  

The preceding discussion appears to suggest that the PERM is a single-chain technique.  This is not true.  We just quote two recent examples.  The PERM was utilized to simulate the denaturation transition of a simple model for double-stranded DNA (two SAW's).\cite{CausoEtal:PRE2000}  A truly multi-chain system was studied in Ref.~\citen{FrauenkronGrassberger:JCP1997}.  This work is concerned with the phase diagram of semidilute polymer solutions for $T \leq \Theta$ (see Fig.~\ref{fig:phaseDiagram}).  For a review of these and other applications see Ref.~\citen{Grassberger:CPC2002}.

\subsection{Configurational-Bias Monte Carlo and Recoil-Growth Algorithm}
\label{subsub:CBMC_RG}

An alternative multi-chain MC scheme, incorporating the RR method, is the Con\-fi\-gu\-ra\-tional-Bias Monte Carlo (CBMC) algorithm.\footnote{CBMC was introduced by Siepmann and Frenkel in the 1990s.  It can be applied to lattice and off-lattice models.  The initial off-lattice applications have demonstrated the power of the algorithm for the study of a large variety of problems in polymer physics.  Therefore, CBMC has become an important and widely used simulation technique.  A comprehensive and very pedagogical account of the method, including flowcharts of the algorithm and examples, is given in the textbook by Frenkel and Smit.\cite{FrenkelSmitBook}}  Contrary to the PERM, CBMC builds up a new chain step-by-step without controlling the weights $\mathsfsl{w}_i$ ``on the go''.  It is only after a successful construction that the resultant bias is removed:  The new chain is accepted according to the Metropolis criterion with a probability dependent on the total Rosenbluth weights of the new and the old chain configurations.  This additional test warrants sampling from the Boltzmann distribution.

A recent extension of CBMC is the ``Recoil-Growth (RG) algorithm''.\footnote{The RG algorithm was introduced in Refs.~\citen{ConstaEtal:JCP1999,ConstaEtal:MolPhys1999}.  A detailed description may be found in Chap.~13.7 of the textbook by Frenkel and Smit.\cite{FrenkelSmitBook}  For practical applications, it is very helpful that the \texttt{FORTRAN} codes of the ``Case Studies'' may be downloaded from \texttt{http://molsim.chem.uva.nl/frenkel\_smit}.}  Contrary to the RR method, which only looks ahead one step while constructing a chain, the RG algorithm uses a more sophisticated growth procedure.  It places a long retractable feeler at the head of a growing chain.  The feeler spys out the environment to find favorable pathways for the chain construction.  The efficiency of the method resides in the fact that the growth does not terminate if the feeler encounters a trap.  It merely recoils back from the trap and pursues its search in a different direction.  After the construction is completed, the new chain replaces the old one, just as in CBMC, with a probability determined by their respective weights according to the Metropolis criterion.    

\subsubsection{Configurational-Bias Monte Carlo (CBMC)}
\label{subsubsec:CBMC}  
To illustrate the CBMC scheme in more detail we consider a solution of SAW's on a lattice (Fig.~\ref{fig:cbmc}):
\begin{figure}
\begin{center}
\epsfysize=42mm
\epsfbox{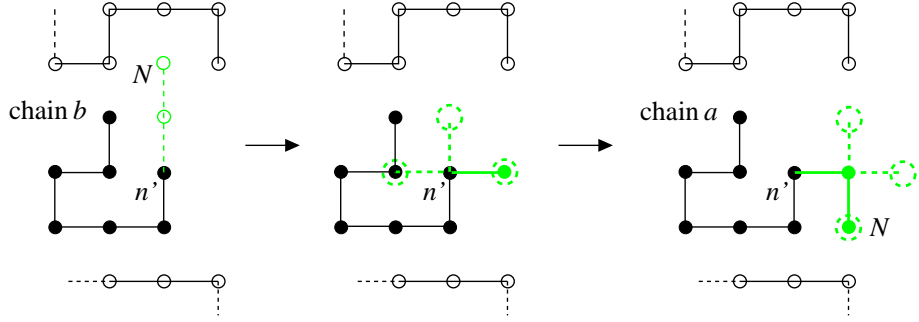}
\end{center}
\caption[]{Schematic of a CBMC move executed in a polymer solution.  The initial configuration $\boldsymbol{x'}$ of the solution is shown in the left figure.  From all chains of $\boldsymbol{x'}$ the chain $b$ and one of its monomers, $n'$, are chosen.  Here, $n'=N-2$.  The chain portion $n',\ldots,N$ is deleted (dashed grey lines connecting the open grey circles) and reconstructed step-by-step.  Upon completion we obtain a new chain configuration, ``chain $a$'', and so a new configuration, $\boldsymbol{x}$, of the solution.  In the construction, the new bond $\vec{b}_{i-1}$ from monomer $i\!-\!1$ to monomer $i$ ($=N-1,N$) is selected from the set of bond vectors $\{\vec{b}\}$ according to the local Boltzmann factor $P_{\text{s},i}^a(\boldsymbol{x})$ [Eq.~\eqref{eq:PCBMC_forth}].  For lattice models this set, which defines the number of trial directions at each step, is finite and closely related to lattice structure:  E.g.\ for the hypercubic lattice, it coincides with the number of lattice directions $z$ (or $z-1$ if backfolds shall be excluded a priori).  For the BFM the trial directions may be taken as the ensemble of allowed bonds [Eq.~\eqref{eq:BFMbonds}].  For a continuum model there is a priori an infinite number of possible directions, from which a suitable finite number of trial directions must be selected so that new chain configurations are efficiently generated (see Chap.~13.3 of Ref.~\citen{FrenkelSmitBook} for details).  If only excluded volume interactions are present, as assumed in the figure, $P_{\text{s},i}^a(\boldsymbol{x})=1/2$ for $i=N-1$ and $P_{\text{s},i}^a(\boldsymbol{x})=1/3$ for the last monomer $N$.  Thus, the total weight $W^a(\boldsymbol{x})$ of chain $a$ is 6, whereas that of the chain portion $n'=N-1,N$ of the old chain $b$ is $W^b(\boldsymbol{x'})=4$.  Thus, $W^a(\boldsymbol{x})/W^b(\boldsymbol{x'})=1.5$, and chain $a$ will be accepted according to Eq.~\eqref{eq:metropolis_CBMC}.}
\label{fig:cbmc}
\end{figure}
\begin{enumerate}
\item Given the initial configuration $\boldsymbol{x'}$ of the system we randomly select a chain and one of its monomers.  Let this be the monomer $n'$ of chain $b$ ($1 \leq n' \leq N$).  The confi\-guration of the chain portion $i=n',\ldots, N$ is characterized by the sequence of bonds $\vec{b}_{n'-1}, \ldots, \vec{b}_{N-1}$.  Each bond represents a specific choice from the set of all possible bond vectors $\{ \vec{b} \}$.  So, we can write the Rosenbluth weight of the monomers $n', \ldots, N$ of chain $b$ as 
\begin{equation}
W^b(\boldsymbol{x'}) = \prod_{i=n'}^N \mathsfsl{w}_i^b(\boldsymbol{x'})\;,
\quad \mathsfsl{w}_i^b(\boldsymbol{x'}) = \text{e}^{-\beta 
u_i^b(\vec{b}_{i-1})} + \sum_{\vec{b}\neq \vec{b}_{i-1}} 
\text{e}^{-\beta u_i^b(\vec{b})} \;.
\label{eq:W_CBMC_back}
\end{equation}
Here, $u_{n'}^b(\vec{b}_{n'-1})$ is the energy of monomer $n'$ at its actual position in the chain.  It includes the interactions with the monomers of all other chains and with the monomers $i=1,\ldots,n'-1$ of its own chain [Eq.~\eqref{eq:UbondRR}].  The monomers $n'+1$ to $N$ have to be omitted because one thinks of the chain $b$ as being (re-) constructed step-by-step via the RR procedure with probability [Eq.\eqref{eq:PsmethB}]
\begin{equation} 
P_{\text{s},i}^b(\boldsymbol{x'})=\frac{\text{e}^{-\beta u_i^b(\vec{b}_{i-1})}}
{\mathsfsl{w}_i^b (\boldsymbol{x'})} \;.
\label{eq:Psi_CBMC_back}
\end{equation}
Thus, the potential energy of the chain portion $n',\ldots,N$ is given by
\begin{equation}
U^b(\boldsymbol{x'}) = \sum_{i=n'}^N  u_i^b(\vec{b}_{i-1}) \;,
\label{eq:U_CBMC_back}
\end{equation}
and the probability to propose the chain portion may be expressed as 
\begin{equation}
P_\text{s}^b(\boldsymbol{x'}) = \prod_{i=n'}^N 
P_{\text{s},i}^b(\boldsymbol{x'})= 
\frac{\text{e}^{-\beta U^b(\boldsymbol{x'})}}
{W^b(\boldsymbol{x'})} \;.
\label{eq:CBMC_back}
\end{equation}
\item The monomers $n'$ to $N$ of chain $b$ are deleted.  This corresponds to a ``shrinkage-growth'' implementation of the algorithm, which is (presumably) more efficient than a ``growth-shrinkage'' procedure (see Sec.~\ref{subsubsec:snake}).
\item To obtain a new configuration $\boldsymbol{x}$ of the system the chain $b$ is fully ($n'=1$) or only in part ($n'>1$) reconstructed:
\begin{itemize}
\item If $n'=1$, we start building a new chain by randomly placing the first monomer somewhere in system.  Let this new chain be labeled ``chain $a$''. At the position, where the monomer is inserted, it interacts with the other chains of the system.  It has an energy $u_{n=1}^a(\vec{b}_{0})$ [$= U(\vec{r}_1)$, see Eq.~\eqref{eq:UbondRR}], giving rise to the weight $\mathsfsl{w}_1=\exp[-\beta u_{1}^a(\vec{b}_{0})]$.  
\item If $n'> 1$, we rebuild the chain portion $i=n',\ldots,N$ monomer-by-monomer via the RR method.  This growing chain, comprising the initial portion of chain $b$ ($i=1,\ldots, n'-1$) and the newly attached monomers ($i=n,\ldots,N$) will also be called ``chain $a$''.  In the following we suppress the prime ($^\prime$) to indicate that a new chain conformation is obtained, even if the first $n\!-\!1$ monomers are identical with those of chain $b$.  To add the $n$th monomer we proceed as described in the next step.
\end{itemize}
\item A monomer may be attached to the growing chain via a bond from the set of possible bond vectors $\{\vec{b}\}$.  Out of these trial directions we choose one according to Eq.~\eqref{eq:PsmethB}.  This implies for the monomer $i$ ($=n,\ldots,N$)
\begin{equation}
P_{\text{s},i}^a(\boldsymbol{x})=\frac{\text{e}^{-\beta u_i^a(\vec{b}_{i-1})}}
{\sum_{\{\vec{b}\}} \text{e}^{-\beta u_i^a(\vec{b})}} = 
\frac{\text{e}^{-\beta u_i^a(\vec{b}_{i-1})}}
{\mathsfsl{w}^a_i(\boldsymbol{x})} \;,
\label{eq:PCBMC_forth}
\end{equation}
where $u_i^a(\vec{b}_{i-1})$ is the change in potential energy of the system due to the addition of the new bond $\vec{b}_{i-1}$ from monomer $i\!-\!1$ to monomer $i$ [Eq.~\eqref{eq:UbondRR}].
\item The preceding step is repeated until the construction of the chain is completed.  Thus, the new chain configuration occurs with probability [Eq.~\eqref{eq:WmethB}]
\begin{equation}
P_\text{s}^a(\boldsymbol{x}) = \prod_{i=n}^N P_{\text{s},i}^a(\boldsymbol{x}) 
= \frac{\text{e}^{-\beta U^a(\boldsymbol{x})}}{W^a(\boldsymbol{x})} \;,
\label{eq:CBMC_forth}
\end{equation}
where $W^a(\boldsymbol{x})$ and $U^a(\boldsymbol{x})$ are defined analogously to Eqs.~(\ref{eq:W_CBMC_back},\ref{eq:U_CBMC_back}).
\item Now, we recognize that $P_\text{s}^a(\boldsymbol{x})$ may be interpreted as the probability of proposing a transition from the old configuration $\boldsymbol{x'}$ to the new configuration $\boldsymbol{x}$.  Correspondingly, $P_\text{s}^b(\boldsymbol{x'})$ is the probability for the reverse step.  That is,
\begin{equation}
P_\text{s}^a(\boldsymbol{x}) = P_\text{pro}(\boldsymbol{x'}\rightarrow\boldsymbol{x}) \quad \text{and} \quad 
P_\text{s}^b(\boldsymbol{x'}) = P_\text{pro}(\boldsymbol{x}\rightarrow\boldsymbol{x'})  \;.
\label{eq:CBMC_forth+back}
\end{equation}
Finally, we insert Eq.~\eqref{eq:CBMC_forth+back} into Eq.~\eqref{eq:metropolis} to obtain the acceptance probability for the new configuration 
\begin{align}
\text{acc}(\boldsymbol{x'}\rightarrow \boldsymbol{x}) 
& = \min \bigg (1,\frac{P_\text{pro}(\boldsymbol{x}
\rightarrow \boldsymbol{x'})} {P_\text{pro}(\boldsymbol{x'}\rightarrow 
\boldsymbol{x})}\,\text{e}^{-\beta\left[U^a(\boldsymbol{x}) - 
U^b(\boldsymbol{x'}) \right]} \bigg) \nonumber \\
& = \min \bigg (1,\frac{W^a(\boldsymbol{x})}
{W^b(\boldsymbol{x'})} \bigg) \;.
\label{eq:metropolis_CBMC}
\end{align} 
\end{enumerate}

\paragraph{Discussion.}  An important feature of the CBMC method is its non-local character:  A successful CBMC construction implies a large-scale configurational change.  Either a new chain is inserted somewhere in the system --this may be employed very efficiently to study phase equilibria of polymer solutions in the bulk\cite{dePablo:AnnRevPhysChem1999} and in thin films\cite{MullerEtal:IJMP2001}-- or part\footnote{This part need not necessarily start at monomer $n'+1$ and terminate at the chain end, as assumed in our previous discussion.  It can also comprise an inner portion of the chain, say the monomers $n'+1,\ldots,n'+m'-1 <N$.  In this case, the reconstruction has to satisfy the additional condition that it starts at monomer $n'$ and ends at the position of monomer $n'+m'$ (see Chap.~13.4 of Ref.~\citen{FrenkelSmitBook}).  This variant of CBMC may be used to relax e.g.\ ring polymers, which have no free ends so that the method described above could not be applied.} of a chain is regrown.  This leads to a rapid decorrelation of chain configurations and to efficient sampling, provided the system has a low or moderate density and the chains are not too long.  When dealing with long chains and/or dense melts the following problems occur:
\begin{itemize}
\item Chain construction in CBMC is based on the Rosenbluth method, yielding a distribution of configurations that differs from the Boltzmann distribution.  This difference becomes more pronounced with increasing $N$, even for an isolated chain at the $\Theta$-point (see Secs.~\ref{subsub:rosenbluth},\ref{subsub:PERM}).  As CBMC does not control the weights while synthesizing the chain, contrary to the PERM, the acceptance rate for chain reconstructions falls exponentially in $N$ for large chain lengths.    
\item Another problem results from the ``shortsightedness'' of the RR algorithm.  By looking only one step ahead, the chain construction can run into traps.  For isolated chains this trapping implies that there is still an exponential attrition in $N$ for long chains, albeit with an attrition constant much smaller than $\lambda$ (see Eq.~\eqref{eq:attrition} and footnote on page~\pageref{pg:trapping}). 
\item In a dense system, in addition to trapping, a further problem occurs.  If a (part of a) new chain is inserted, it is fairly likely to be constructed just in the space originally occupied by the (part of the) old chain which was removed.  This can lead to strong correlations between the new and old chain configurations so that sampling becomes inefficient.\cite{LeontidisEtal:1995}  In this situation, it might be best to combine CBMC with the slithering-snake algorithm (Sec.~\ref{subsubsec:snake}).  That is, to try to regrow just the terminal bond at the other chain end according to Eq.~\eqref{eq:metropolis_CBMC}.  In a dense melt, it thus appears as if CBMC cannot be expected to decorrelate chain configurations more efficiently than the slithering-snake algorithm.
\end{itemize}

%
% Reinitialize footnote counter due to overflow
% ---------------------------------------------
\setcounter{footnote}{0}
%----------------------------------------------
%

\subsubsection{Recoil-Growth (RG) Algorithm}
\label{subsubsec:rg}  
The RG algorithm was suggested as an alternative to CBMC,\footnote{The RG algorithm was introduced for SAW's on a cubic lattice in Ref.~\citen{ConstaEtal:JCP1999}.  Reference~\citen{ConstaEtal:MolPhys1999} extends this study to continuum models.  A comprehensive discussion of these works may be found in the textbook by Frenkel and Smit (Chap.~13.7).\cite{FrenkelSmitBook}} exhibiting two major changes:
\begin{enumerate}
\item Instead of looking one step ahead the RG algorithm scans the environment via a retractable ``feeler''.  The feeler consists of a self-avoiding chain portion having at most $N_\text{recoil}$ monomers.\footnote{Of course, in general $N_\text{recoil} > 2$, implying that the feeler is longer than one bond.  Otherwise, the ``shortsightedness'' of CBMC is not removed.  The idea to improve the RR method by looking several steps ahead is not new.  It is embodied e.g.\ in the ``scanning method'' of Meirovitch.\cite{scanningmethod}  This method still uses a one-step growth, but chooses a new bond $\vec{b}_{i}$ according to the probability that a SAW of $N_\text{scan}$ monomers can be constructed in direction of $\vec{b}_{i}$.  As this implies an enumeration of all possible SAW's of length $N_\text{scan}$ starting at some monomer $i$, the scanning parameter $N_\text{scan}$ is usually much smaller than $N$.  Thus, trapping cannot be avoided completely.  This would only be the case if $N_\text{scan}=N-i$, i.e., if one scanned all possible ways to complete the chain up to monomer $N$ in direction of $\vec{b}_{i}$.  For a comparative discussion of the RG algorithm  and the scanning method see Ref.~\citen{ConstaEtal:JCP1999}.}  The ability of the feeler to shrink and to grow helps to circumvent dense regions.  This allows for the search of suitable pathways to complete the construction of the chain.  
\item Contrary to CBMC, the incremental weights $\mathsfsl{w}_i$ for each newly added monomer are not calculated ``on the fly'', but only after a new chain has been successfully constructed.  Thus, the computation of the weights is carried out only once.  In CBMC, it can happen that a lot of time is spent to calculate the weights of a partially grown chain which must then be discarded because the construction has run into a trap before completion.  
\end{enumerate}
\paragraph{Description of the RG Algorithm.} These differences show that the RG algorithm comprises two independent steps: a construction step using the retractable feeler and an acceptance step including the weight calculation.  We discuss these steps separately.  In our discussion we assume that a whole chain is inserted in the system after a randomly chosen old chain has been removed (``shrinkage-growth procedure'').  As before, the new chain is called ``chain $a$'', the old ``chain $b$'',  and the respective new and old configurations of the system are denoted by $\boldsymbol{x}$ and $\boldsymbol{x'}$.

\vspace{4mm} 

\noindent Construction step (Fig.~\ref{fig:rg}):
\begin{enumerate}
\item We place the first monomer of chain $a$ at the random trial position $\vec{r}_1$ and determine its energy $U(\vec{r}_1)=u^a_1(\vec{b}_0)$.  To decide whether the position is accessible (``open'') we accept it with probability (see e.g.\ Ref.~\citen{FrenkelSmitBook} for further discussion of this point)
\begin{equation}
P^\text{open}_1(\boldsymbol{x}) = \min \Big (1,
\text{e}^{-\beta u^a_1(\vec{b}_0)} \Big) 
\;\; \stackrel{\text{SAW}}{\longrightarrow} \;\;
\begin{cases}
1 & \text{if $\vec{r}_1$ is unoccupied}\;, \\
0 & \text{otherwise} \;,
\end{cases}
\label{eq:RGopen}
\end{equation}
where ``SAW'' means that there are only hard-core interactions.\footnote{In the SAW-limit, the decision of whether the position is open is not probabilistic, but deterministic.  One just checks on overlaps with other monomers:  If no overlap occurs, the position is open.   For continuous potentials, however, the decision becomes probabilistic.  We compare $P^\text{open}_1(\boldsymbol{x})$ to a random number $\zeta$, uniformly distributed between 0 and 1.  The position is open if $\zeta < P^\text{open}_1(\boldsymbol{x})$.  Otherwise, it is closed.  This means that we accept the position with probability $P^\text{open}_1(\boldsymbol{x})$.}  If the position is ``open'', we continue with step 2. Otherwise, step 1 is repeated.
\begin{figure}
\begin{center}
\epsfysize=45mm
\epsfbox{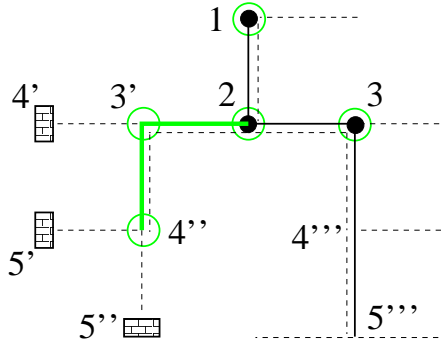}
\end{center}
\caption[]{Illustration of the recoil-growth procedure for a SAW on a square lattice, using $k_\text{RG}=2$ and $N_\text{recoil}=3$.  The first monomer is placed at an empty (``open'') lattice site.  At each site that the growing chain visits, it has $k_\text{RG}$ trial directions for the next step (indicated by the dashed lines: there are 2 for each monomer).  These directions are chosen at random from the $z\!-\!1$ forward lattice directions.  Starting at monomer 1 the chain first tries the path $1\, 2\, 3'$.  At that point, the first monomer ($3-N_\text{recoil}+1$) becomes fixed (indicated by $\bigcirc \rightarrow \bullet$ in the figure).  Then, the chain attempts to grow to position $4'$ and finds it blocked.  Thus, it recoils to monomer $3'$ and tries the remaining direction leading to monomer $4''$.  At that point, the monomer $4 - N_\text{recoil}+1=2$ becomes fixed ($\bigcirc \rightarrow \bullet$), as the ``feeler'' $2\, 3'4''$ has attained its full length $N_\text{recoil}=3$ (grey thick solid line).  However, the next two trial directions, $4'\rightarrow 5'$ and $4'\rightarrow 5''$, are found blocked.  So, the feeler first recoils to $3'$, where it realizes that $k_\text{RG}$ trial directions have been exhausted so that it must fall back its maximum length $N_\text{recoil}$ to monomer 2.  From there, it finds the open path $2\,3\,4''',5'''$ so that monomer 3 becomes permanently attached to the chain.  In this way, the construction goes on up to monomer $N-N_\text{recoil}+1$.  If this monomer was fixed, the construction stops, as a feeler to monomer $N$ exists.  Figure adapted from Ref.~\citen{ConstaEtal:JCP1999}.}
\label{fig:rg}
\end{figure}
\item Assume that the chain construction has arrived at monomer $i$.  We randomly choose a new bond $\vec{b}_{i}$ from monomer $i$ to monomer $i+1$, compute $P^\text{open}_{i+1}(\boldsymbol{x})$ for the trial position of $i\!+\!1$, and decide whether it is open or not according to Eq.~\eqref{eq:RGopen}.  If it is closed, we keep choosing new bonds up to a maximum number of $k_\text{RG}$ trial directions.\footnote{In practical implementations, the number of trial directions need not be the same for every monomer $i$.  It may depend on $i$.  For instance, one could choose for half of the monomers 2 trial directions and for the other half 3 directions so that $k_\text{RG}=2.5$ on average.  Thus, $k_\text{RG}$ need not be integer.}  As soon as the first open direction is found, we proceed to step 3.

Otherwise, a recoil step is performed.  That is, the chain moves back to monomer $i-1$ where it searches for an open direction, using the $k_\text{rest}=k_\text{RG}-k_\text{checked}$ previously unchecked directions.  The first open direction found is used to place (again) monomer $i$.  If the chain fails to find an open direction from the $k_\text{rest}$ remaining ones at monomer $i-1$, it recoils to $i-2$ and checks the previously unused directions for a possibility to grow from there.  In difficult situations the chain keeps falling back until a maximum number of $N_\text{recoil}$ recoil steps has been performed.  If this number is exceeded, the construction of chain has to resume from step 1.

This shows that the RG algorithm is characterized by two parameters: $k_\text{RG}$ and $N_\text{recoil}$.  Tuning of these parameters is important to optimize the efficiency of the method.
\item If the construction has successfully placed the $n$th monomer, we attach monomer $n-N_\text{recoil}+1$ permanently to the chain, as recoiling can only fall back to $n-N_\text{recoil}+1$.  If this occurs and still no open direction is found at monomer $n-N_\text{recoil}+1$, the growth process terminates and we must go back to step 1.  Thus, the $N_\text{recoil}$ monomers, $i=n-N_\text{recoil}+1,\ldots,n$, may be regarded as ``retractable feeler'' of maximum length $N_\text{recoil}$ which probes the territory ahead of monomer $n-N_\text{recoil}$. 
\item Repeat steps 2 and 3 until the monomer $N-N_\text{recoil}+1$ was attached to the chain.  This implies that the feeler attained the chain end.  A complete chain of length $N$ has thus been constructed.  
\end{enumerate}

\vspace{4mm} 

\noindent Acceptance step:  If chain $a$ has been successfully constructed, we have to determine its weight $W^a(\boldsymbol{x})$ and that of the old chain $b$, $W^b(\boldsymbol{x'})$, which chain $a$ attempts to replace.  This may be done as follows:
\begin{enumerate}
\item In the construction step each monomer $i$ disposes of (at most) $k_\text{RG}$ trial directions into which the growth of a feeler of maximum length $N_\text{recoil}$ may be attempted.  Assume that we have checked $k=1,\ldots,k_\text{checked}$ of these directions and found that the $k_\text{checked}$th direction is the first allowing us to complete the construction of the feeler up to length $N_\text{recoil}$.  We now test the remaining $k_\text{rest} =  k_\text{RG} - k_\text{checked}$ directions to find out how many ``feelers'' of length $N_\text{recoil}$ can be grown.  (For the last monomers, $N-N_\text{recoil}+1 < i \leq N$, the length of the feeler is shortened by one bond, as $i$ advances step-by-step toward $N$.)  For monomer $i$ let $m_i(\boldsymbol{x})$ denote the number of directions where a feeler of full length can be grown.  We have $1 \le m_i(\boldsymbol{x}) \le 1 + k_\text{rest}$.  As monomer $i$ is only irrevocably added to the chain, if its position was initially open [Eq.~\eqref{eq:RGopen}] and if at least one feeler of length $N_\text{recoil}$ may be grown from it, the probability with which we propose to place monomer $i$ is given by [see Eq.~\eqref{eq:PCBMC_forth}]
\begin{equation}
P_{\text{s},i}^a(\boldsymbol{x}) = P^\text{open}_i (\boldsymbol{x}) \cdot
\frac{1}{m_i(\boldsymbol{x})} = \frac{1}{\mathsfsl{w}_i^a (\boldsymbol{x})} \;.
\label{eq:Psia_RG}
\end{equation}
With respect to CBMC $P_{\text{s},i}^a(\boldsymbol{x})$ differs by the absence of the Boltzmann factor $\exp[-\beta u_i^a(\vec{b}_{i-1})]$, since only excluded-volume interactions are taken into account during the construction.  So, the new chain $a$ is proposed with probability
\begin{equation}
P_\text{s}^a(\boldsymbol{x}) = \prod_{i=1}^N 
\frac{1}{\mathsfsl{w}_i^a (\boldsymbol{x})} =
\frac{1}{W^a(\boldsymbol{x})} 
\quad (m_N(\boldsymbol{x})=1) \;.
\label{eq:RG_forth}
\end{equation}
\item The weight calculation of the old chain $b$ has two ingredients.  First, we calculate $P^\text{open}_i(\boldsymbol{x'})$ for each monomer $i$ according to Eq.~\eqref{eq:RGopen}.\footnote{For the old chain we only calculate $P^\text{open}_i(\boldsymbol{x'})$.  This is different from the construction step, where we decide whether the position is open or closed according to probability $P^\text{open}_i(\boldsymbol{x})$.  That is, we first calculate $P^\text{open}_i(\boldsymbol{x})$ and then compare it to a random number $\zeta$ uniformly distributed between 0 and 1. If $\zeta < P^\text{open}_i(\boldsymbol{x})$, the position is open.  Otherwise, it is closed [see Eq.~\eqref{eq:RGopen}].}  Second, we attempt to grow feelers in $k_\text{RG}-1$ randomly chosen directions and we count the number of feelers that attain length $N_\text{recoil}$.  This number plus 1 yields $m_i(\boldsymbol{x'})$ (``plus 1'', since one feeler of length $N_\text{recoil}$ already exists along the backbone of chain $b$).  In analogy to chain $a$, we thus write
\begin{equation}
P_\text{s}^b(\boldsymbol{x'}) = \prod_{i=1}^N 
\frac{P^\text{open}_i(\boldsymbol{x'})}{m_i(\boldsymbol{x'})}
=\prod_{i=1}^N \frac{1}{\mathsfsl{w}_i^b (\boldsymbol{x})} =
\frac{1}{W^b(\boldsymbol{x'})}
\quad (m_N(\boldsymbol{x'})=1) \;.
\label{eq:RG_back}
\end{equation}
\item Finally, the total potential energy of chains $a$ and $b$, $U^a(\boldsymbol{x})$ and $U^b(\boldsymbol{x'})$, must be computed before the new chain can be accepted with probability [see Eqs.~(\ref{eq:CBMC_forth+back},\ref{eq:metropolis_CBMC})]   
\begin{equation}
\text{acc}(\boldsymbol{x'}\rightarrow \boldsymbol{x})
= \min \bigg (1,\frac{W^a(\boldsymbol{x})}{W^b(\boldsymbol{x'})}
\,\text{e}^{-\beta\left[U^a(\boldsymbol{x}) -
U^b(\boldsymbol{x'}) \right]} \bigg) \;.
\label{eq:metropolis_RG}
\end{equation}
\end{enumerate}
The RG algorithm has two adjustable parameters: $k_\text{RG}$, the (average) number of trial directions of a monomer, and $N_\text{recoil}$, the length of the retractable feeler.  Intuitively, one expects that $N_\text{recoil}$ should be large, whereas $k_\text{RG}$ should be small.  The main advantage of the RG algorithm is its ability to avoid traps by probing the environment with a feeler.  A short feeler will strongly reduce this ability and thus the rate of successful chain construction.  On the other hand, the value of $k_\text{RG}$ should not be too large because $1 \leq m_i(\boldsymbol{x}) \leq k_\text{RG}$.  (Remember that $m_i(\boldsymbol{x})$ denotes the number of feelers of length $N_\text{recoil}$, which may be grown from monomer $i$.)  A large $k_\text{RG}$ allows for many different values of $m_i(\boldsymbol{x})$, as we go along the chain to calculate the weight $W^a(\boldsymbol{x})$.  The spread in $m_i(\boldsymbol{x})$ leads to a wide distribution of $W^a(\boldsymbol{x})$, which will strongly deviate from the Boltzmann distribution (see Sec.~\ref{subsub:rosenbluth}).  

\paragraph{Applications of the RG Algorithm.}  These expectations are confirmed by applications of the RG algorithm to 3D lattice\cite{ConstaEtal:JCP1999} and 3D off-lattice\cite{ConstaEtal:MolPhys1999} models of polymer solutions.  For the chain lengths studied ($N \leq 100$) good choices are $2 \lesssim k_\text{RG} \lesssim 3$ and $3 \lesssim N_\text{recoil} \lesssim 10$, with the need to have larger values for both parameters if the density of the solution increases.  Furthermore, it was observed that, for high densities and long chains, the RG algorithm may be an order of magnitude more efficient than CBMC. 

Our preliminary results\cite{AicheleEtal:InPrep} on 2D polymer solutions, simulated with the Kremer-Grest model (see Sec.~\ref{subsec:continuum}), appear to confirm the trends observed in 3D.  Presumably due to the fact that the risk of trapping is more severe in 2D than in 3D,\cite{BatoulisKremer:JPhysA1988,HemmerHemmer:PRA1986} we found that, even for an isolated chain, chain lengths $N \gtrsim 100$ are very difficult to simulate via CBMC.  Here, the RG algorithm provides a powerful alternative.  While the performance of the algorithm depends only weakly on the choice of $N_\text{recoil}$, provided a long feeler is employed, e.g.\ $N_\text{recoil}=N/2$, $k_\text{RG}$ must be optimized carefully. 
\begin{figure}
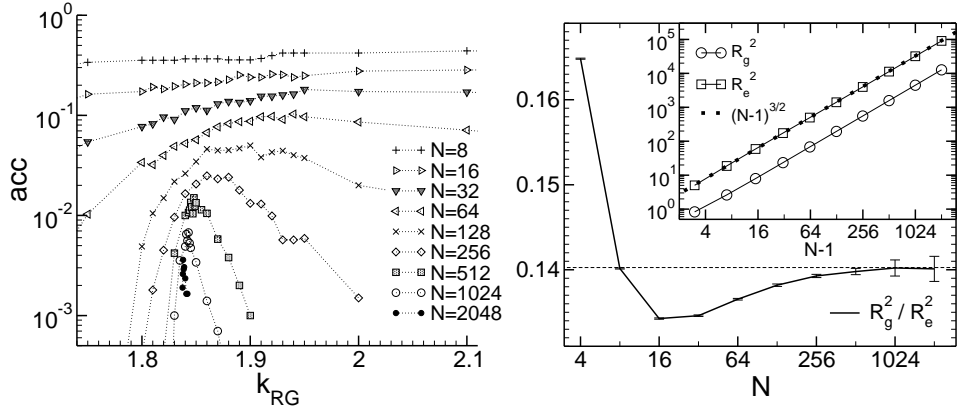

\begin{center}
\vspace*{30mm}
\begin{picture}(125,54)
\unitlength=1mm
\put(-41,0){%
\epsfysize=54mm
\epsfbox{timings_Nrec_eq_half_N_allN_single_chain.eps}
}
\put(26,1.25){%
\epsfysize=50mm
\epsfbox{Rg_over_Re_vs_N.eps}
}
\end{picture}
\end{center}
\caption[]{Left: Acceptance probability versus $k_\text{RG}$ for $8 \leq N \leq 2048$ obtained from simulations of isolated 2D chains (``Kremer-Grest model'') via the RG algorithm.  For all cases, $N_\text{recoil}=N/2$.  Note the sharp peak of $\text{acc}(k_\text{RG})$ for long chains.  The value of $k_\text{RG}$ at the peak yields the optimum choice $k_\text{RG}^\text{opt}(N)$ for this chain length.  For $k_\text{RG}^\text{opt}$ we find: $\text{acc}\propto 1/N$.  Right:  Plot of $R^2_\text{g}/R^2_\text{e}$ versus $N$ (main figure) and of $R^2_\text{e}$ and $R^2_\text{g}$ versus the number of bonds $N-1$ (inset).   The dashed horizontal line in the main figure represents the large-$N$ limit of $R^2_\text{g}/R^2_\text{e}$ [Eq.~\eqref{eq:char_ratio_2D}].  The dotted line in the inset shows the theoretically predicted power law $R^2_\text{e} \propto R^2_\text{g} \sim N^{2\nu}$ with $\nu=3/4$.}
\label{fig:acc_and_radii_2D_KremerGrest}
\end{figure}
Figure~\ref{fig:acc_and_radii_2D_KremerGrest} shows that the acceptance probability develops a pronounced peak for long chains.  While for $N \lesssim 64$ the acceptance probability is fairly insensitive to the precise choice of $k_\text{RG}$, if $k_\text{RG} \approx 2$, it has to be adjusted to 4 significant digits for $N=2048$.  In this case, the optimum value is $k_\text{RG}^\text{opt}=1.838$.  This means that on average there are four monomers with two trial directions, followed by one monomer with only one trial direction.\footnote{We arrive at that conclusion in the following way.  Suppose we allow the monomers to have either 1 or 2 trial directions.  Let $p$ denote the probability that a monomer has two trial directions.  Then, we pose $k_\text{RG}=2 \cdot p + 1\cdot (1-p)$ so that $p=0.838$ with $k_\text{RG}^\text{opt}=1.838$ for $N=2048$.  Thus, out of 5 monomers, roughly four monomers have two trial directions, leaving one trial direction for the remaining monomer.} 

To exemplify that the RG algorithm produces correct statistical properties Fig.~\ref{fig:acc_and_radii_2D_KremerGrest} also shows the radius of gyration $R_\text{g}^2$, the end-to-end distance $R^2_\text{e}$, and the ratio $R_\text{g}^2/R^2_\text{e}$.  For isolated, two-dimensional chains one expects to find the critical exponent $\nu=3/4$~\cite{PelissettoVicari:PhysRep2002} and the ratio\cite{CaraccioloEtAl:JPA1990}
\begin{equation}
\lim_{N \to \infty} \frac{R_\text{g}^2}{R_\text{e}^2} = 0.14026 \pm 0.00011 \;.
\label{eq:char_ratio_2D}
\end{equation}
The figure illustrates that both predictions are well borne out by the simulation data.  Apparently, the 2D Kremer-Grest model quickly converges to the large-$N$ limit, i.e, to $R^2_\text{e} \propto R^2_\text{g} \sim N^{3/2}$ and to Eq.~\eqref{eq:char_ratio_2D}.  Deviations are only visible when investigating the ratio $R^2_\text{g}/R^2_\text{e}$ for $N \lesssim 512$.  For longer chains corrections-to-scaling are small.  In this respect, the Kremer-Grest model agrees with the results obtained for the SAW on a square lattice.\cite{CaraccioloEtAl:JPA1990}

%%%%%%%%%%%%%%%%%%%%%%%%%%%%%%%%%%%%%%%%%%%%%%%%%%%%%%%%%%%%%%%%%%%%%%%%
\section{Conclusions}
\label{sec:conc}

Behind the title of this chapter ``Monte Carlo Simulation of Polymers: Coarse-Grained Models'' a topic of a large breadth is hidden.  So, a selection is necessary.  We have employed several criteria in this selection.  

First, we concentrate on simple generic models retaining only basic features of a polymer chain (chain connectivity, excluded volume, etc.; see Sec.~\ref{sec:models}).  Coarse-grained models derived from specific polymers are only touched upon briefly (Appendix~\ref{app:cg}), although this is an important current research topic.\cite{MullerPlathe:2002,JBETal:AdvPolySci2000}  

Second, we define a generic model as one consisting of monomers with the simplest imaginable structure.  They are identified with sites on a lattice or with Lennard-Jones spheres in the continuum.  The monomers in the chain are all the same and uncharged (``neutral homopolymers'').  Their interaction is either purely repulsive or consists of a short-range repulsion supplemented by an attractive potential at larger distances (see Sec.~\ref{sec:models}).  Thus, nor did we consider specific interactions, such as electrostatic interactions, H-bonds, interactions between different species of monomers (e.g.\ binary mixtures, block-copolymers), etc. --this will be done in other chapters of this book-- neither did we discuss current coarse-graining approaches which do not model a chain as a concatenation of monomers, but represent the whole polymer as a soft ellipsoidal\cite{MuratKremer:JCP1998,EurichMaass:JCP2001} or spherical\cite{LouisEtal:Physica2002} particle.

Within the scope of these generic models we presented various algorithms.  What is the upshot of this discussion for applications?  Here are some suggestions:
\begin{itemize}
\item Dilute solutions:
\begin{itemize}
\item For isolated chains, exempt of strong monomer-monomer interactions, the pivot algorithm is currently considered as being the most efficient method to study global properties related to the chain extension ($R_\text{g}$, $R_\text{e}$, the exponent $\nu$, etc.).  To study properties related to the partition function (e.g.\ the exponent $\gamma$) other algorithms are better suited (see footnote on page~\pageref{pg:gamma}).
\item To initialize the simulation via the pivot algorithm or for the analysis of local properties the slithering-snake algorithm represents an attractive alternative (see discussion on page~\pageref{pg:snake_for_pivot}).
\item If attractive monomer-monomer interactions are present, as it is the case close to the $\Theta$-point, it appears as if the pruned-enriched Rosenbluth method (PERM) is currently the most efficient algorithm (Sec.~\ref{subsub:PERM}).
\end{itemize}
\item From dilute solutions to dense melts:
\begin{itemize}
\item As the density of the solution increases, the large-scale pivot moves become quickly inefficient.  Here, either bilocal updating schemes, such as the slithering-snake algorithm, or non-local chain reconstructions via configurational-bias Monte Carlo (CBMC) are better suited.  CBMC has become a well established tool, particularly in its grand-canonical formulation for the study of phase diagrams.  It transpires that its range of applicability may be extended by the recently proposed recoil-growth (RG) algorithm (see Sec.~\ref{subsub:CBMC_RG}).  When the solution becomes more and more concentrated, the probability to renew the configuration of large chain portions becomes small.  In this limit, the CBMC and RG methods reduce to the slithering-snake algorithm.  
\item In dense melts consisting of long chain ($N \gtrsim 10^3$) also the slithering-snake algorithm faces serious problems to equilibrate the system (Sec.~\ref{subsubsec:snake}). Here, it is better to use connectivity-altering moves 
%(see Chap.~{\bf Zitat auf Doros-Theodorous-Vortrag}) 
instead of attempting to regrow (parts of) a chain.  An example is the recently proposed double-pivot algorithm which appears to be very efficient in equilibrating dense melts (see Sec.~\ref{subsubsec:pivot}).
\end{itemize}
\end{itemize}
The previous points only refer to the study of conformational and structural properties of polymer systems as well as to an efficient equilibration of the system. If the focus of interest are dynamic properties, local moves should be employed because they mimic best the physical dynamics (if solvent-mediated hydrodynamic interactions are absent; 
%see Chap.~{\bf Zitat auf Burkhard-D\"unwegs-Vortrag}).
see Ref.~\citen{duenweg2004}).    

The suggestions made above are a result of the discussion given in this chapter.  Certainly, our discussion suffers from omissions.  Personally, we feel that the most serious one are generalized ensemble techniques,\cite{Berg:CPC2002,BruceWilding:Preprint} such as simulated or parallel tempering (see Chap.~14.1 of Ref.~\citen{FrenkelSmitBook} for an introduction).  In the context of polymer physics, these methods have been applied e.g.\ to determine the chemical potential of polymer chains (simulated tempering)\cite{WildingMuller:JCP1994}, to accelerate the equilibration of dense polymer melts (parallel tempering)\cite{BunkerDuenweg:PRE2000} or to the simulation of phase transitions\cite{YanDePablo:JCP2000}.  

We apologize for this review-like end of our report, but hope that its content, together with the bibliography, will be helpful for those interested in MC simulations of polymer models.

%%%%%%%%%%%%%%%%%%%%%%%%%%%%%%%%%%%%%%%%%%%%%%%%%%%%%%%%%%%%%%%%%%%%%%%%
% Appendices
%%%%%%%%%%%%%%%%%%%%%%%%%%%%%%%%%%%%%%%%%%%%%%%%%%%%%%%%%%%%%%%%%%%%%%%%
%\appendix
\begin{appendix} 
\section{Realistic Models and Coarse-Graining of Real Polymers}
\label{app:cg}
%%%%%%%%%%%%%%%%%%%%%%%%%%%%%%%%%%%%%%%%%%%%%%%%%%%%%%%%%%%%%%%%%%%%%%%%
The presentation of this chapter concentrated on generic coarse-grained models (Sec.~\ref{sec:models}).  They are particularly useful for the study of the large-scale behavior of polymers.  For these generic models several simulation algorithms were discussed.  If one is now interested in properties of a specific molecule or material, the presented methods can still be applied.  But their efficiency must be tested for the particular application.\cite{LeontidisEtal:1995}  The simulation of real materials must use models which reflect the underlying chemical structure of the polymer, even at the coarse-grained level.  So, some mapping between a detailed, chemically realistic and a coarse-grained model must be established.\cite{MullerPlathe:2002,JBETal:AdvPolySci2000,MullerPlathe:2003}  The following paragraphs explain the need and the main steps of such a coarse-graining procedure and briefly describe an application to modeling a melt of poly(vinyl alcohol).  

\paragraph{From Realistic to Coarse-Grained Models.} 
An ideal simulation approach would consist in studying systems of long chains ($N \gtrsim 10^3$) with potentials being calculated from the simultaneous motion of all nuclei and electrons (via the Car-Parrinello method \cite{CarParrinello}).  However, already for an isolated chain such an approach is not feasible due to the large spread of subatomic and molecular time scales.\cite{Binder_MCMD1995,Kremer:MacroChemPhys2003}  For $N\!\sim \!10^3$ the relaxation time of a polymer in dilute solution is about $1\ \umu\text{s}$, whereas the inclusion of the electrons requires a time step of $\sim\!\!10^{-17}$ s in a MD simulation.  This disparity of 11 orders of magnitude cannot be covered in present day simulations.  A further restriction results from the system sizes that may be simulated.  Typically, the total number of particles (electrons and nuclei) is limited to $\sim\!\!10^3$.

Thus, simplifications are necessary.\footnote{Here, we assume that the whole chain is simulated at the same level of modeling.  Of course, it is also possible to treat some part of the chain in atomistic detail (e.g.\ via Car-Parrinello), whereas others are modeled on a coarse-grained level.  A recent example of such a multiscale approach is the selective adsorption of polycarbonate on a nickel surface.\cite{DelleSiteEtal:PRL2002}}  
There are several levels to this.\cite{MullerPlathe:2003}  In a first step, the ``on the fly'' calculated quantum-mechanical potentials can be replaced by empirical potentials for the bond lengths, the bond angles, the torsional angles and the non-bonded interactions between distant monomers along the backbone of the chain (``quantum level $\rightarrow$ atomistic level'').  After careful optimization of the parameters of these potentials quantitative agreement between simulation and experimental results can be obtained.\cite{MullerPlathe:2002,SmithEtal:ChemPhys2000}  If the system is to stay in thermal equilibrium on both the local monomeric scale and the global scale of the chain, such comparisons between simulation and experiments are limited to high temperatures up to now.  Extensions to low temperatures where crystallization or sluggish glass-like relaxation occur still represent a great challenge for simulations on atomistic scale.\footnote{An important contribution to this field are the ``amorphous cell'' simulations pioneered by Theodorou and Suter \cite{TheodorouSuter:Macro1985a,WeberHelfand:JCP1979}.  Here, the idea is to fold an infinitely long chain in the simulation box such that the monomer density is close to the experimental one and the resulting chain structure is reasonable.  With this approach mechanical properties of the amorphous, glassy melt have been studied successfully.}  To tackle these problems the models must be further simplified.  In a second step of simplification, fast degrees of freedom (bond length and bond angle vibrations, etc.) may therefore be eliminated by a coarse-graining procedure (``atomistic level $\rightarrow$ coarse-grained level'').  This procedure leads either to generic models, as those which were discussed in the previous sections, or to a coarse-grained model for a specific polymer (see Sec.~\ref{sec:models}).  Recent approaches to the latter case have been reviewed in Refs.~\citen{MullerPlathe:2002,JBETal:AdvPolySci2000,MullerPlathe:2003}. 

The first step, from the quantum to the atomistic level, may be interpreted as an ab-initio approach or a bottom-up construction of the model. When pushing the simplifications of the model beyond the atomistic to the coarse-grained level it becomes more and more apparent that the empirical potentials are just simulation parameters. So, instead of the bottom-up construction, one may also choose a more pragmatic top-down procedure.  That is, given some (measured) properties, what are the potentials which can reproduce them?~\footnote{One such method for the solution of the inverse problem, known as ``reverse Monte Carlo'', is actually used to interpret neutron and x-ray scattering data.\cite{McGreevy,Soper1996}}  In practice, one often proceeds as follows:  By the bottom-up method a first guess of coarse-grained potentials is constructed. In a second step, these potentials are optimized with respect to some experimentally or theoretically known properties.  With the final potentials one may then (try to) predict quantities which are not easily measurable.  In the following we will sketch one such pragmatic approach.

\begin{figure}[tb]
\begin{center}
\epsfxsize=12cm
\epsfbox{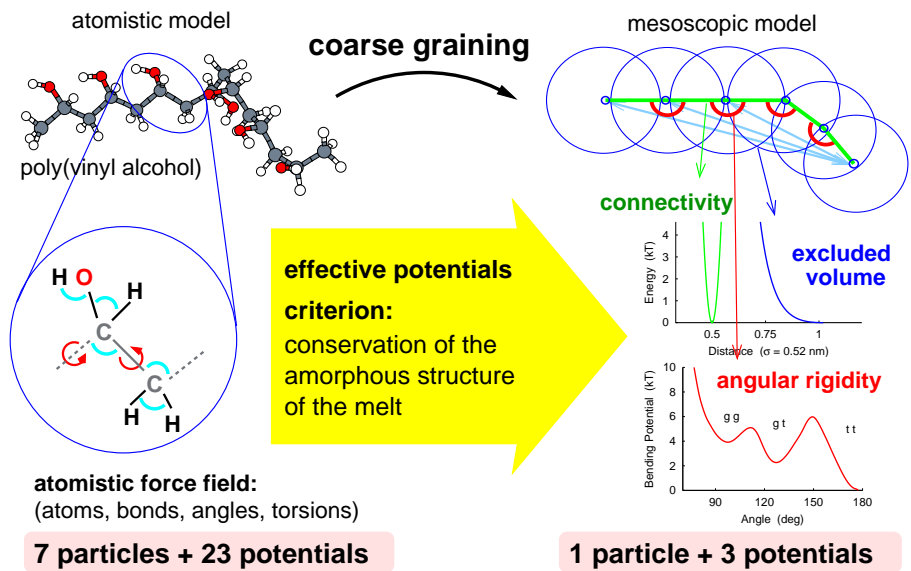}
\caption[]{Schematic illustration of the coarse-graining procedure developed in Ref.~\citen{ReithEtal:Macro2001}.  Starting from an atomistically detailed model of poly(vinyl alcohol) a bead-spring model is constructed.  A monomer of the atomistic model is represented by a sphere on the coarse-grained level.  The spheres interact by a harmonic bond potential and a purely repulsive LJ-like potential [Eq.~\eqref{eq:LJ12-6REP}].  The parameters of these potentials are optimized in the following way.  In an atomistic simulation at high temperature, the structure of a melt of poly(vinyl alcohol) and the local conformation of a polymer are determined: the structure by the pair-distribution function $g(r)$~\cite{HansenMcDonaldBook} and the conformation by the distribution of the angle between two vectors formed by connecting the C-atom of the CHOH-group of the $(i-1)$th monomer with that of the $i$th monomer (first vector) and that of the $i$th monomer with that of the $(i+1)$th monomer (second vector).  From the angular distribution a potential of mean force\cite{HansenMcDonaldBook} is calculated. In this case, it could directly be used as the bending potential for the bond angle in the coarse-grained model.  Contrary to that, the parameters of the LJ-potentials are iteratively optimized until $g(r)$ of the poly(vinyl alcohol) melt coincides with that of the coarse-grained model.}
\end{center}
\label{fig:pvaCG}
\end{figure}

\paragraph{An Example of How to Coarse-Grain Real Polymers.}  The idea and the realization of the coarse-graining are illustrated by the example of poly(vinyl alcohol) in Fig.~\ref{fig:pvaCG}.\cite{ReithEtal:Macro2001} 

On the coarse-grained level, the chemical structure of the monomers is suppressed.  They are represented by spheres, bound to each other by a harmonic potential.  The orientation of successive bonds along the chain is determined by a bond angle potential, whereas other non-bonded monomers interact by the repulsive part of a LJ-like potential
\begin{equation}
U_\text{LJ}^\text{rep}(r) = \left \{
\begin{array}{ll}
 \epsilon_n ({\sigma}/{r})^{n}- \epsilon_6 ({\sigma}/{r})^{6} + \epsilon_0
  & \quad \mbox{for} \; r \leq r_\text{min}\;, \\
0 & \quad \mbox{else} \;,
\label{eq:LJ12-6REP}
\end{array}
\right .
\end{equation}
where $r_\text{min}$ is the minimum of the potential and $\epsilon_0$ is chosen such that the minimum is shifted to zero to have a continuous force.  When fitting the potential to real polymers it turned out\cite{MeyerEtal:JCP2000} that for the repulsive term, an exponent smaller than 12 may be better adapted because the coarse-grained units have to be somewhat softer than atoms.\footnote{By doing more radical coarse-graining steps lumping many monomers\cite{PaddingBriels:JCP2002} or even whole chains into one particle\cite{LouisEtal:Physica2002}, one will end up with soft overlapping potentials, as used e.g.\ in ``Dissipative Particle Dynamics'' 
%(see Chap.~{\bf Zitat auf Burkhard-D\"unwegs-Vortrag}).}
(see Ref.~\citen{duenweg2004}). }   
By comparing high-temperature simulations of an atomistically detailed and a coarse-grained model, the potential parameters of the latter are adjusted in such a way that it faithfully reproduces the backbone rigidity of poly(vinyl alcohol) and the local packing of its monomers in the liquid state. 

The construction of the model was done at one state point. It is a priori not guaranteed how far from this reference conditions the model still matches the behavior of the original one. It turns out that in the present case, the crystallization at lower temperatures is well reproduced.\cite{MeyerMP:Macro2002}

%---------------------------------------------------------------------------
% Acknowledgments
\section*{Acknowledgments}

The results reported here, as far as it concerns our own work, were obtained in fruitful collaboration with M. Aichele, J.-L. Barrat, O. Biermann, K. Binder, R. Faller, B. Lobe, E. Luitjen, M. M\"uller, F. M\"uller-Plathe, L. Mattioni, W. Paul, D. Reith, and M. Wolfgardt.  The authors' research is supported in part by the ESF Supernet programme and by the LEA ``Macromolecules in nanoscopically structured media''.  J.B. is indebted to the IUF for finanical support.  Furthermore, we gratefully acknowledge generous grants of computing time by the IDRIS (Orsay, France).

\end{appendix} 

%---------------------------------------------------------------------------
% References

\epsfclipoff
%---------------------------------------------------------------------------
% End of text
%---------------------------------------------------------------------------
\end{document}